\documentclass[11pt]{article}
\usepackage{slashed,verbatim,subfigure}
\usepackage[numbers,sort&compress]{natbib}
\usepackage{amsmath}
\usepackage{amssymb}
\usepackage{appendix}
\usepackage{graphicx}
\usepackage{dcolumn}
\usepackage{bm}
\usepackage[colorlinks=true,linktocpage=true,
linkcolor=blue,citecolor=blue]{hyperref}
\usepackage[a4paper]{geometry}
\catcode`@11
\def\seceqaa{\@addtoreset{equation}{section}
	\def\theequation{A\arabic{equation}}}
\def\seceqbb{\@addtoreset{equation}{section}
	\def\theequation{B\arabic{equation}}}
\def\seceqcc{\@addtoreset{equation}{section}
	\def\theequation{C\arabic{equation}}}
\def\seceqdd{\@addtoreset{equation}{section}
	\def\theequation{D\arabic{equation}}}
\def\seceqee{\@addtoreset{equation}{section}
	\def\theequation{E\arabic{equation}}}
\catcode`@11
\newcommand{\be}{\begin{eqnarray}}
\newcommand{\ee}{\end{eqnarray}}
\begin{document}
\large
\title{\bf{ M-Theory Exotic Scalar Glueball Decays to  Mesons at Finite Coupling}}
\author{Vikas Yadav\footnote{email- viitr.dph2015@iitr.ac.in}~~and~~Aalok Misra\footnote{email-aalokfph@iitr.ac.in}\vspace{0.1in}\\
Department of Physics,\\
Indian Institute of Technology Roorkee, Roorkee 247667, India}
\date{}
\maketitle
\begin{abstract}
Using the pull-back of the perturbed type IIA metric corresponding to the perturbation of \cite{MQGP}'s M-theory uplift of \cite{metrics}'s UV-complete top-down type IIB holographic dual of large-$N$ thermal QCD, at finite coupling, we obtain the interaction Lagrangian corresponding to exotic scalar glueball($G_E$)-$\rho/\pi$-meson interaction, linear in the exotic scalar glueball and up to quartic order in the $\pi$ mesons. In the Lagrangian the coupling constants are determined as (radial integrals of) \cite{MQGP}'s M-theory uplift's metric components and six radial functions appearing in the M-theory metric perturbations. Assuming $M_G>2M_\rho$, we then compute $\rho\rightarrow2\pi, G_E\rightarrow2\pi, 2\rho, \rho+2\pi$ decay widths as well as the direct and indirect (mediated via $\rho$ mesons) $G_E\rightarrow4\pi$ decays. For numerics, we choose $f0[1710]$ and compare with previous calculations. We emphasize that our results can be made to match PDG data (and improvements thereof) exactly by appropriate tuning of some constants of integration appearing in the solution of the M-theory metric perturbations and the $\rho$ and $\pi$ meson radial profile functions - a flexibility that our calculations permits.
\end{abstract}

\newpage
\tableofcontents

\newpage

\section{Introduction}
The non-abelian nature of QCD makes it possible to form color-neutral bound states of gauge bosons known as glueballs (gg, ggg, etc). In pure Yang-Mills theory these are the only possible particle states. Glueballs are represented by quantum numbers $J^{PC}$, where J denotes total angular momentum, P denotes parity,
 and C denotes charge conjugation. Their spectrum has been studied in detail in lattice gauge theory. Despite the theoretical proof of existence of glueballs their experimental identification
remains difficult. This  difficulty in the identification arises mainly becasue of lack of information about coupling of glueballs with quark-antiquark states in strongly coupled QCD. Lattice
simulation of QCD provides a reliable means of studying
the glueballs, but lattice simulation of QCD with dynamical quarks are notoriously difficult. Lattice QCD predicts the mass of the lightest scalar glueball to be around 1600-1800 MeV.

In this paper we have obtained the decay width for `exotic' scalar glueball by explicitly computing the couplings between scalar glueballs and mesons by using \cite{MQGP}'s M-theory uplift of \cite{metrics}'s type IIB holographic dual of large-$N$ thermal QCD at finite gauge coupling. In the past few decades
AdS/CFT correspondence \cite{maldacena} and its generalization - gauge/gravity duality - has been used extensively to study non-supersymmetric gauge field theories. The AdS/CFT(gauge-gravity duality) establishes a map between
correlation functions of gauge invariant composite operators with large $N_c$ and large 't Hooft coupling to perturbations of certain backgrounds in classical(super-)gravity. Gauge/gravity duality has been used to compute glueball and meson spectra in large $N_c$ QCD.

In the past decade, (glueballs and) mesons have been studied extensively to gain new insights into the non-perturbative regime of QCD. Various holographic setups such as soft-wall model, hard wall model, modified soft wall model, etc. amongst the bottom-up approaches and the Sakai-Sugimoto model, have been used to obtain the glueball and meson spectra\cite{Sakai-Sugimoto-1} and study interaction between them. Let us very briefly summarize the recent work done by our group in this context.
In \cite{Sil_Yadav_Misra-EPJC-17}, we initiated a top-down $G$-structure holographic large-$N$ thermal QCD phenomenology at {\it finite} gauge coupling and {\it finite} number of colors, in particular from the vantage point of the M theory uplift of the delocalized SYZ type IIA mirror of the top-down UV complete holographic dual of large-$N$ thermal QCD of \cite{metrics}, as constructed in \cite{MQGP}. We calculated up to (N)LO in $N$, masses of $0^{++}, 0^{--}, 0^{-+}, 1^{++}$ and $2^{++}$ glueballs, and found very good agreement with some of the lattice results on the same. In \cite{Yadav_Misra-Sil-EPJC-17} we continued  by evaluating the spectra of (pseudo-)vector and (pseudo-)scalar mesons and compared our results with \cite{Sakai-Sugimoto-1}, \cite{Dasgupta_et_al_Mesons} and \cite{PDG}. In this paper, we look at two-, three- and four-body (`exotic' scalar) glueball decays into $\rho$ and $\pi$ mesons, and show that our results permit obtaining an exact match with PDG data including any improvements upon the same expected to be obtained in the future. The same in the context of the type IIA Sakai-Sugimoto model were first considered in \cite{Hashimoto-glueball}, and more recently in \cite{Brunner_Hashimoto-results}. To our knowledge, a top-down study of holographic glueball-to-meson decays in the context of the M-theory uplift of a UV-complete type IIB holographic dual at finite coupling, was entirely missing in the literature. This paper fills in this gap.

The remainder of the paper is organized as follows.  In Section {\bf 2}, we give a brief introduction to a variety of topics like \cite{metrics}'s type IIB holographic dual of large-$N$ thermal QCD, its SYZ type IIA mirror and its subsequent M-theory uplift as worked out in \cite{MQGP}, $SU(3)$-structure of type IIB/A and $G_2$ structure of M theory uplift and a discussion on why in the MQGP limit the gauge theory is essentially 2+1 dimensional with gluonic bound states (glueballs) and the lightest vector and pseudo-scalar mesons. Section {\bf 3} is on obtaining the EOMs and their solutions for the six scalar functions relevant to exotic $0^{++}$ glueball M theory metric perturbations. In Section {\bf 4} via two sub-sections, mesons arising from the Kaluza-Klein reduction of gauge fields on the world volume of the flavor $D6$-branes, and in particular their radial profile functions appearing in the same, are discussed. Section {\bf 5} is devoted to obtaining the exotic scalar glueball-meson interaction Lagrangian up to linear order in the glueball and up to quartic order in the meson fields. In Section {\bf 6}, decay widths corresponding to $G_E\rightarrow2\pi, 2\rho, \rho+2\pi, 4\pi^0, 2\pi^a+2\pi^b$ as well as $\rho\rightarrow2\pi$ are obtained. Finally, Section {\bf 7} has a discussion on the results obtained. There are two appendices to supplement the main text: appendix {\bf A} gives the metric components of the M-theory uplift of \cite{MQGP} near $\theta_1\sim N^{-\frac{1}{5}}, \theta_2\sim N^{-\frac{3}{10}}$ and appendix {\bf B} is the potential appearing when the $\rho$-meson's radial profile function's equation of motion is rewritten as a Schr\"{o}dinger-like equation in the MQGP limit.

\section{Background: Large-$N$ Thermal QCD at Finite Gauge Coupling from M-Theory}

In this section, we will provide a lightning review of the type IIB background of \cite{metrics} - a UV complete holographic dual of large-$N$ thermal QCD -
discuss the 'MQGP' limit of \cite{MQGP} along with the motivation for considering this limit,
 issues as discussed in \cite{MQGP} pertaining to construction of delocalized S(trominger) Y(au) Z(aslow) mirror and approximate supersymmetry along with (an appendix-supplemented) discussion on the SYZ mirror in fact being independent of angular delocalization,
construction explicit $SU(3)$ and $G_2$ structures respectively of type IIB/IIA and M-theory uplift as constructed for the first time in \cite{NPB}, \cite{EPJC-2}.


Let us start with the UV-complete holographic dual of large-$N$ thermal QCD as constructed in  Dasgupta-Mia et al \cite{metrics}.
To include fundamental quarks at non-zero temperature in the context of type IIB string theory, the authors of \cite{metrics} considered  $N$ $D3$-branes placed at the tip of six-dimensional conifold, $M\ D5$-branes wrapping the vanishing $S^2$ and $M\ \overline{D5}$-branes  distributed along the resolved $S^2$ placed at anti-podal points relative to the $M$ $D5$-branes. Let us denote the average $D5/\overline{D5}$ separation  by ${\cal R}_{D5/\overline{D5}}$. On the gravity side, the domain of the radial coordinate, in \cite{metrics}, is divided into the IR, the IR-UV interpolating region and the UV with the $\overline{D5}$-branes placed at the outer boundary  of the IR-UV interpolating region/inner boundary of the UV region. Roughly, $r>{\cal R}_{D5/\overline{D5}}$, would be the UV.    The $N_f\ D7$-branes are holomorphically embedded via Ouyang embedding in the resolved conifold geometry in the brane construction. They are present  in the UV, the IR-UV interpolating region and they dip into the (confining) IR (but do not touch the $D3$-branes with the shortest $D3-D7$ string corresponding to the lightest quark). In addtion, $N_f\ \overline{D7}$-branes are present in the UV and the UV-IR interpolating region. This brane construct ensures UV conformality and chiral symmetry breaking in the IR. Let us understand this in some more detail.  In the UV, one has $SU(N+M)\times SU(N+M)$ color gauge group and $SU(N_f)\times SU(N_f)$ flavor gauge group. There occurs a partial Higgsing of $SU(N+M)\times SU(N+M)$ to $SU(N+M)\times SU(N)$ as one goes from $r>{\cal R}_{D5/\overline{D5}}$  to $r<{\cal R}_{D5/\overline{D5}}$ \cite{K. Dasgupta et al [2012]}. The reason is that in the IR, the $\overline{D5}$-branes are integrated out resulting in the reduction of the rank of one of the product gauge groups (which is $SU(N + {\rm number\ of}\ D5-{\rm branes})\times SU(N + {\rm number\ of}\ \overline{D5}-{\rm branes})$; the number of $\overline{D5}$-branes drops off in the IR to zero). By the same token, the $\overline{D5}$-branes are integrated in the UV resulting in the conformal Klebanov-Witten-like $SU(M+N)\times SU(M+N)$ color gauge group \cite{KW}.  The two gauge couplings, $g_{SU(N+M)}$ and $g_{SU(N)}$, were shown in \cite{KS} to flow  logarithmically  and oppositely via: $4\pi^2\left(\frac{1}{g_{SU(N+M)}^2} + \frac{1}{g_{SU(N)}^2}\right)e^\phi \sim \pi;\
 4\pi^2\left(\frac{1}{g_{SU(N+M)}^2} - \frac{1}{g_{SU(N)}^2}\right)e^\phi \sim \frac{1}{2\pi\alpha^\prime}\int_{S^2}B_2$. One thus sees that $\int_{S^2}B_2$, in the UV, is the obstruction to obtaining conformality which is why $M$ $\overline{D5}$-branes were included in \cite{metrics} to cancel the net $D5$-brane charge in the UV. Further, the $N_f$ flavor $D7$-branes which appear in the dilaton profile, enter the RG flow of the gauge couplings. This therefore needs to be annulled by $N_f\ \overline{D7}$-branes which is the reason for their inclusion in the UV in \cite{metrics}. The RG flow equations for the gauge coupling $g_{SU(N+M)}$ - corresponding to the gauge group of a relatively higher rank - can be used to show that the same flows towards strong coupling, and the $SU(N)$ gauge coupling flows towards weak coupling. One can show that the strongly coupled
  $SU(N+M)$ is Seiberg dual to weakly coupled  $SU(N-(M - N_f))$; the addition of the flavor branes hence decelerates the reduction in the rank of the gauge group under Seiberg duality.  One then performs a Seiberg duality cascade such that $N$ decreases to 0 but there is a finite $M$ left at the end. One will thus be left with an $SU(M)$ gauge theory with $N_f$ flavors which confines in the IR. It was the finite temperature version of this $SU(M)$ gauge theory that was looked at by the authors of \cite{metrics}. So, at the end of the duality cascade in the IR, number of colors $N_c$ is identified with $M$, which in the `MQGP limit' can be tuned to equal 3. The number of colors $N_c = N_{\rm eff}(r) + M_{\rm eff}(r)$, where $N_{\rm eff}(r) = \int_{\rm Base\ of\ Resolved\ Warped\ Deformed\ Conifold}F_5$ and $M_{\rm eff} = \int_{S^3}\tilde{F}_3$ (the $S^3$ being dual to $\ e_\psi\wedge\left(\sin\theta_1 d\theta_1\wedge d\phi_1 - B_1\sin\theta_2\wedge d\phi_2\right)$, wherein $B_1$ is an asymmetry factor defined in \cite{metrics}, and $e_\psi\equiv d\psi + \cos ~\theta_1~d\phi_1 + \cos ~\theta_2~d\phi_2$) where $\tilde{F}_3 (\equiv F_3 - \tau H_3)\propto M(r)\equiv 1 - \frac{e^{\alpha(r-{\cal R}_{D5/\overline{D5}})}}{1 + e^{\alpha(r-{\cal R}_{D5/\overline{D5}})}}, \alpha\gg1$  \cite{IR-UV-desc_Dasgupta_etal}.  Further, the flavor group $SU(N_f)\times SU(N_f)$, is broken in the IR to $SU(N_f)$ because the IR has only $N_f$ $D7$-branes.  The gravity dual is given by a  resolved warped deformed conifold wherein the $D3$-branes and the $D5$-branes are replaced by fluxes in the IR, and the back-reactions are included in the warp factor and fluxes.

It was argued in \cite{NPB} that  the length scale on the gravity side in the IR will be given by:
\begin{eqnarray}
\label{length-IR}
& & L\sim\sqrt{M}N_f^{\frac{3}{4}}\sqrt{\left(\sum_{m\geq0}\sum_{n\geq0}N_f^mM^nf_{mn}(\Lambda)\right)}\left(\sum_{l\geq0}\sum_{p\geq0}N_f^lM^p g_{lp}(\Lambda)\right)^{\frac{1}{4}}g_s^{\frac{1}{4}}\sqrt{\alpha^\prime}\nonumber\\
& & \equiv N_f^{\frac{3}{4}}\sqrt{\left(\sum_{m\geq0}\sum_{n\geq0}N_f^mM^nf_{mn}(\Lambda)\right)}\left(\sum_{l\geq0}\sum_{p\geq0}N_f^lM^p g_{lp}(\Lambda)\right)^{\frac{1}{4}} L_{\rm KS},
\end{eqnarray}
which implies that  in the IR, relative to KS, there is a color-flavor enhancement of the length scale. Hence,  in the IR, even for $N_c^{\rm IR}=M=3$ and $N_f=2$ (light flavors) upon inclusion of higher order terms in $M$ and $N_f$, $L\gg L_{\rm KS}(\sim L_{\rm Planck})$ in the MQGP limit involving $g_s\stackrel{\sim}{<}1$, implying that the stringy corrections are suppressed and one can trust supergravity calculations.

Hence, the type IIB model of \cite{metrics} make it an ideal holographic dual of thermal QCD because: (i) it is UV conformal (Landau poles are absent), (ii) it is IR confining with required chiral symmetry breaking in the IR, (iii) the quarks transform in the fundamental representation of flavor and color groups, and (iv) it  is defined for the full range of temperature - both low and high.



 In \cite{MQGP}, the authors considered the following limit:
\begin{eqnarray}
\label{limits_Dasguptaetal-ii}
& & \hskip -0.17in  {\rm MQGP\ limit}: \frac{g_sM^2}{N}\ll  1, g_sN\gg1, {\rm finite}\
 g_s, M.
\end{eqnarray}
The motivation for considering the MQGP limit which was discussed in detail in \cite{NPB}, is summarized now. The usual AdS/CFT limit involves $g_{\rm YM}\rightarrow0, N\rightarrow\infty$ such that the 't Hooft coupling $g_{\rm YM}^2N$ is very large. However, for strongly coupled thermal systems like sQGP, this limit is not relevant as it is expected that $g_{\rm YM}$ is finite, and $N_c=3$ \cite{Natsuume}. From the discussion in the paragraph preceding (\ref{length-IR}), one recollects that at the end of the Seiberg duality cascade in the IR, $N_c=M$. Note that in the MQGP limit (\ref{limits_Dasguptaetal-ii}), $M$  can be set to equal 3. Further, in the MQGP limit,  $g_s\stackrel{<}{\sim}1$. The finiteness of $g_s$ requires one to construct the M theory uplift of \cite{metrics}. These were precisely the reasons for coining `MQGP limit' in \cite{MQGP}. In fact this was the reason why the type IIA mirror was first constructed in \cite{MQGP} a la delocalized Strominger-Yau-Zaslow mirror symmetry, and then its M-theory uplift obtained in the same paper.



In order to be able to implement quantum mirror symmetry a la SYZ \cite{syz}, one needs a special Lagrangian (sLag) $T^3$ fibered over a large base. Defining delocalized T-duality/local $T^3(x,y,z)$ coordinates \cite{MQGP}:
\begin{equation}
\label{xyz defs}
x = \sqrt{h_2}h^{\frac{1}{4}}sin\langle\theta_1\rangle\langle r\rangle \phi_1,\ y = \sqrt{h_4}h^{\frac{1}{4}}sin\langle\theta_2\rangle\langle r\rangle \phi_2,\ z=\sqrt{h_1}\langle r\rangle h^{\frac{1}{4}}\psi,
\end{equation}
 it was shown in \cite{transport-coefficients,EPJC-2} that the aforementioned $T^3$ is the $T^2$-invariant sLag of \cite{M.Ionel and M.Min-OO (2008)} for a deformed/resolved conifold.   Hence,  the local $T^3$ of (\ref{xyz defs}) is the  sLag needed to effect the construction of the SYZ mirror.

In the `delocalized limit' \cite{M. Becker et al [2004]}  $\psi=\langle\psi\rangle$, under the transformation:
\begin{equation}
\label{transformation_psi}
\left(\begin{array}{c} \sin\theta_2 d\phi_2 \\ d\theta_2\end{array} \right)\rightarrow \left(\begin{array}{cc} \cos \langle\psi\rangle  & \sin \langle\psi\rangle \\
- \sin \langle\psi\rangle & \cos \langle\psi\rangle
\end{array}\right)\left(\begin{array}{c}
\sin\theta_2 d\phi_2\\
d\theta_2
\end{array}
\right),
\end{equation}
and an appropriate shift in $\psi$, it was shown in \cite{MQGP} that one introduces a local isometry along $\psi$ in the resolved warped deformed conifold in the gravity dual in \cite{metrics}; of course this is not true globally.

 Now, to be able to construct the SYZ mirror, one also needs to ensure a large base  of the $T^3(x,y,z)$ fibration. This is effected via:
\cite{F. Chen et al [2010]}:
\begin{eqnarray}
\label{SYZ-large base}
& & d\psi\rightarrow d\psi + f_1(\theta_1)\cos\theta_1 d\theta_1 + f_2(\theta_2)\cos\theta_2d\theta_2,\nonumber\\
& & d\phi_{1,2}\rightarrow d\phi_{1,2} - f_{1,2}(\theta_{1,2})d\theta_{1,2},
\end{eqnarray}
for appropriately chosen large values of $f_{1,2}(\theta_{1,2})$.  The guiding priciple behind choosing such large values of $f_{1,2}(\theta_{1,2})$, as given in \cite{MQGP}, is that one requires the metric obtained after SYZ-mirror transformation applied to the non-K\"{a}hler  resolved warped deformed conifold to be like like a non-K\"{a}hler warped resolved conifold at least locally.
 This was explicitly demonstrated in \cite{NPB} and appropriate values of $f_{1,2}(\theta_{1,2})$ obtained therein.

The aforementioned delocalization procedure used to construct the type IIA mirror of the UV-complete \cite{metrics}'s type IIB holographic dual of large-$N$ thermal QCD  {\it a la} SYZ triple-T-duality prescription  and its  M-theory uplift as worked out in \cite{MQGP}, is in fact, {\it not} restricted to fixed-$\psi$ mirrors. To understand this, let us look at the example of the mirror of a $D5$-brane wrapping the resolved $S^2$ with fluxes as studied in \cite{SYZ-free-of-delocalization} - in paricular sections 5 and 6 therein. In the large-complex structure limit and after a fixed-$\psi$ coordinate rotation, the SYZ mirror was found in \cite{SYZ-free-of-delocalization} to be $D6$-brane wrapping a non-K\"ahler deformed conifold. As shown in (section 6 of) \cite{SYZ-free-of-delocalization},  an explicit $G_2$ structure can be constructed in terms of which the M-theory uplift of the type IIA mirror could be rewritten, which is valid $\forall\psi$.  Hence, the type IIA mirror in Sec. {\bf 6} of \cite{SYZ-free-of-delocalization} obtained from arbitrary-$\psi$ M theory metric, will be the same as the fixed-$\psi$ type IIA mirror of Sec. {\bf 5} of \cite{SYZ-free-of-delocalization} that was obtained using delocalization. Thus, the fixed $\psi$ value chosen to effect the abovementioned delocalized SYZ mirror, could simply be replaced by an arbitrary $\psi$, implying  the type IIA mirror is effectively free of delocalization. 

Let us understand what SYZ mirror transformation via triple T-duality does to the brane construct.
A single T-duality along a direction orthogonal to the $D3$-branes world volume, e.g., $z$ of $T^3(x,y,z)$, yields $D4$ branes that are straddling a pair of $NS5$-branes with world-volume coordinates, let us say, denoted by $(\theta_1,x)$ and $(\theta_2,y)$. A second T-duality along $x$ and a third T-duality along $y$ would yield a Taub-NUT space  from each of the two $NS5$-branes \cite{T-dual-NS5-Taub-NUT-Tong}. The $D7$-branes yield $D6$-branes which get uplifted to Kaluza-Klein monopoles in M-theory \cite{KK-monopoles-A-Sen} which also involve Taub-NUT spaces. Globally, one expects the eleven-dimensional uplift would involve a seven-fold of $G_2$-structure, analogous to the uplift of $D5$-branes wrapping a two-cycle in a resolved warped conifold \cite{Dasguptaetal_G2_structure}.


We will now briefly review  $G=SU(3), G_2$-structures of the holographic type IIB dual of \cite{metrics}, its delocalized type IIA SYZ mirror and its M-theory uplift constructed in \cite{MQGP}. In \cite{transport-coefficients}, it was shown that the five $SU(3)$ structure torsion classes, in the MQGP limit, were given by (schematically):
\begin{eqnarray}
\label{T-IIB-i}
& & T_{SU(3)}^{\rm IIB}\in W_1 \oplus W_2 \oplus W_3 \oplus W_4 \oplus W_5 \sim \frac{e^{-3\tau}}{\sqrt{g_s N}} \oplus \left(g_s N\right)^{\frac{1}{4}} e^{-3\tau}\oplus \sqrt{g_s N}e^{-3\tau}\oplus -\frac{2}{3} \oplus -\frac{1}{2},
\end{eqnarray}
wherein $(r\sim e^{\frac{\tau}{3}})$, such that:
 \begin{equation}
 \label{T-IIB-ii}
 \frac{2}{3}W^{\bar{3}}_5=W^{\bar{3}}_4
 \end{equation}
  in the UV-IR interpolating region/UV, implying a Klebanov-Strassler-like supersymmetry
  \cite{Butti et al [2004]}. Locally, around $\theta_1\sim\frac{1}{N^{\frac{1}{5}}}, \theta_2\sim\frac{1}{N^{\frac{3}{10}}}$, the type IIA torsion classes of the delocalized SYZ type IIA mirror metric, were worked out in \cite{NPB} to be:
 \begin{eqnarray}
 \label{T-IIA}
 T_{SU(3)}^{\rm IIA} &\in& W_2 \oplus W_3 \oplus W_4 \oplus W_5 \sim \gamma_2g_s^{-\frac{1}{4}} N^{\frac{3}{10}} \oplus g_s^{-\frac{1}{4}}N^{-\frac{1}{20}} \oplus g_s^{-\frac{1}{4}} N^{\frac{3}{10}} \oplus g_s^{-\frac{1}{4}} N^{\frac{3}{10}}\approx \gamma W_2\oplus W_4\oplus W_5\nonumber\\
 & & \stackrel{\scriptsize\rm fine\ tuning:\gamma\approx0}{\longrightarrow}\approx W_4\oplus W_5.
 \end{eqnarray}
   Further,
   \begin{equation}
   W_4\sim \Re e W_5,
   \end{equation}
    indicative of supersymmetry after constructing the delocalized SYZ mirror.

Apart from quantifying the departure from $SU(3)$ holonomy due to intrinsic contorsion arising from the NS-NS three-form $H$, via the evaluation of the $SU(3)$ structure torsion classes, to our knowledge for the first time in the context of holographic thermal QCD {\bf at finite gauge coupling} in \cite{NPB}: \\
(i) the existence of approximate supersymmetry of the type IIB holographic dual of \cite{metrics} in the MQGP limit near the coordinate branch $\theta_1=\theta_2=0$ was explicitly shown, which apart from the existence of a special Lagrangian three-cycle,  is essential for construction of the local SYZ type IIA mirror;\\
 (ii)   it was shown that the large-$N$ suppression of the deviation of the type IIB resolved warped deformed conifold from being a complex manifold, is lost on being duality-chased to type IIA, and that a fine tuning   in $W_2^{\rm IIA}$ can ensure that the local type IIA mirror is complex;\\
(iii)  for the local type IIA $SU(3)$ mirror,  the possibility of surviving approximate supersymmetry was explicitly shown which is essential as SYZ mirror is supersymmetric.

We can get a one-form type IIA potential from the triple T-dual (along $x, y, z$) of the type IIB $F_{1,3,5}$ in \cite{MQGP} and using which the following $D=11$ metric was obtained in \cite{MQGP} ($u\equiv\frac{r_h}{r}$):
\begin{eqnarray}
\label{Mtheory met}
& &\hskip -0.6in   ds^2_{11} = e^{-\frac{2\phi^{IIA}}{3}} \left[g_{tt}dt^2 + g_{\mathbb{R}^3}\left(dx^2 + dy^2 + dZ^2\right) +  g_{uu}du^2  +   ds^2_{IIA}({\theta_{1,2},\phi_{1,2},\psi})\right] \nonumber\\
& & \hskip -0.6in+ e^{\frac{4{\phi}^{IIA}}{3}}\Bigl(dx_{11} + A^{F_1}+A^{F_3}+A^{F_5}\Bigr)^2.
\end{eqnarray}

The torsion tensor associated with the $G_2$ structure of a seven fold, possesses 49 components and  can be split into torsion components
as:
\begin{equation}
T=T _{1}g+T _{7}\lrcorner \varphi +T _{14}+T _{27}
\label{torsioncomps}
\end{equation}
where $T _{1}$ is a function and gives the $\mathbf{1}$ component of $T$. We also have $T _{7}$, which is a $1$-form and hence gives the $\mathbf{7}$ component, and, $T _{14}\in \Lambda _{14}^{2}$ gives the $\mathbf{14}$
component. Further, $T _{27}$ is traceless symmetric, and gives the $\mathbf{27}$
component. Writing $T_i$ as $W_i$, we can split $W$ as
\begin{equation}
W=W_{1}\oplus W_{7}\oplus W_{14}\oplus W_{27}.  \label{Wsplit}
\end{equation}
From \cite{G2-Structure}, we see that a $G_2$ structure can be defined as:
\begin{equation}
\label{G_2_i}
\varphi_0 = \frac{1}{3!}{f}_{ABC}e^{ABC} = e^{-\phi^{IIA}}{f}_{abc}e^{abc} + e^{-\frac{2\phi^{IIA}}{3}}J\wedge e^{x_{10}},
\end{equation}
where $A,B,C=1,...,6,10; a,b,c,=1,...,6$, and ${f}_{ABC}$ are the structure constants of the imaginary octonions.
Using the same, the $G_2$-structure torsion classes were worked out in \cite{NPB} around $\theta_1\sim\frac{1}{N^{\frac{1}{5}}}, \theta_2\sim\frac{1}{N^{\frac{3}{10}}}$ (schematically):
  \begin{equation}
  \label{G2}
  T_{G_2}\in W_2^{14} \oplus W_3^{27} \sim \frac{1}{\left(g_sN\right)^{\frac{1}{4}} }\oplus \frac{1}{\left(g_sN\right)^{\frac{1}{4}}}.
  \end{equation}
   Hence, the approach of the seven-fold, locally, to having a $G_2$ holonomy ($W_1^{G_2}=W_2^{G_2}=W_3^{G_2}=W_4^{G_2}=0$)  is accelerated in the MQGP limit.

As stated earlier, the global uplift to M-theory of the type $IIB$ background of \cite{metrics} is expected to involve a seven-fold of $G_2$ structure (not $G_2$-holonomy due to non-zero M theory four-form fluxes). It is therefore extremely important to be able to see this, at least locally. It is in this sense that the results of \cite{MQGP} are of great significance as one explicitly sees in the context of  holographic thermal QCD {\bf at finite gauge coupling},  though locally, the aforementioned $G_2$ structure having worked out the non-trivial $G_2$-structure torsion classes.

Let us now argue that in the MQGP limit,  apart from the gluon-bound states, i.e. glueballs, and the light ($\rho/\pi$) mesons, all other scalar mesons are integrated out. As per \cite{witten}, supersymmetry can be broken by imposing anti-periodic boundary conditions for fermions along the $x^0$-circle (which at finite temperature has periodicity given by the reciprocal of the Hawking temperature). This is expected to generate fermionic masses of the order of the reciprocal of the $S^1_{t}$ radius $R_{r_h}$ and scalar masses of the order of $\frac{g_s^2N}{R_{r_h}}$. We will now argue that $R_{r_h}$ is very small implying scalar mesons (apart from the lightest $\rho$-vector and pionic pseudo-scalar mesons) are very heavy and are hence integrated out, and effectively the 3+1-dimensional QCD-like theory thus reduces to 2+1 dimensions. From (\ref{M-th components}), one sees that working with a near-horizon coordinate $\chi: r = r_h + \chi, \frac{\chi}{r_h}\ll1$, $G^M_{rr}dr^2 = \frac{d\chi^2}{\chi}F_{rr}(r_h) \equiv d\rho^2$ or
$\chi = \frac{\rho^2}{4F_{rr}(r_h)}$. Thus:
\begin{equation}
\label{radius-x0}
-G^M_{tt}dt^2 + G^M_{rr}dr^2 = -G^M_{tt}\ ^\prime(r_h)\frac{\rho^2}{4F_{rr}(r_h)}dt^2 + d\rho^2 \equiv - 4\pi^2 R_{r_h}^2 dt^2 + d\rho^2.
\end{equation}
We therefore read off the radius of the temporal direction:
\begin{eqnarray}
\label{radius-t}
& & R_{r_h} = \sqrt{\frac{1+9b^2}{1+6b^2}}\frac{r_h}{\pi L^2}\rho \sim T\rho\sim T \sqrt{\chi}\sqrt{F_{rr}(r_h)}\sim\sqrt{\chi}T^{\frac{3}{2}}.
\end{eqnarray}
One hence sees that $R_{r_h}$ is very small and hence the assertion.

\section{Glueballs from M-theory metric perturbations}

To start off our study of glueball decays into meson, we first need to understand how glueballs are obtained in the M-theory background. Glueballs are gauge invariant composite states in the Yang-Mills theory and their duals corresponds to the supergravity fluctuations in the near horizon geometry of brane solutions. The M-theory metric for D=11 was written out in (\ref{Mtheory met}). Here $g^{IIA}_{MN}$ and $\phi^{IIA}$ corresponds to the metric components and dilaton in type IIA string background respectively; $A$'s are the one form potential in type IIA background. The M-theory metric components up to NLO in $N$ near $\theta_{1}=\alpha_{\theta_{1}}N^{-\frac{1}{5}}$, $\theta_{2}=\alpha_{\theta_{2}}N^{-\frac{3}{10}}$, $\phi_{1,2}=0/2\pi$, whereat an explicit $G_2$ structure was worked out in \cite{NPB}, are given in (\ref{M-th components}).
 The general M-theory metric fluctuations corresponding to `exotic' scalar glueball  with $J^{PC}=0^{++}$ in terms of the three dimensional spacetime $x^{1},x^{2},x^{3}$ can be written following \cite{Constable_Myers},\cite{Hashimoto-glueball} as:
\begin{eqnarray}
\label{M-th perturbation}
h_{tt}&=&-q_{1}(r)G^{M}_{tt}G_{E}(x^{1},x^{2},x^{3})\nonumber\\
h_{rr}&=&-q_{2}(r)G^{M}_{rr}G_{E}(x^{1},x^{2},x^{3})\nonumber\\
h_{ra}&=&q_{3}(r)G^{M}_{aa}\frac{\partial_{a}G_{E}(x^{1},x^{2},x^{3})}{M_g^2}, \ \ a=1,2,3\nonumber\\
h_{ab}&=&G^{M}_{ab}\Bigg(q_{4}(r)\eta_{ab}-q_{5}(r)\frac{\partial_{a}\partial_{b}}{M_g^2}\Bigg)G_{E}(x^{1},x^{2},x^{3}),\ \ b=1,2,3\nonumber\\
h_{11,11}&=&q_{6}(r)G^{M}_{11,11}G_{E}(x^{1}x^{2}x^{3})\nonumber\\
\end{eqnarray}
Here $G_E(x^{1},x^{2},x^{3})$ is the glueball field in the 2+1 dimensional spacetime and, $M_g$ is the mass of the glueball. The explicit expression for functions $q_{i={1,2,...,6}}$ can be obtained by solving their EOM's obtained from 11-D action. The 11-d action, using $\int C_3\wedge G_4\wedge G_4=0$ \cite{MQGP}, is given as:
$${\rm S}_{11}=\int d^{11}x\sqrt{-{\rm det}{\rm g}}\ \Bigg(R-\frac{1}{2 \times 4!}|G_4|^2\Bigg),$$
the first order perturbation of whose EOM yields:
\begin{equation}
\label{Einstein-fluxes}
R_{{M}{N}}^{(1)}=\frac{1}{12}\left(-3{G_{{M}}^{\ \ {P}_2}}_{{Q}{R}}G_{{N}}^{\ \ {{P}_{3}}{Q}{R}}h_{{{P}_{2}}{{P}_{3}}}+\frac{1}{3}G^{{P}_{2}}_{\ \ {N}{P}{Q}}g^{{N}{M}}G^{{{P}_{3}}{N}{P}{Q}}h_{{P}_{2}{P}_{3}}G^{M}_{{M}{N}}-\frac{G^2}{12}
h_{{M}{N}}\right).
\end{equation}
Here, hatted letters like ${M}, {N}$ etc go from 0 to 10 while, $R_{{M}{N}}^{(1)}$ is perturbed part of the Ricci tensor. Putting in the expressions for each of the components following coupled eom's were obtained{\footnote{here term $\delta R[{M},{N}]$ represents the EOM corresponding to coordinates ${M},{N}$.}},

\noindent $\bullet\delta {\rm R[t,t]}$\\
{
\begin{eqnarray}
\label{EOM11}
& & \hskip -0.9in {q_1}''(r)+{q_1}'(r) \Biggl(\frac{2 \left(6 a^2 \log (r)+r^2\right)}{r (2
	\log (r)+1) \left(r^2-3 a^2\right)}\nonumber\\
& & \hskip -0.9in -\frac{1}{52488 \pi ^{3/2} r^2 (2 \log (r)+1) \alpha _{\theta _1}^4 \alpha _{\theta _2}^3 \left(r^2-a^2\right)}\Biggl\{\left(\frac{1}{N}\right)^{2/5} \Biggl[38416 \pi ^{3/2} r \alpha _{\theta _2}^7
\left(6 a^2 \log (r)+r^2\right)\nonumber\\
& & \hskip -0.9in +177147 \sqrt{6} {g_s}^{3/2} M {N_f} \alpha _{\theta _1}^8 \Biggl(12 a^2 \log ^2(r) \left(-27 a^2+15 r^2+r\right)-3 a^2 r +72 \log ^3(r) \left(9 a^4 -a^2 r^2
\right)\nonumber\\
& & \hskip -0.9in +\log (r) \left(-216 a^4+72 a^2 r^2 -6 a^2 r +4
r^3\right)+r^3\Biggr)\Biggr]\Biggr\} -\frac{3 a^2 r}{54 a^4+15 a^2 r^2 +r^4}+\frac{4 {r_h}^4}{r^5-r
	{r_h}^4}+\frac{5}{r}\Biggr)\nonumber\\
& & \hskip -0.9in +{q_1}(r) \Biggl(\frac{4 \pi  {g_s} (K^1)\ ^2 N \left(6 a^2+r^2\right)}{r^4 \left(9 a^2
	+r^2\right)\left(1-\frac{r_h^4}{r^4}\right)}  -\frac{3 {g_s}^3 (K^1)\ ^2 \log N  M^2 {N_f} \log (r) \left(6 a^2+r^2\right)}{4
	\pi  (r^4-{r_h}^4) \left(9 a^2+r^2\right)}\Biggr) =0.
\end{eqnarray}
}
Defining:
\begin{eqnarray}
\label{a1-b1-a2-defs}
& & {a_1}=\frac{243 \sqrt{\frac{3}{2}} b^2 \left(9 b^2-1\right) {g_s}^{3/2} M \left(\frac{1}{N}\right)^{2/5} {N_f} \alpha _{\theta _1}^4 \log ^2({r_h})}{\pi ^{3/2} \left(3
   b^2-1\right) \alpha _{\theta _2}^3}+\frac{\left(\frac{12}{1-3 b^2}-\frac{6}{54 b^4+15 b^2+1}\right) b^2+5}{2 {r_h}}\nonumber\\
& & {a_2}=\frac{\left(6 b^2+1\right) {g_s} {K^1}^2 \left(16 \pi ^2 N-3{g_s}^2 {\log (N)} M^2 {N_f} \log ({r_h})\right)}{16 \pi  \left(9 b^2+1\right) {r_h}^3}\nonumber\\
& & b_1 =1,
\end{eqnarray}
the $q_1(r)$ EOM, near $r=r_h$, can be written as:
\begin{equation}
\label{q1-near_rh}
{q_1}''(r) + \left({a_1}+\frac{1}{ (r-{r_h})}\right) {q_1}'(r)+{a_2} {q_1}(r)=0,
\end{equation}
whose solution is given by:
\begin{eqnarray}
\label{solution-q1}
& & \hskip -0.3in q_1(r) =  e^{- {a_1}r} \Biggl(c_{1\ q_1}
U\left(1-\frac{a_2}{a_1},1,a_1(r-{r_h})\right)+c_{2\ q_1}L_{\frac{{a_2}}{{a_1}}-1}({a_1} (r-{r_h}))\Biggr).
\end{eqnarray}
Since
\begin{eqnarray}
\label{Neumann-q1(r)-1}
& & \hskip -0.8in q_1'(r) = -\frac{{c_{1\ {q_1}}} e^{-{a_1} {r_h}}}{(r-{r_h}) \Gamma \left(1-\frac{{a_2}}{{a_1}}\right)}+\frac{e^{-{a_1} {r_h}}}{\Gamma \left(1-\frac{{a_2}}{{a_1}}\right)} \Biggl({c_{1\ {q_1}}} \Biggl({a_2} \log
   ({a_1})+{a_1} \psi ^{(0)}\left(1-\frac{{a_2}}{{a_1}}\right)+({a_2}-{a_1}) \psi ^{(0)}\left(2-\frac{{a_2}}{{a_1}}\right)\nonumber\\
   && \hskip -0.8in +2 {a_1}+{a_2}
   \log (r-{r_h})-{a_2}+2 \gamma  {a_2}\Biggr)-{a_1} c_2 \left(L_{\frac{{a_2}}{{a_1}}-1}(0)+L_{\frac{{a_2}}{{a_1}}-2}^1(0)\right) \Gamma
   \left(1-\frac{{a_2}}{{a_1}}\right)\Biggr)+O\left((r-{r_h})^1\right),
\end{eqnarray}
we conclude that to be able to impost Neumann boundary condition at $r=r_h$: $q_1'(r_h)=0$, one requires to set $c_{2\ q_1}=0$ and
\begin{equation}
\label{Neumann-q1(r)-2}
 \left(-\frac{{a_2}}{a_1}+1\right)=-n, n\in\mathbb{Z}^+\cup\left\{0\right\}.
\end{equation}
We shall choose $n=1$, implying $a_1 = \frac{a_2}{2}$,\\
So:
\begin{eqnarray}
\label{q1-EOM-Nbc_n=1}
& & \hskip -1in q_1(r\sim r_h) = \frac{1}{2} e^{1-\frac{{a_2} r}{2}} \left(c_2 e^{\frac{{a_2} {r_h}}{2}} ({a_2} ({r_h}-r)+2) {Ei}\left(\frac{1}{2} {a_2} (r-{r_h})\right)+4 c_1
   e^{\frac{{a_2} {r_h}}{2}} (-{a_2} r+{a_2} {r_h}+2)+2 c_2 e^{\frac{{a_2} r}{2}}\right).\nonumber\\
& &
\end{eqnarray}
Again, setting $c_{2\ q_1}=0$:
\begin{equation}
\label{q1_solution_near_rh}
q_1(r \sim r_h) = -\frac{1}{3} a_2^3 c_{1_{{q1}}} (r-{r_h})^3+\frac{3}{2} a_2^2 c_{1_{{q1}}} (r-{r_h})^2-4 a_2 c_{1_{{q1}}} (r-{r_h})+4 c_{1_{{q1}}}+O
   (r-{r_h})^4.
\end{equation}
Further, using (\ref{EOM55}),  $c_{1\ {q_1}}=0$, i.e.:
\begin{equation}
\label{q1_solution_near_rh1}
q_1(r \sim r_h) = 0.
\end{equation}
In the UV, defining:
\begin{eqnarray}
&& \alpha=5-\frac{27 \sqrt{\frac{3}{2}} {g^{UV}_s}^{3/2} M^{UV} \left(\frac{1}{N}\right)^{2/5} {N^{UV}_f} \alpha _{\theta _1}^4}{2 \pi ^{3/2} \alpha _{\theta _2}^3}\nonumber\\
&&\beta=\frac{1}{16} {m_0}^2 {r_h}^2 \left(16-\frac{3 {g^{UV}_s}^2 {\log(N)} {\log(r)} {M^{UV}}^2 {N^{UV}_f}}{\pi ^2 N}\right),
\end{eqnarray}
(\ref{EOM11}) reduces to:
\begin{eqnarray}
\label{EOM11UV}
{q_1}''(r)+\frac{\left| \alpha \right|  {q_1}'(r)}{r}+\frac{\left| \beta \right|  {q_1}(r)}{r^4}=0,
\end{eqnarray}
whose solution is given as:
{\small
\begin{eqnarray}
\label{q1solution_UV}
&&q_1(r)=\left(\frac{1}{r}\right)^{\frac{1}{2} (\left| \alpha \right| +\left| \left| \alpha \right| -1\right| -1)} \Biggl(c_2 e^{-\frac{i \sqrt{\left| \beta \right| }}{r}} \,
   _1F_1\left(\frac{1}{2} (\left| \left| \alpha \right| -1\right| +1);\left| \left| \alpha \right| -1\right| +1;\frac{2 i \sqrt{\left| \beta \right| }}{r}\right)\nonumber\\
   &&+\frac{c_1
   2^{-\frac{\left| \left| \alpha \right| -1\right| }{2}} \left| \beta \right| ^{-\frac{\left| \left| \alpha \right| -1\right| }{4}} \left(\frac{i}{r}\right)^{-\frac{\left|
   \left| \alpha \right| -1\right| }{2}} K_{\frac{\left| \left| \alpha \right| -1\right| }{2}}\left(\frac{i \sqrt{\left| \beta \right| }}{r}\right)}{\sqrt{\pi }}\Biggr).
\end{eqnarray}
}
We conclude that for the solution to vanish in the UV region one requires $c_1=0$, then the solution can be approximated as:
\begin{eqnarray}
\label{q1solution_r_rUV}
&&q_1(r\rightarrow r_{UV})={c^{UV}_{2_{q_1}}} \left(\frac{-\frac{3 \sqrt{\frac{3}{2}} {g^{UV}_s}^{3/2} {m_0}^2 {M^{UV}} {N^{UV}_f} {r_h}^2 \alpha _{\theta _1}^4}{16 \pi ^{3/2}
   N^{2/5} \alpha _{\theta _2}^3}-\frac{{m_0}^2 {r_h}^2}{12}}{r^6}+\frac{1}{r^4}\right).
\end{eqnarray}\\\\\\
\noindent$\bullet\delta {\rm R[x^1,x^1]}$\\\\\\
{\small
\begin{eqnarray*}
& & \hskip -1.1in {q_5}''(r)+\left(\frac{100 {g_s} N \pi  \left(r^2+6 a^2 \right)}{r^4 \left(r^2+9 a^2 \right)\left(1-\frac{{r_h}^4}{r^4}\right)}-\frac{75 {g_s}^3
	\log N  M^2 {N_f} \left(r^2+6 a^2 \right) \log (r)}{4 \pi  \left(r^4-{r_h}^4\right) \left(r^2+9 a^2 \right)}\right)
{q_1}(r)\nonumber\\\\
& & \hskip -1.1in +\left(\frac{75 {g_s}^3 \log N  M^2 {N_f} \left(r^2+6 a^2 \right) \log (r)}{2 \pi  \left(r^4-{r_h}^4\right)
	\left(r^2+9 a^2\right)}-\frac{200 {g_s} N \pi  \left(r^2+6 a^2\right)}{\left(r^4-{r_h}^4\right) \left(r^2+9 a^2\right)}\right) {q_4}(r)\nonumber\\\\
& & \hskip -1.1in +\Biggl(-\frac{12 \sqrt{6} a^4 {g_s}^{3/2} M^3 \sqrt[5]{\frac{1}{N}} \left(r^2+6 a^2\right) \left(\frac{68260644 \left(54 a^2+5\right) \left({r_h}^4-10000\right) \log (10)}{\left(100-3 a^2
		\right)^4}-\frac{30876125 \left(12 a^2 +1\right) \left({r_h}^4-6561\right) \log (9)}{9 \left(a^2
	-27\right)^4}\right) }{5 \log N ^5 M_g^2 \pi ^{3/2} r^2 \left(r^2+9 a^2 \right)\left(1-\frac{{r_h}^4}{r^4}\right)}
	\nonumber\\\\
	& & \hskip -1.1in -\frac{1}{r^2 \left(r^2-3 a^2\right)
		\left(r^2+6 a^2 \right) \left(r^2+9 a^2 \right) \left(r^4-{r_h}^4\right) (2 \log (r)+1)}\nonumber\\
	& & \hskip -1.1in \times\Biggl\{2 \Biggl[r^2 \left(648 a^6 r^2-9 a^4 \left(17 r^4-13 {r_h}^4\right)+a^2 \left(27 r^2 {r_h}^4-75 r^6\right)-6 r^8+2 r^4 {r_h}^4\right)\nonumber\\
& & \hskip -1.1in +2 \log (r) \left(324
   a^6 \left(r^4+{r_h}^4\right)+a^4 \left(99 r^2 {r_h}^4-135 r^6\right)+a^2 \left(3 r^4 {r_h}^4-51 r^8\right)-4 r^{10}\right)\Biggr]\Biggr\}\Biggr)
{q_5}(r)\nonumber\\
& & \hskip -1.1in +{q_3}(r) \Biggl[\frac{8 {r_h}^4}{r^5 \left(1-\frac{{r_h}^4}{r^4}\right)}-\frac{\left(r^2+9 a^2\right) \left(\frac{2 r}{r^2+9 a^2}-\frac{2 r \left(r^2+6 a^2 \right)}{\left(r^2+9 a^2
		\right)^2}\right)}{r^2+6 a^2 } +\frac{4 \left(r^2+6 a^2  \log (r)\right)}{r \left(r^2-3 a^2
	\right) (2 \log (r)+1)}\nonumber\\
& &  \hskip -1.1in -\frac{1}{13122 \sqrt{2} \pi ^{3/2} r^2 \left(r^2-3 a^2 \right) (2 \log (r)+1) \alpha _{\theta _1}^4 \alpha _{\theta
		_2}^3}\Biggl\{\left(\frac{1}{N}\right)^{2/5} \Biggl(177147 \sqrt{3} {g_s}^{3/2} M {N_f}
	\biggl(r^3-3 a^2  r+72 \left(9 a^4 -a^2 r^2 \right) \log ^3(r)\nonumber\\
	& & \hskip -1.1in +12 a^2 \left(15 r^2+r-27
	a^2\right) \log ^2(r)+\left(-216 a^4+72 a^2 r^2 -6 a^2 r +4 r^3\right) \log
	(r)\biggr) \alpha _{\theta _1}^8\nonumber\\
	& & \hskip -1.1in +19208 \sqrt{2} \pi ^{3/2} r \left(r^2+6 a^2  \log (r)\right) \alpha _{\theta
		_2}^7\Biggr)\Biggr\}+\frac{12}{r}\Biggr]+\left(\frac{75 {g_s}^2 \log N  {N_f} M^2}{64 (K^1)\ ^2 N \pi ^2 r}+\frac{25}{(K^1)\ ^2
	r}\right) {q_1}'(r)\nonumber\\
& & \hskip -1.1in +2 {q_3}'(r)+\Biggl(\frac{25 \left(-\frac{8 {r_h}^4}{r^5-r {r_h}^4}+\frac{6 a^2 r}{r^4+15 a^2 r^2+54 a^4 }-\frac{4 \left(r^2+6 a^2 \log (r)\right)}{r \left(r^2-3 a^2\right) (2 \log (r)+1)}-\frac{16}{r}\right)}{2 (K^1)\ ^2}\nonumber\\
	& & \hskip -1.1in +\frac{1}{26244 \sqrt{2} (K^1)\ ^2 \pi ^{3/2} r^2 \left(r^2-3 a^2 \right) (2 \log
			(r)+1) \alpha _{\theta _1}^4 \alpha _{\theta _2}^3}\Biggl\{25 \left(\frac{1}{N}\right)^{2/5}
		\nonumber\\
& & \hskip -1.1in \times		 \Biggl(177147
	\sqrt{3} {g_s}^{3/2} M {N_f} \Biggl[r^3-3 a^2  r+72 \left(9 a^4 -b^2 r^2 {r_h}^2\right) \log
	^3(r)+12 a^2  \left(15 r^2+r-27 a^2 \right) \log ^2(r)\nonumber\\
& & \hskip -1.1in +\left(-216 a^4+72 a^2 r^2 -6
	a^2 r +4 r^3\right) \log (r)\Biggr] \alpha _{\theta _1}^8+19208 \sqrt{2} \pi ^{3/2} r \left(r^2+6 a^2  \log
	(r)\right) \alpha _{\theta _2}^7\Biggr)\Biggr\}\Biggr) {q_4}'(r)\nonumber\\
	& & \hskip -1.1in +\Biggl(\frac{1}{12} \left(\frac{48 {r_h}^4}{r^5
	\left(1-\frac{{r_h}^4}{r^4}\right)}-\frac{6 \left(r^2+9 b^2 {r_h}^2\right) \left(\frac{2 r}{r^2+9 b^2 {r_h}^2}-\frac{2
		r \left(r^2+6 b^2 {r_h}^2\right)}{\left(r^2+9 b^2 {r_h}^2\right)^2}\right)}{r^2+6 b^2 {r_h}^2}+\frac{24 \left(r^2+6 b^2
	{r_h}^2 \log (r)\right)}{r \left(r^2-3 b^2 {r_h}^2\right) (2 \log
	(r)+1)}+\frac{9}{r}\right)\nonumber\\
\end{eqnarray*}
\begin{eqnarray}
\label{EOM22}
& & \hskip -1.1in -\frac{1}{26244
	\sqrt{2} \pi ^{3/2} r^2 \left(r^2-3 a^2 \right) (2 \log (r)+1) \alpha _{\theta _1}^4 \alpha _{\theta _2}^3}\Biggl\{\left(\frac{1}{N}\right)^{2/5}\nonumber\\
& & \hskip -1.1in \times \Biggl(177147 \sqrt{3} {g_s}^{3/2} M {N_f} \biggl[r^3-3 a^2 r+72 \left(9 a^4 -a^2 r^2 \right) \log ^3(r)+12 a^2  \left(15 r^2+r-27 a^2
	\right) \log ^2(r)\nonumber\\
	& & \hskip -1.1in +\left(-216 a^4+72 a^2 r^2 -6 a^2 r +4 r^3\right) \log (r)\biggr]
	\alpha _{\theta _1}^8 +19208 \sqrt{2} \pi ^{3/2} r \left(r^2+6 a^2  \log (r)\right) \alpha _{\theta _2}^7\Biggr)\Biggr\}\Biggr)
{q_5}'(r) -\frac{25 {q_4}''(r)}{(K^1)\ ^2}  = 0.\nonumber\\
& &
\end{eqnarray}}

Defining:
{\small
\begin{eqnarray}
\label{gamma51---gamma56_defs}
& & \gamma_{32} \equiv -\frac{2 {\alpha_4}}{{\beta_3}}\nonumber\\
& & \gamma_{33} \equiv -\frac{{\alpha_4} \left(\frac{243 \sqrt{6} b^2 \left(9 b^2-1\right) {g_s}^{3/2} M \left(\frac{1}{N}\right)^{2/5} {N_f} \alpha _{\theta _1}^4 \log ^2({r_h})}{\pi
   ^{3/2} \left(3 b^2-1\right) \alpha _{\theta _2}^3}+\frac{-486 b^6+261 b^4+90 b^2+7}{-162 b^6 {r_h}+9 b^4 {r_h}+12 b^2 {r_h}+{r_h}}\right)}{{\beta_3}}\nonumber\\
& & \gamma_{51} \equiv \frac{100 {a_2} {c_{1_{q4}}}}{{K^1}^2}\nonumber\\
& & \gamma_{52} \equiv  \frac{25 {a_2} {c_{1_{q4}}} \left(-3 {a_2}-\frac{2 \left(\frac{-1134 b^6+297 b^4+138 b^2+11}{54 b^4 {r_h}+15 b^2 {r_h}+{r_h}}-\frac{243 \sqrt{6}
   b^2 \left(9 b^2-1\right) {g_s}^{3/2} M \left(\frac{1}{N}\right)^{2/5} {N_f} \alpha _{\theta _1}^4 \log ^2({r_h})}{\pi ^{3/2} \alpha _{\theta _2}^3}\right)}{3
   b^2-1}\right)}{{K^1}^2}\nonumber\\
& & \gamma_{55} \equiv \frac{25 \left(6 b^2+1\right) {g_s} {C_{1_{q4}}} \left(3 {g_s}^2 {\log (N)} M^2 {N_f} \log ({r_h})-16 \pi ^2 N\right)}{2 \left(9 \pi  b^2+\pi \right)
   {r_h}^3}\nonumber\\
& & \gamma_{56} \equiv \frac{25 {g_s} {c_{1_{q4}}} \left(54 b^4 (2 {a_2} {r_h}+3)+b^2 (30 {a_2} {r_h}+33)+2 {a_2} {r_h}+3\right) \left(16 \pi ^2 N-3 {g_s}^2
   {\log (N)} M^2 {N_f} \log ({r_h})\right)}{4 \pi  \left(9 b^2+1\right)^2 {r_h}^4},\nonumber
\end{eqnarray}}
(\ref{EOM22}) near $r=r_h$ can be written as:
\begin{eqnarray}
\label{q5-near-rh}
& & \hskip -0.3in{q_5}''(r)+\frac{{q_5}'(r)}{ (r-{r_h})}+\frac{2{q_5}(r)}{{r_h} (r-{r_h})} +\gamma_{52}+\gamma_{56}+\gamma_{33}+\frac{\gamma_{51}+\gamma_{55}+\gamma_{32}}{r-{r_h}}=0,\nonumber
\end{eqnarray}
whose solution is given by:
{\small
\begin{eqnarray}
\label{q5-solution-near-rh}
& & \hskip -1in q_5(r\sim r_h) = \frac{1}{2} \left(2 \sqrt{2} {c_{1_{q5}}}-{\gamma_{51}} {r_h}-{\gamma_{55}} {r_h}-{\gamma_{32}} {r_h}\right)+\frac{1}{4} (r-{r_h})^2 \left(\frac{4
   \sqrt{2} {c_{1_{q5}}}}{{r_h}^2}-{\gamma_{52}}-{\gamma_{56}}-{\gamma_{33}}\right)-\frac{2 \sqrt{2} {c_{1_{q5}}} (r-{r_h})}{{r_h}}.\nonumber\\
   &&
\end{eqnarray}}
So, to be able impose Neumann boundary condition $q_5'(r=r_h)=0$, one needs to set $c_2=0$ and $c_{1_{q5}}=N^{-\alpha_5}$, $\alpha_5>0$,
In the UV region($r>r_h$), (\ref{EOM22}) can be approximated as:
{\small
\begin{eqnarray}
\label{EOM22_UV}
&&{q_5}''(r)+\frac{0.75}{r}{q_5}'(r)-\frac{192019. b^4 {g^{UV}_s}^{5/2} {M^{UV}}^3 N^{4/5} {r_h}^2}{{\log(N)}^5 {m_0}^2 r^2}{q_5}(r)+\frac {12. {c^{UV} _ {1 _ {q_ 3}}}} {r} - \frac {1256.64 {g^{UV} _s} \
N {c^{UV} _ {2 _ {q_ 1}}}} {{m_ 0}^2 r^6 {r_h}^2} \nonumber\\
&&+ \frac {3769.91 \
{g^{UV} _s} N {c^{UV}_ {2 _ {q_ 4}}}} {{m_ 0}^2 r^6 {r_h}^2}=0
\end{eqnarray}}
whose solution after a large large r and large N expnasion can be written as:
{\small
\begin{eqnarray}
\label{q5solution_UV}
&&q_5(r\rightarrow r_{UV})=\frac{{\log(N)}^5 \sqrt[5]{N} (0.0196\  {c^{UV}_{2_{q_4}}}-0.0065\  {c^{UV}_{2_{q_1}}})}{b^4 {g^{UV}_s}^{3/2} {M^{UV}}^3 r^4 {r_h}^4}
\end{eqnarray}}
\noindent$\bullet\delta {\rm R[x^1,r]}$
{\footnotesize
\begin{eqnarray}
\label{EOM25}
& & \hskip -1in {q_1}'(r) \left(\frac{200 \pi  {g_s} N r\left(6 b^2 {r_h}^2+r^2\right)}{21 \left(r^4-{r_h}^4\right) \left(9 a^2 +r^2\right)}-\frac{25
	{g_s}^3 \log N  M^2 {N_f} r(4 \log (r)+1) \left(6 a^2+r^2\right)}{56 \pi  \left(r^4-{r_h}^4\right) \left(9 a^2+r^2\right)}\right)\nonumber\\
& & \hskip -1in +{q_1}(r) \left(\frac{400 \pi  {g_s} N {r_h}^4 \left(6 a^2 +r^2\right)}{21
	\left(r^4-{r_h}^4\right)^2 \left(9 a^2 +r^2\right)}-\frac{25 {g_s}^3 \log N  M^2 {N_f} {r_h}^4 (4 \log
	(r)+1) \left(6 a^2+r^2\right)}{28 \pi   \left(r^4-{r_h}^4\right)^2 \left(9 a^2
	+r^2\right)}\right)\nonumber\\
& & \hskip -1in +{q_4}'(r) \left(\frac{25 {g_s}^3 \log N  M^2 {N_f}r (4 \log (r)+1) \left(6 a^2+r^2\right)}{28 \pi  \left(r^4-{r_h}^4\right) \left(9 a^2 +r^2\right)}-\frac{400 \pi  {g_s} N r\left(6 a^2
	+r^2\right)}{21 \left(r^4-{r_h}^4\right) \left(9 a^2+r^2\right)}\right)\nonumber\\
& & \hskip -1in +{q_3}(r) \Biggl[\frac{8 \sqrt{6} a^4 {g_s}^{3/2}
	M^3 \sqrt[5]{\frac{1}{N}}  \left(6 a^2+r^2\right) \left(\frac{68260644 \left({r_h}^4-10000\right) \log
		(10) \left(54 a^2+5\right)}{\left(100-3 a^2 \right)^4}-\frac{30876125 \left({r_h}^4-6561\right) \log (9)
		\left(12 a^2 +1\right)}{9 \left(a^2-27\right)^4}\right)}{35 \pi ^{3/2} \log N ^5 M_g^2 r \left(9 a^2
	+r^2\right)\left(1-\frac{{r_h}^4}{r^4}\right)}\nonumber\\
& & \hskip -1in +\frac{1}{21 r \left(r^4-{r_h}^4\right) (2 \log (r)+1) \left(r^2-3 a^2\right) \left(6 a^2 +r^2\right) \left(9 a^2+r^2\right)}\nonumber\\
& & \hskip -1in\times\Biggl\{-7614 a^6 r^4-17820 a^6 r^4  \log (r)+20412 a^6 {r_h}^{4} \log (r)+8910
	a^6 {r_h}^{4}+144 a^4 r^6 +432 a^4 r^6  \log (r)\nonumber\\
	& & \hskip -1in -216 a^4 r^2 {r_h}^4-576 a^4 r^2 {r_h}^4 \log
	(r)+393 a^2 r^8 +978 a^2 r^8  \log (r)-489 a^2 r^4 {r_h}^4-1170 a^2 r^4 {r_h}^4 \log (r)+39 r^{10}\nonumber\\
	& & \hskip -1in +94
	r^{10} \log (r)-47 r^6 {r_h}^4-110 r^6 {r_h}^4 \log (r)\Biggr\}\Biggr] +{q_3}'(r) = 0.
\end{eqnarray}}
Defining:
\begin{eqnarray}
\label{alpha1+alpha4+beta1+beta3-defs}
& & \alpha_1 \equiv  -\frac{25 {a_2} \left(6 b^2+1\right) {g_s} {c_{1_{q_1}}} \left(16 \pi ^2 N-3 {g_s}^2 {\log (N)} M^2 {N_f} \log ({r_h})\right)}{42 \pi  \left(9
   b^2+1\right) {r_h}^2}\nonumber\\
& & \alpha_4 \equiv-\frac{25 {a_2} \left(6 b^2+1\right) {g_s} {c_{1_{q4}}} \left(3 {g_s}^2 {\log (N)} M^2 {N_f} \log ({r_h})-16 \pi ^2 N\right)}{21 \pi  \left(9
   b^2+1\right) {r_h}^2}\nonumber\\
& & \beta_1 \equiv \frac{25 \left(6 b^2+1\right) {g_s} {c_{1\ {q_1}}} \left(16 \pi ^2 N-3 {g_s}^2 {\log (N)} M^2 {N_f} \log ({r_h})\right)}{84 \left(9 \pi  b^2+\pi
   \right) {r_h}^2}\nonumber\\
   &&\gamma_1 \equiv \frac{25 {g_s} {c_{1\ {q_1}}} \left(54 b^4 ({a_2} {r_h}+3)+3 b^2 (5 {a_2} {r_h}+13)+{a_2} {r_h}+3\right) \left(3 {g_s}^2 {\log (N)}
   M^2 {N_f} \log ({r_h})-16 \pi ^2 N\right)}{84 \pi  \left(9 b^2+1\right)^2 {r_h}^3}\nonumber\\
& & \beta_3 = \frac{13}{42},
\end{eqnarray}
(\ref{EOM25}) near $r=r_h$ can be written as:
\begin{equation}
\label{q3-EOM-near-rh}
{q_3}'(r) + \frac{({\alpha_1} + {\alpha_4}+\gamma_1)}{ (r-{r_h})}+\frac{{\beta_1}}{(r-{r_h})^2}+\frac{{\beta_3}
	{q_3}(r)}{r-{r_h}}=0,
\end{equation}
whose solution is given by:
\begin{equation}
\label{q3-solution-near-rh}
q_3(r\sim r_h) = -\frac{({\alpha_1}+{\alpha_4}+{\gamma_1})}{\beta_3}-\frac{{\beta_1}}{(1-{\beta_3})(r-{r_h})}+c_{q_3}
(r-{r_h})^{-{\beta_3}}.
\end{equation}
To be able to impose Neumann boundary condition at $r=r_h$, one needs to set $c_{q_3}=0$.
For ${c_{1\ {q_1}}}$ only $\alpha_4$, $\beta_3$, and $\gamma_3$ gives a non-zero value,
\begin{equation}
\label{q3-solution-near-rh-1}
q_3(r\sim r_h) = -\frac{{\alpha_4}}{\beta_3}.
\end{equation}
In the UV region ($r>r_h $), the (\ref{EOM25}) can be approximated as:
{\small
\begin{eqnarray}
\label{q3EOM_UV}
&&\hskip -1inr^4 \Biggl(r^7 \alpha _{\theta _2}^3 \left(18287.5 b^4 {g^{UV}_s}^{5/2} {M^{UV}}^3 N^{4/5} {r_h}^2 {q_3}(r)+1. {\log(N)}^5 {m_0}^2 r {q_3}'(r)\right)+{g_s}
   {\log(N)}^5 {m_0}^2 N {c^{UV}_{2_{q_4}}}\nonumber\\
   &&\hskip -1in \left(239.359 \alpha _{\theta _2}^3-177.683 {g^{UV}_s}^{3/2} M \left(\frac{1}{N}\right)^{2/5} {N^{UV}_f} \alpha
   _{\theta _1}^4\right) r^{\frac{27 \sqrt{\frac{3}{2}} {g^{UV}_s}^{3/2} M \left(\frac{1}{N}\right)^{2/5} {N^{UV}_f} \alpha _{\theta _1}^4}{2 \pi ^{3/2} \alpha _{\theta
   _2}^3}}\Biggr)\nonumber\\
   &&\hskip -1in +{g^{UV}_s} {\log(N)}^5 {m_0}^2 N {c^{UV}_{2_{q_1}}} r^{\frac{27 \sqrt{\frac{3}{2}} {g^{UV}_s}^{3/2} M \left(\frac{1}{N}\right)^{2/5}
   {N^{UV}_f} \alpha _{\theta _1}^4}{2 \pi ^{3/2} \alpha _{\theta _2}^3}} \left(88.8414 {g^{UV}_s}^{3/2} M \left(\frac{1}{N}\right)^{2/5} {N^{UV}_f} r^4 \alpha _{\theta
   _1}^4+\alpha _{\theta _2}^3 \left(59.8399 {r_h}^4-119.68 r^4\right)\right)\nonumber\\
   &&
\end{eqnarray}}

whose solution after taking an expansion around large r and large n can be written as:
\begin{eqnarray}
\label{q3solution_UV}
&&q_3(r\rightarrow r_{UV})=\frac{0.00654434\  {\log(N)}^5 {m_0}^2 \sqrt[5]{N} ({c^{UV}_{2_{q_1}}} -2 {c^{UV}_{2_{q_4}}} )}{b^4 {g^{UV}_s}^{3/2} {M^{UV}}^3 r^7
   {r_h}^2}+{c^{UV}_{1_{q_3}}}
\end{eqnarray}
\noindent$\bullet\delta {\rm R[x^3,x^3]}$\\
This yields an EOM for $q_4(r)$ which is identical to that for $q_1(r)$ for both UV and IR region.

%

\noindent $\bullet\delta {\rm R[r,r]}$\\
\begin{eqnarray}
\label{EOM55}
& & {q_1}'(r)=0.
\end{eqnarray}
This along with the $\delta {\rm R[t,t]}$ EOM implies that $c_{1\ q_1}$ is vanishingly small. In Section {\bf 6}, we set it to zero while calculating decay widths associated with decays of the exotic scalar glueball.

\noindent$\bullet\delta {\rm R[\theta_1,\theta_1]}$
\begin{eqnarray}
\label{EOM66}
& & {q_2}(r)-\frac{49 \pi ^3 N^{3/5} r^2 \alpha _{\theta _2}^2 \left(6 a^2+r^2\right)\left(1-\frac{{r_h}^4}{r^4}\right)}{216 {g_s}^3 M^2 {N_f}^2 \left(9 a^2+r^2\right) \log ^2(r) \left(108 a^2+r\right)^2}q_6(r) = 0.
\end{eqnarray}




\noindent$\bullet\delta {\rm R[\theta_2,\theta_2]}$\\
\begin{eqnarray}
\label{EOM77}
& & {q_6}(r)\Bigg( -\frac{3 {g_s}^2 {\log N} M^2 {N_f} \log (r)\left(1-\frac{{r_h}^4}{r^4}\right)}{32 \pi ^2 N}+\frac{{r_h}^4}{r^4}-1\Bigg)+{q_2}(r)=0.
\end{eqnarray}

\noindent$\bullet\delta {\rm R[\theta_1,\theta_2]}$\\
\begin{eqnarray}
\label{EOM67}
& & -\frac{49 \sqrt{3} \pi ^{3/2} \sqrt[5]{N} r \left(6 a^2+r^2\right) \left(1-\frac{{r_h}^4}{r^4}\right)\left(36 a^2 \log (r)+r\right){q_6}(r)}{\sqrt{2}32 {g_s}^{3/2} M {N_f} \alpha _{\theta _2} \left(9
   a^2+r^2\right) \log (r) \left(108 a^2 +r\right)^2}+{q_2}(r) = 0.
\end{eqnarray}



We see that (\ref{EOM66}) - (\ref{EOM67}) are identically satisfied by setting $q_2(Z) = q_6(Z) = 0$.

\noindent $\bullet$ All other remaining equations $\delta {\rm R[m,n]}$ for $(m,n)\in\left\{\theta_{1,2},x,y,z,x^{11}\right\}$ are automatically satisfied provided:
\begin{equation}
\label{EOMS-allothercompact}
\frac{1}{2} (K^1)\ ^2 {q_3}(r)+\frac{1}{4} (K^1)\ ^2 {q_5}'(r)+{q_1}'(r)-3 {q_4}'(r)=0.
\end{equation}
In the IR, near $r=r_h$, by substituting solutions for $q_{3,4,5}(Z)$, one sees that (\ref{EOMS-allothercompact}) is identically satisfied.

\section{Meson Sector}

To start off our study of glueball-meson interaction in the type IIA background we first have to understand how the mesons are obtained in the theory. The meson sector in the type IIA dual background of top-down holographic type IIB setup\cite{MQGP} is given by the flavor D6-branes action. We first need to understand how the D6 branes are embedded in the mirror(constructed in \cite{MQGP}) of the resolved warped deformed conifold of \cite{metrics}. To obtain the pullback metric and the pullback NS-NS flux on the D6 branes, we choose the first branch of the Ouyang embedding where $(\theta_{1},x)=(0,0)$ and we consider the `z' coordinate as a function of r, i.e z(r)\cite{Dasgupta_et_al_Mesons}. In \cite{Yadav_Misra-Sil-EPJC-17} a diagonal metric $\{t,x^1,x^2,x^3,r,\theta_1,\theta_2,\tilde{x},\tilde{y},\tilde{z}\}$was used to obtain the mirror of the Ouyang embedding, but it turns out that the embedding conditions remains same even with the nondiagonal basis $\{t,x^1,x^2,x^3,r,\theta_1,\theta_2,x,y,z\}$ . For $\theta_{1}=\alpha_{\theta_{1}}N^{-\frac{1}{5}}$ and $\theta_{2}=\alpha_{\theta_{2}}N^{-\frac{3}{10}}$ one will assume that the embedding of the D6-brane will be given by $\iota :\Sigma ^{1,6}\Bigg(t, R^{1,2},r,\theta_{2}\sim\frac{\alpha_{\theta_{2}}}{N^{\frac{3}{10}}},y\Bigg)\hookrightarrow M^{1,9}$, effected by: $z=z(r)$. As obtained in \cite{Yadav_Misra-Sil-EPJC-17} one sees that z=constant is still a solution and by choosing $z=\pm {\cal C}\frac{\pi}{2}$, one can choose the $D6/\bar{D6}$-branes to be at ``antipodal" points along the z coordinate.\\
As done in \cite{Dasgupta_et_al_Mesons} after redefining (r,z) in terms of new variables (Y,Z):
\begin{eqnarray}
\label{coord change}
&&r=r_{h}e^{\sqrt{Y^{2}+Z^{2}}}\nonumber\\
&&z={\cal C}\arctan\frac{Z}{Y}
\end{eqnarray}
the constant embedding of the $D6(\bar{D6})$-branes corresponds to $z=\frac{\pi}{2}$ for ${\cal C}=1$ for D6-branes and $z=-\frac{\pi}{2}$ for ${\cal C}=-1$ for $\bar{D6}$-branes, both corresponding to $Y=0$. Vector mesons are obtained by considering gauge fluctuations of a background gauge field along the world volume of the embedded flavor D6 branes. Turning on a gauge field fluctuation $F\frac{\sigma^3}{2}$ about a small background gauge field $F_0\frac{\sigma^3}{2}$ and the backround $i^*(g+B)$. This implies:
{\footnotesize
\begin{eqnarray}
\label{DBI expansion}
& & {\rm Str}\left.\sqrt{-{\rm det}_{t,\mathbb{R}^{1,2},Z,\theta_2,y}\left(i^*(g+B) + (F_0 + F)\frac{\sigma^3}{2}\right)}\right|_{Y=0}\delta\left(\theta_2 - \frac{\alpha_{\theta_2}}{N^{\frac{3}{10}}}\right)\nonumber\\
& & = \sqrt{-{\rm det}_{\theta_2,y}\left(i^*(g+B)\right)}\left. {\rm Str}\sqrt{{\rm det}_{t,\mathbb{R}^{1,2},Z}\left(i^*(g+B) + (F_0 + F)\frac{\sigma^3}{2}\right)}\right|_{Y=0}\delta\left(\theta_2 - \frac{\alpha_{\theta_2}}{N^{\frac{3}{10}}}\right)
\nonumber\\
& & = \left.\sqrt{-{\rm det}_{\theta_2,y}\left(i^*(g+B)\right)}\sqrt{{\rm det}_{t,\mathbb{R}^{1,2},Z}(i^*g)} {\rm Str}\left({\bf 1}_2 - \frac{1}{2}\left[(i^*g)^{-1}\left((F_0 + F)\frac{\sigma^3}{2}\right)\right]^2 + ....\right)\right|_{Y=0}\delta\left(\theta_2 - \frac{\alpha_{\theta_2}}{N^{\frac{3}{10}}}\right).
\nonumber\\
\end{eqnarray}}
Picking up terms quadratic in $\tilde{F}$:
{\small
\begin{equation}
\label{DBI action}
\hskip -0.6in {\rm S}^{IIA}_{D6}=\frac{T_{D_{6}}(2\pi\alpha\prime)^{2}}{4}\left(\frac{1}{T_h}\right){\rm Str}\int d^{3}xdZd\theta_{2}dy \delta\Bigg(\theta_{2}-\frac{\alpha_{\theta_{2}}}{N^{\frac{3}{10}}}\Bigg) e^{-\Phi}\sqrt{-{\rm det}_{\theta_{2}y}(\iota^*(g+B))}\sqrt{{\rm det}_{t,{\mathbb  R}^{1,2},Z}(\iota^*g)}g^{\mu\nu}F_{\nu\rho}g^{\rho\sigma}F_{\sigma\mu},
\end{equation}
}
$T_h$ - the horizon temperature being given by \cite{EPJC-2}:
\begin{eqnarray}
\label{T-RC}
& & T_h = \frac{\partial_rG^{\cal M}_{00}}{4\pi\sqrt{G^{\cal M}_{00}G^{\cal M}_{rr}}}\nonumber\\
& & = {r_h} \left[\frac{1}{2 \pi ^{3/2} \sqrt{{g_s} N}}-\frac{3 {g_s}^{\frac{3}{2}} M^2 {N_f} \log ({r_h}) \left(-\log
   {N}+12 \log ({r_h})+\frac{8 \pi}{g_sN_f} +6-\log (16)\right)}{64 \pi ^{7/2} N^{3/2}} \right]\nonumber\\
   & & + a^2 \left(\frac{3}{4 \pi ^{3/2} \sqrt{{g_s}} \sqrt{N} {r_h}}-\frac{9 {g_s}^{3/2} M^2 {N_f} \log ({r_h})
   \left(\frac{8 \pi }{{g_s} {N_f}}-\log (N)+12 \log ({r_h})+6-2 \log (4)\right)}{128 \pi ^{7/2} N^{3/2}
   {r_h}}\right).\nonumber\\
   & &
\end{eqnarray}
Here $\iota^*g$ and $\iota^*B$ are the pulled back metric and NS-NS $B$ on the $D6$-brane respectively. Writing the Klauza-Klein modes for the gauge fields in a 2+1 dimensional minkowski spacetime consisting of $x^{1,2,3}$ as,
\begin{eqnarray}
\label{AZ+Amu}
& & A_\mu(x^\nu,Z) = \sum_{n=1} \rho_\mu^{(n)}(x^\nu)\psi_n^{\mu}(Z)\ \ \ \mu =1,2,3\nonumber\\
& & A_Z(x^\nu,Z) = \sum_{n=0} \pi^{(n)}(x^\nu)\phi_n(Z),
\end{eqnarray}
one obtains:
\begin{eqnarray}
\label{full-expansion}
& & -\frac{V}{4}\int d^3x dZ \sum_{nm}\Biggl({\cal V}_2(Z)\tilde{F}^{(n)}_{\mu\nu}\tilde{F}^{(m)\mu\nu}\psi_m(Z)\psi_n(Z) + {\cal V}_1(Z)\rho^{(m)}_\mu \rho^{(n)\mu }\dot{\psi}_m\dot{\psi}_n+ {\cal V}_1(Z)\partial_{\mu}\pi^{n}\partial^{\mu}\pi^{m}\phi_{n}\phi_{m}\nonumber\\
&&- {\cal V}_1(Z)\partial_{\mu}\pi^{n}{\rho^{(m)}}^{\mu}\phi_{n}\dot{\psi}_{m}- {\cal V}_1(Z)\partial_{\mu}\pi^{m}{\rho^{n}}^{\mu}\phi_{m}\dot{\psi}_{n}\Biggr).
\end{eqnarray}
The terms quadratic in $\psi/\dot{\psi}$ in (\ref{full-expansion}) are given as:
\begin{eqnarray}
\label{Ftildesq-i}
& & -\frac{V}{4}\int d^3x dZ \sum_{nm}\left({\cal V}_2(Z)\tilde{F}^{(n)}_{\mu\nu}\tilde{F}^{(m)\mu\nu}\psi_m(Z)\psi_n(Z) + {\cal V}_1(Z)\rho^{(m)}_\mu \rho^{(n)\mu }\dot{\psi}_m\dot{\psi}_n\right),
\end{eqnarray}
where:
\begin{eqnarray}
\label{V_V1_V2-defs}
&& V = -T_{D_{6}}2(2\pi\alpha')^2\Bigg(\frac{1}{T_h}\Bigg)\int dyd\theta_{2}\delta\Bigg(\theta_{2}-\frac{\alpha_{\theta_{2}}}{N^{3/10}}\Bigg) \nonumber\\
&&{\cal V}_{1}(Z)=2\sqrt{h}g^{ZZ}e^{-\Phi}\sqrt{-{\rm det}_{\theta_2,y}\left(i^*(g+B)\right)}\sqrt{{\rm det}_{\mathbb{R}^{1,2},t,Z}(i^*g)}\nonumber\\
&&{\cal V}_{2}(Z)=h e^{-\Phi}\sqrt{-{\rm det}_{\theta_2,y}\left(i^*(g+B)\right)}\sqrt{{\rm det}_{\mathbb{R}^{1,2},t,Z}(i^*g)}.
\end{eqnarray}

Now, $F_{\mu\nu}(x^\rho,|Z|) = \sum_n\partial_{[\mu}\rho_{\nu]}^{(n)}\psi_n(Z)\equiv \tilde{F}_{\mu\nu}^{(n)}\psi_n(Z)$. The EOM satisfied by $\rho_\mu(x^\nu)^{(n)}$ is: $\partial_\mu \tilde{F}^{\mu\nu}_{(n)} + \partial_\mu\log\sqrt{g_{t,\mathbb{R}^{1,2},|Z|}}\tilde{F}^{\mu\nu}_{(n)} = \partial_\mu\tilde{F}^{\mu\nu}_{(n)} = {\cal M}_{(n)}^2\rho^\nu_{(n)}$. After integrating by parts once, and utilizing the EOM for $\rho^{(n)}_\mu$, one writes :
\begin{eqnarray}
\label{Ftildesq-ii}
& & \int d^3x dZ\  \left(-2 {\cal V}_2(Z) {\cal M}_{(m)}^2\psi_n^{\rho_\mu}\psi_m^{\rho_\mu} + {\cal V}_1(Z)\dot{\psi}_n^{\rho_\mu}\dot{\psi}_m^{\rho_\mu}\right)\rho^{\mu (n)}\rho_{\mu}^{(m)},
\end{eqnarray}
which yields the following equations of motion:
\begin{eqnarray}
\label{eoms_psi_n_rhomu}
& & \psi^{\mu}_{(m)}: \frac{d}{dZ\ }\left({\cal V}_1(Z) \dot{\psi}_{(m)}^{\mu}\right) + 2 {\cal V}_2(Z){\cal M}_{(m)}^2\psi^{\mu}_m = 0.
\end{eqnarray}

The normalization condition of $\psi_{n}$ are given as
\begin{eqnarray}
\label{norm_psi}
&&V\int dZ\ {\cal V}_{2}(Z)\ \psi_{n}\psi_{m}=\delta_{nm}\nonumber\\
&&\frac{V}{2}\int dZ\ {\cal V}_{1}(Z)\ \partial_{Z}\psi_{n} \partial_{Z}\psi_{m}=m_{n}^2\delta_{nm}.
\end{eqnarray}
Thus the action for vector meson part for all $n\ge 1$can be wriiten as
\begin{eqnarray}
& & -\int d^3x  \sum_{n}\left(\frac{1}{4}\tilde{F}^{(n)}_{\mu\nu}\tilde{F}^{(n)\mu\nu}+\frac{m_{n}^2}{2}\rho^{(n)}_\mu \rho^{(n)\mu }\right).
\end{eqnarray}

To normalize the kinetic term for $\pi ^{n}$, we impose the normalization condition for all n corresponding to $\pi ^{n}$ which ranges from 0 to $\infty $
\begin{eqnarray}
\label{norm_scalar}
&&\frac{V}{2}\int dZ\ {\cal V}_{1}(Z)\ \phi_{n}\phi_{m}=\delta_{nm}.
\end{eqnarray}
From \ref{norm_psi}, it is seen that we can choose $\phi_{n}=m_{n}^{-1}\dot{\psi}_{n}$ for all $n\ge 1$. For $n=0$ corresponding to $\phi_0$ we choose its form such as it is orthogonal to $\dot{\psi}_{n}$ for all $n\ge 1$. By writing $\phi_{0}=\frac{C}{{\cal V}_{1}(Z)}$, we have
$$(\phi_{0},\phi_{n})\propto\int\ dZ\ C\partial_{Z}\psi\ =0.$$
Thus the cross component in (\ref{full-expansion}) vanishes for $n=0$, and the remaining cross components can be absorbed in the $\rho_{\mu}^{n}$ by following a specific gauge transformation given as,
$$\rho_{\mu}^{n}\rightarrow\rho_{\mu}^{n}+m_{n}^{-1}\partial_{\mu}\pi^{n}.$$
Then the action becomes:
\begin{eqnarray}
& & -\int d^3x  \left[\frac{1}{2}\partial_{\mu}\pi^{0}\partial^{\mu}\pi^{0} + \sum_{n\ge 1}\left(\frac{1}{4}\tilde{F}^{(n)}_{\mu\nu}\tilde{F}^{(n)\mu\nu}+\frac{m_{n}^2}{2}\rho^{(n)}_\mu \rho^{(n)\mu }\right)\right].
\end{eqnarray}

\subsection{Radial Profile Function $\psi_1(Z)$ for $\rho$-Meson}

Up to NLO in $N$:
{\footnotesize
\begin{eqnarray}
\label{V1NLO}
& & \hskip -0.4in {\cal V}_1(Z) = \frac{1}{108 \pi ^2 {\log N} \alpha _{\theta _1}^3 \alpha _{\theta _2}^2}\Biggl\{M \sqrt[5]{\frac{1}{N}} {N_f} e^{-4 Z} {e^{4 Z}-1} \left(2 \sqrt[5]{\frac{1}{N}} \alpha _{\theta _2}^2+81 \alpha _{\theta
		_1}^2\right)\nonumber\\
	& & \hskip -0.4in \log \left({r_h} e^Z\right) \Biggl(3 \log \left({r_h} e^Z\right) \left(3 a^2 \left({g_s} {N_f} \left(8 {\log (N)} {r_h} e^Z-1\right)+32 \pi  {r_h} e^Z\right)-2 {g_s} {N_f}
   {r_h}^2 e^{2 Z}\right)+3 a^2 ({g_s} ({\log (N)}-3) {N_f}+4 \pi )\nonumber\\
   &&\hskip -0.4in-216 a^2 {g_s} {N_f} {r_h} e^Z \log ^2\left({r_h} e^Z\right)+2
   {r_h}^2 e^{2 Z} ({g_s} {\log (N)} {N_f}+4 \pi )\Biggr)\Biggr\},
\end{eqnarray}}
and
\begin{eqnarray}
\label{V2NLO}
& &\hskip -0.4in {\cal V}_2(Z) = \frac{1}{54 \pi  {\log N} {r_h}^2 \alpha _{\theta _1}^3 \alpha _{\theta _2}^2}\Biggl\{{g_s} M N^{3/5} {N_f} \left(81 \sqrt[5]{N} \alpha _{\theta _1}^2+2 \alpha _{\theta _2}^2\right)\nonumber\\
& & \hskip -0.4in \log
	\left({r_h} e^Z\right) \Biggl(3 a^2 e^{-2 Z} \Biggl((3 \log \left({r_h} e^Z\right) \left({g_s} {N_f} \left(8 {\log (N)} {r_h} e^Z+1\right)+32 \pi  {r_h} e^Z\right)-{g_s}
   ({\log (N)}+3) {N_f}\nonumber\\
   &&\hskip -0.4in-72 {g_s} {N_f} {r_h} e^Z \log ^2\left({r_h} e^Z\right)-4 \pi \Biggr)+2 {r_h}^2 \left({g_s} {\log (N)} {N_f}-3
   {g_s} {N_f} \log \left({r_h} e^Z\right)+4 \pi \right)\Biggr)\Biggr\}.
\end{eqnarray}
Hence the Schr\"{o}dinger-like equation satisfied by $g(Z)\equiv\sqrt{{\cal V}_1(Z)}\psi_1(Z)$ will have a potential given by (\ref{V-psi1}). Near the horizon, $Z=0$, and the aforemetioned Schr\"{o}dinger-like equation can be written as:
\begin{equation}
\label{near_Z=0-EOM-redefined_psi1}
g''(Z)+g(Z) \left(\frac{\omega_1}{Z}+\omega_2+\frac{1}{4 Z^2}\right)=0,
\end{equation}
wherein:
{\footnotesize
\begin{eqnarray}
\label{omega_1-and-2_defs}
& & \omega_1\equiv \frac{1}{4} \left({m_0}^2-3 b^2 \left({m_0}^2-2\right)\right)+18 b^2 {r_h} \log
   ({r_h})-\frac{3 b \gamma  {g_s} M^2 \left({m_0}^2-2\right) \log ({r_h})}{2 N}+\frac{36 b
   \gamma  {g_s} M^2 {r_h} \log ^2({r_h})}{N},\nonumber\\
& & \omega_2\equiv -\frac{4}{3}+\frac{3}{2} b^2 \left({m_0}^2+72 {r_h}-4\right)-36 b^2 {r_h} \log ({r_h})+\frac{3 b \gamma
   {g_s} M^2 \left({m_0}^2-4\right) \log ({r_h})}{N}-\frac{72 b \gamma  {g_s} M^2 {r_h}
   \log ^2({r_h})}{N}.\nonumber\\
& &
\end{eqnarray}}
The solution to (\ref{near_Z=0-EOM-redefined_psi1}) is given by:
\begin{eqnarray}
\label{solution-redefined-psi1-near_Z=0}
 & & g(Z) = \tilde{c}_{1\ \psi_1} M_{-\frac{i {\omega_1}}{2 \sqrt{{\omega_2}}},0}\left(2 i \sqrt{{\omega_2}} Z\right) + \tilde{c}_{2\ \psi_1}
   W_{-\frac{i {\omega_1}}{2 \sqrt{{\omega_2}}},0}\left(2 i \sqrt{{\omega_2}} Z\right).
\end{eqnarray}
Now,
{\footnotesize
\begin{eqnarray}
\label{1oversqrtV1}
& &  \frac{1}{\sqrt{{\cal V}_1}} = \frac{2 \pi  \sqrt[10]{N}}{\sqrt{3} \sqrt{{\cal D}}} + {\cal O}\left(N^{-\frac{1}{10}}\right)\nonumber\\
		& & =\frac{\pi  \sqrt[10]{N}}{\sqrt{3} \sqrt{Z} \sqrt{\frac{M {N_f} {r_h}^2 \log ({r_h}) \left(3 \log
   ({r_h}) \left({g_s} {N_f} \left(3 b^2 (8 {\log (N)} {r_h}-1)-2\right)+96 \pi  b^2
   {r_h}\right)+3 b^2 ({g_s} ({\log (N)}-3) {N_f}+4 \pi )-216 b^2 {g_s} {N_f} {r_h}
   \log ^2({r_h})+2 {g_s} {\log (N)} {N_f}+8 \pi \right)}{{\log (N)} \alpha _{\theta _1} \alpha
   _{\theta _2}^2}}} \nonumber\\
				& &  +	{\cal O}\left(N^{-\frac{1}{10}},Z^{\frac{3}{2}}\right),
\end{eqnarray}}
where:
{\small
\begin{eqnarray}
\label{cal D-def}
& &  {\cal D} \equiv  \frac{1}{{\log N} \alpha _{\theta _1} \alpha _{\theta _2}^2}\Biggl\{M {N_f} {r_h}^2 e^{-4 Z} \sqrt{e^{4 Z}-1} (\log(e^Zr_h))\nonumber\\
& & \Biggl(3 (\log ({r_h})+Z) \left(3 b^2 \left({g_s} {N_f} \left(8 {\log (N)} {r_h} e^Z-1\right)+32 \pi
    {r_h} e^Z\right)-2 {g_s} {N_f} e^{2 Z}\right)+3 b^2 ({g_s} ({\log (N)}-3) {N_f}+4 \pi
   )\nonumber\\
   &&-216 b^2 {g_s} {N_f} {r_h} e^Z (\log ({r_h})+Z)^2+2 e^{2 Z} ({g_s} {\log (N)}
   {N_f}+4 \pi )\Biggr)\Biggr\}
\end{eqnarray}}
Thus:
\begin{eqnarray}
\label{solution-redefinied-psi1-near-Z=0}
& & \psi_1(Z) = Z^{-\frac{1}{2}}\Biggl[c_{1\ \psi_1} M_{-\frac{i {\omega_1}}{2 \sqrt{{\omega_2}}},0}\left(2 i \sqrt{{\omega_2}} Z\right)+c_{2\ \psi_1}
   W_{-\frac{i {\omega_1}}{2 \sqrt{{\omega_2}}},0}\left(2 i \sqrt{{\omega_2}} Z\right)\Biggr],
\end{eqnarray}
which yields:
\begin{eqnarray}
\label{psi1'}
& & \hskip -0.4in\psi_1^\prime(Z) =\frac{1}{{\sqrt{2} \sqrt{i \sqrt{{\omega_2}}} Z \Gamma \left(\frac{i {\omega_1}}{2
   \sqrt{{\omega_2}}}-\frac{1}{2}\right) \Gamma \left(\frac{i {\omega_1}}{2
   \sqrt{{\omega_2}}}+\frac{1}{2}\right)}}\Biggl\{i c_2 \Biggl(2 \sqrt{{\omega_2}} \Gamma \left(\frac{i {\omega_1}}{2
   \sqrt{{\omega_2}}}+\frac{1}{2}\right) \Biggl(\psi ^{(0)}\left(\frac{i {\omega_1}}{2
   \sqrt{{\omega_2}}}-\frac{1}{2}\right)\nonumber\\
   &&\hskip -0.4in+\log \left(2 i \sqrt{{\omega_2}}\right)+\log (Z)+2 \gamma
   \Biggr)+\left(\sqrt{{\omega_2}}-i {\omega_1}\right) \Gamma \left(\frac{i {\omega_1}}{2
   \sqrt{{\omega_2}}}-\frac{1}{2}\right) \Biggl(\psi ^{(0)}\left(\frac{i {\omega_1}}{2
   \sqrt{{\omega_2}}}+\frac{1}{2}\right)+\log \left(2 i \sqrt{{\omega_2}}\right)+\log (Z)\nonumber\\
   &&\hskip -0.4in+2 \gamma
   \Biggr)\Biggr)\Biggr\}+ ....
\end{eqnarray}
To ensure that the coefficient of the $\frac{1}{Z}$ term in $\psi_1^\prime(Z\sim0)$ vanishes, we set:
\begin{equation}
\label{quantization}
-\frac{1}{2}+\frac{i \omega_1}{2 \sqrt{\omega_2}}=-1,
\end{equation}
that implies $\omega_1= i \sqrt{\omega_2}$, and:
\begin{eqnarray}
\label{m_0}
& & m_0 = 2.479 +2.911 {r_h} \log ({r_h}) -\frac{0.289 \gamma  {g_s} M^2 \log ({r_h})}{N}.
\end{eqnarray}
Further for well-behaved $\psi_1'(Z)$ near $Z=0$ one requires to set $c_{2\ \psi_1}=0$. Therefore:
\begin{eqnarray}
\label{psi1-quant}
& & \psi_1(Z) = -\frac{{c_{\psi_1}} \sqrt{i \sqrt{{\omega_2}}} {\omega_2} Z^2}{\sqrt{2}}-\sqrt{2} {c_{\psi_1}} \left(i
   \sqrt{{\omega_2}}\right)^{3/2} Z+\sqrt{2} {c_{\psi_1}} \sqrt{i \sqrt{{\omega_2}}},
\end{eqnarray}
and
\begin{eqnarray}
\label{psi1'-quant}
& & \psi_1^\prime(Z) = -\sqrt{2} {c_{\psi_1}} {\omega_2} \sqrt{i \sqrt{{\omega_2}}} Z-\sqrt{2} {c_{\psi_1}} \left(i
   \sqrt{{\omega_2}}\right)^{3/2}.
\end{eqnarray}
To satisfy Neumann boundary condition at $Z=0$, one will hence set: $ c_{1_{\ \psi_1}} = c_{\psi_1} = N^{-\Omega_\psi}, \Omega_\psi>1$. Also, for $b=0.57,$ $|\omega_2|={\cal O}\left(r_h\log r_h, \frac{g_s M^2}{N}r_h(\log r_h)^2\right)<<1$.

\subsection{Radial Profile Function $\phi_0(Z)$ for $\pi$-Meson}

Near $Z=0$:
{\footnotesize
\begin{eqnarray}
\label{phi0-1}
& &  \phi_0(Z) = \frac{{\cal C}_{\phi_0}}{{\cal V}_1(Z)} = \frac{\sqrt[5]{N} \left(\frac{\pi ^2 {\cal C}_{\phi_0} \alpha _{\theta _1} \alpha _{\theta _2}^2}{3 \left(3 b^2+2\right) {g_s} M
   {N_f}^2 {r_h}^2 \log ({r_h})}-\frac{\pi ^2 {\cal C}_{\phi_0} \alpha _{\theta _1} \alpha _{\theta _2}^2}{\left(3
   b^2+2\right) {g_s} {\log (N)} M {N_f}^2 {r_h}^2}\right)}{Z}\nonumber\\
   & &  +\frac{\frac{2 \pi ^2 {\cal C}_{\phi_0} \alpha _{\theta _2}^4}{81 \left(3 b^2+2\right) {g_s} {\log(N)} M {N_f}^2 {r_h}^2
   \alpha _{\theta _1}}+\frac{2 \pi ^2 {\cal C}_{\phi_0} \alpha _{\theta _2}^4}{243 \left(3 b^2+2\right) {g_s} M {N_f}^2
   {r_h}^2 \alpha _{\theta _1} \log ({r_h})}}{Z}\nonumber\\
& &  +\sqrt[5]{N} \left(\frac{\pi ^2 {\cal C}_{\phi_0} \alpha _{\theta _1} \alpha _{\theta _2}^2}{72 b^2 {g_s}^2 {\log(N)} M {N_f}^2
   {r_h}^3 \log ({r_h})}+\frac{2 \pi ^2 b^2 {\cal C}_{\phi_0} \alpha _{\theta _1} \alpha _{\theta _2}^2}{\left(3 b^2+2\right)^2
   {g_s} M {N_f}^2 {r_h}^2 \log ({r_h})}\right)\nonumber\\
   & &  +\Bigg(-\frac{\pi ^2 {\cal C}_{\phi_0} \alpha _{\theta _2}^4}{2916 b^2{g_s}^2 {\log(N)} M {N_f}^2 {r_h}^3 \alpha _{\theta _1} \log
   ({r_h})} -\frac{4 \pi ^2 b^2 {\cal C}_{\phi_0} \alpha _{\theta _2}^4}{81 \left(3 b^2+2\right)^2 {g_s} M {N_f}^2 {r_h}^2
   \alpha _{\theta _1} \log ({r_h})}\Bigg)\nonumber\\
   & &  +\frac{\pi ^2 {\cal C}_{\phi_0} \sqrt[5]{N} Z \alpha _{\theta _1} \alpha _{\theta _2}^2 \left(9 b^4 \left(4 {\log(r_h)}^2-6
   {\log(r_h)}+3\right)-12 b^2 \left(5 {\log(r_h)}^2+3 {\log(r_h)}-3\right)-8 {\log(r_h)}^2+12\right)}{9 \left(3 b^2+2\right)^3
   {g_s}^{7/2} M {N_f}^2 \log ^3({r_h})}\nonumber\\
   &&  -\frac{\pi ^2 {\cal C}_{\phi_0}\sqrt[5]{N} Z^2 \alpha _{\theta _1} \alpha _{\theta _2}^2 \left(432 b^4 {\log(r_h)}^3-18 \left(3
   b^2+2\right)^2 b^2 {\log(r_h)}+3 \left(3 b^2+2\right)^3+4 \left(3 b^2+2\right) \left(9 b^4-15 b^2-2\right) {\log(r_h)}^2\right)}{9
   \left(3 b^2+2\right)^4 {g_s} {\log(r_h)}^4 M {N_f}^2 {r_h}^2}.\nonumber\\
   & &
\end{eqnarray}}
Let $Z\rightarrow\epsilon\rightarrow0$. Then, in (\ref{phi0-1}), one can show that the sum of all terms except that of ${\cal O}(N^{\frac{1}{5}}Z^{2})$, assuming $\log N > 3 |\log r_h|$, vanishes provided:
\begin{eqnarray}
\label{alpha_theta_2-2}
& & \alpha _{\theta _2}=\frac{9 \sqrt[10]{N} \alpha _{\theta _1} \sqrt{\log ^2({r_h}) ({\log(N)}+3 \log ({r_h}))}}{\sqrt{2} \sqrt{\log ^2({r_h})
   ({\log(N)}-3 \log ({r_h}))}}.
\end{eqnarray}
Therefore:
\begin{equation}
\label{phi0-3}
\phi_0(Z\sim0) =-\frac{0.682249 {\cal C}_{\phi _0} \sqrt[5]{N} Z^2 \alpha _{\theta _1} \alpha _{\theta _2}^2}{{g_s} {|\log(r_h)|} M {N_f}^2
   {r_h}^2}.
\end{equation}

\section{Glueball-Meson Interaction Lagrangian}

The couplings appearing in the DBI action after ignoring the derivatives and possible indices can be written as:
\begin{equation}
\label{vertices}
G_E Tr(\pi^{2}),\ \ G_E Tr(\pi,[\pi,\rho]),\ \ G_E Tr([\pi,\rho]^{2}),\ \ G_E Tr(\rho^{2}),\ \  G_E Tr(\rho,[\rho,\rho]),\ \ G_E Tr([\rho,\rho]^{2}).
\end{equation}
The interaction terms written above are generic results for single glueball case. The flavor structure remains same for the case involving multi-glueball vertices. In subsequent sections we will be considering the $n=1,0$ modes respectively in the KK expansion of $A_\mu, A_Z$. Substituting all the fluctuations for the metric in the D-6 brane action gives us the glueball-meson couplings. We only consider the interaction terms that are linear in glueball field $G_{E}$, since we are interested in glueball decays.

The DBI action for D6 branes is written in terms of the 10 dimensional type-IIA metric and dilaton field. The glueball modes and dilaton field for type-IIA background were obtained in terms of 11-D M theory metric perturbations using witten's relation. The perturbed type-IIA field components and dilaton are given as:
\begin{eqnarray}
\label{type IIA components}
&& g^{IIA}_{tt}=\sqrt{G^{M}_{11,11}}\left[\left(1+\frac{h_{11,11}}{2G^{M}_{11,11}}\right)G^{M}_{tt}+h_{tt}\right]\nonumber\\
&& g^{IIA}_{rr}=\sqrt{G^{M}_{11,11}}\left[\left(1+\frac{h_{11,11}}{2G^{M}_{11,11}}\right)G^{M}_{rr}+h_{rr}\right]\nonumber\\
&& g^{IIA}_{ab}=\sqrt{G^{M}_{11,11}}\left[\left(1+\frac{h_{11,11}}{2G^{M}_{11,11}}\right)G^{M}_{ab}+h_{ab}\right]\nonumber\\
&& g^{IIA}_{ra}=\sqrt{G^{M}_{11,11}}\left[\left(1+\frac{h_{11,11}}{2G^{M}_{11,11}}\right)G^{M}_{ra}+h_{ra}\right]\nonumber\\
&& g^{IIA}_{yy}=\sqrt{G^{M}_{11,11}}\left[\left(1+\frac{h_{11,11}}{2G^{M}_{11,11}}\right)G^{M}_{yy}\right]\nonumber\\
&& g^{IIA}_{\theta_{2}y}=\sqrt{G^{M}_{11,11}}\left[\left(1+\frac{h_{11,11}}{2G^{M}_{11,11}}\right)G^{M}_{\theta_{2}y}\right]\nonumber\\
&& g^{IIA}_{\theta_{2}\theta_{2}}=G^{M}_{11,11}\sqrt{G^{M}_{11,11}}\left[\left(1+\frac{3h_{11,11}}{2G^{M}_{11,11}}\right){\cal A}_{\theta_{2}\theta_{2}}\right],
\end{eqnarray}
where $a, b$ run from 1 to 3 corresponding to the spatial part of the metric. Substituting all the expressions for the type IIA metric components $g^{IIA}_{MN}$ and the M-theory  perturbations $h_{MN}$ into the D6-brane DBI action and, working only upto linear order we get three different type of terms as:
\begin{equation}
\label{action-types-of-terms}
{\cal L}_{{\cal O}_d(h^0){\cal O}_\phi (h^0){\cal O}_F(h)} + {\cal L}_{{\cal O}_\phi (h^0){\cal O}_F(h^0){\cal O}_d(h)} + {\cal L}_{{\cal O}_d(h^0){\cal O}_F(h^0){\cal O}_\phi (h)}.
\end{equation}
Here ${\cal O}(h^0)$ represent term wthout any perturbation while ${\cal O}(h)$ represents term with linear order in perturbation. In both the terms subscripts {d,F,$\phi$} corresponds to part of the integrand of the DBI action from which they are obtained, ${\cal O}_d$ corresponds to term obtained from $\sqrt{-{\rm det}(\iota^*(g+B))}$, ${\cal O}_\phi$ corresponds to the term $e^{-\phi}$ and, ${\cal O}_F$ corresponds to the term of type $g^{-1}Fg^{-1}F$. Contributions to the interaction lagrangian from these three different terms were obtained as:
\begin{itemize}
\item
${\cal O}_d(h^0){\cal O}_F(h^0){\cal O}_\phi (h):$\\
\end{itemize}
Here:
{\small
\begin{eqnarray}
 &&\hskip -0.7in{\cal O}_{d}(h^0)=\sqrt{-{\cal A}_{\theta_2\theta_2} {G^M_{11,11}}^2 G^M_{yy}-B^{IIA}_{\theta_2y}\ ^2+G^M_{11,11} {G^{M}_{\theta_{2} y} }}^{2}\sqrt{G^M_{11} {G^{M}_{11,11}}^{5/2} G^M_{22} G^M_{33}
   G^M_{rr} G^M_{tt} {r}_h^2 e^{2 Z}}\nonumber\\
 &&\hskip -0.7in{\cal O}_{F}(h^0)=\frac{2  F_{12}^2}{G^M_{11} G^M_{11,11} G^M_{22}}+\frac{2  F_{13}^2}{G^M_{11} G^M_{11,11} G^M_{33}}+\frac{2_{1Z}^2 e^{-2 Z}}{G^M_{11}
   G^M_{11,11} G^M_{rr} {r}_h^2}+\frac{2  F_{23}^2}{G^M_{11,11} G^M_{22} G^M_{33}}+\frac{2  F_{2Z}^2 e^{-2 Z}}{G^M_{11,11} G^M_{22} G^M_{rr}
   {r}_h^2}+\frac{2 F_{3Z}^2 e^{-2 Z}}{G^M_{11,11} G^M_{33} G^M_{rr} {r}_h^2}\nonumber\\
&&\hskip -0.7in{\cal O}_\phi (h)=-\frac{3 h_{11,11}} {4 {G^M_{11,11}}^{7/4}}.
\end{eqnarray}}
Putting everything together:
{\small
\begin{eqnarray}
&& {\cal L}_{{\cal O}_d(h^0){\cal O}_F(h^0){\cal O}_\phi (h)}=\sqrt{-{\cal A}_{\theta_2\theta_2} {G^M_{11,11}}^{2} G^M_{yy}-B^{IIA}_{\theta_2y}\ ^2+G^M_{11,11} {G^M_{\theta_2 y}}^2} \sqrt{G^M_{11} {G^M_{11,11}}^{5/2} G^M_{22} G^M_{33}
   G^M_{rr} G^M_{tt} {r}_h^2 e^{2 Z}} \nonumber\\
 &&\Bigg(-\frac{3 e^{-2 Z} \phi_{0}(Z)^2 q_{6}(Z)\partial_\mu   \pi \partial^{\mu }\pi  G_E(x^{1},x^{2},x^{3})}{2
  G^M_{11} G^M_{rr} {G^M_{11,11}}^{7/4} {r}_h^2}-\frac{3 \iota e^{-2 Z}  \psi_{1}(Z)   \phi_{0}(Z)^2 q_{6}(Z) \partial_{\mu }\pi\left[\pi ,\rho ^{\mu }\right]
   G_E(x^{1},x^{2},x^{3})}{G^M_{11} G^M_{rr} {G^M_{11,11}}^{7/4} {r}_h^2}\nonumber\\
   &&-\frac{3 e^{-2 Z}q_{6}(Z) \psi_{1}'(Z)^2 \rho ^{\mu } \rho _{\mu }
   G_E(x^{1},x^{2},x^{3})}{2 G^M_{11}G^M_{rr} {G^M_{11,11}}^{7/4} {r}_h^2}-\frac{3 \psi_{1}(Z)^2
   q_{6}(Z) \tilde{F}_{\mu\nu} \tilde{F}^{\mu\nu} }{4 {G^M_{11}}^2{G^M_{11,11}}^{7/4}}\Bigg).
\end{eqnarray}}

\begin{itemize}
\item
${\cal O}_\phi (h^0){\cal O}_F(h^0){\cal O}_d (h):$\\
\end{itemize}
Here:
{\small
\begin{eqnarray}
 &&\hskip -0.7in{\cal O}_{\phi}(h^0)={G^M_{11,11}}^{-3/4}\nonumber\\
 &&\hskip -0.7in{\cal O}_{F}(h^0)=\frac{2  F_{12}^2}{G^M_{11} G^M_{11,11} G^M_{22}}+\frac{2  F_{13}^2}{G^M_{11} G^M_{11,11} G^M_{33}}+\frac{2 F_{23}^2}{G^M_{11,11} G^M_{22} G^M_{33}}+\frac{2 F_{1Z}^2 e^{-2 Z}}{G^M_{11}
   G^M_{11,11} G^M_{rr} {r}_h^2}+\frac{2  F_{2Z}^2 e^{-2 Z}}{G^M_{11,11} G^M_{22} G^M_{rr}
   {r}_h^2}+\frac{2  F_{3Z}^2 e^{-2 Z}}{G^M_{11,11} G^M_{33} G^M_{rr} {r}_h^2}\nonumber\\
   &&\hskip -0.7in{\cal O}_d (h)=\sqrt{-{\cal A}_{\theta_2\theta_2} {G^M_{11,11}}^{2} G^M_{yy}-B^{IIA}_{\theta_2y}\ ^2+G^M_{11,11} {G^M_{\theta_{2}y}}^2} \sqrt{G^M_{11} {G^M_{11,11}}^{5/2} G^M_{22} G^M_{33}
   G^M_{rr} G^M_{tt} {r}_h^2 e^{2 Z}} \Bigg(\frac{h_{11}}{2 G^M_{11}}+\frac{h_{22}}{2 G^M_{22}}+\frac{h_{33}}{2 G^M_{33}}+\frac{h_{rr}}{2
   G^M_{rr}}\nonumber\\
   && \hskip -0.7in+\frac{h_{tt}}{2 G^M_{tt}}\Bigg)-\frac{h_{11,11} \left(2 {\cal A}_{\theta_2\theta_2} G^M_{11,11} G^M_{yy}-{G^M_{\theta_{2}y}}^2\right) \sqrt{G^M_{11} {G^M_{11,11}}^{5/2} G^M_{22} G^M_{33} G^M_{rr} G^M_{tt} {r}_h^2 e^{2 Z}}}{2  \sqrt{G^M_{11,11}
   \left({G^M_{\theta_{2}y}}^2-{\cal A}_{\theta_2\theta_2} G^M_{11,11}G^M_{yy}\right)-B^{IIA}_{\theta_2y}\ ^2}}\nonumber\\
   & &
\end{eqnarray}}
implying:
{\small
\begin{eqnarray}
&&\hskip -0.7in {\cal L}_{{\cal O}_\phi (h^0){\cal O}_F(h^0){\cal O}_d (h)}=\nonumber\\
& & \hskip -0.7in\Bigg(\frac{1}{2} \sqrt{-{\cal A}_{\theta_{2}\theta_{2}} {G^M_{11,11}}^{2}  G^M_{yy}-B^{IIA}_{\theta_2y}\ ^2+G^M_{11,11} {G^M_{\theta_{2}y}}^2} \sqrt{{G^M_{x^1x^1}}^3 {G^M_{11,11}}^{5/2} G^M_{rr}
   G^M_{tt} {r}_h^2 e^{2 Z}}\nonumber\\
& & \hskip -0.7in\times \left(3q_{4}(Z)-q_{1}(Z)-q_{2}(Z)-q_{5}(Z)\frac{\partial_{\mu}\partial^{\mu}}{M_g^2}\right)\nonumber\\
   &&\hskip -0.7in -\frac{q_{6}(Z) \left(9 {\cal A}_{\theta_{2}\theta_{2}}
   {G^M_{11,11}}^{2} G^M_{yy}+5 B^{IIA}_{\theta_2y}\ ^2-7 G^M_{11,11} {G^M_{\theta_{2}y}}^2\right) \sqrt{{G^M_{x^1x^1}}^3 {G^M_{11,11}}^{5/2} G^M_{rr} G^M_{tt} {r}_h^2 e^{2
   Z}}}{4 \sqrt{G^M_{11,11} \left({G^M_{\theta_{2}y}}^2-G^{M}_{\theta_{2}\theta_{2}}G^M_{11,11} G^M_{yy}\right)-B^{IIA}_{\theta_2y}\ ^2}}\Bigg)\Bigg(\frac{2 e^{-2 Z}}{ {G^{M}_{11,11}}^{7/4}G^M_{11} G^{M}_{rr} {r}_h^2}\Bigg)\nonumber\\
   &&\hskip -0.7in \Bigg(\phi_{0}(Z)^2 \partial_{\mu }\pi \partial^{\mu }\pi +\psi_{1}'(Z)^2\rho ^{\mu } \rho _{\mu }+2\iota \phi_{0}^{2}(Z)\psi_{1}(Z)\partial_{\mu}\pi[\pi,\rho^{\mu}] \Bigg)G_E(x^1,x^2,x^3)\nonumber
   \end{eqnarray}}
   {\small
   \begin{eqnarray}
   &&\hskip -0.7in+\Bigg(\frac{1}{2} \sqrt{-{\cal A}_{\theta_{2}\theta_{2}} {G^M_{11,11}}^{2} G^M_{yy}-B^{IIA}_{\theta_2y}\ ^2+G^M_{11,11} {G^M_{\theta_{2}y}}^2} \sqrt{{G^M_{x^1x^1}}^3 {G^M_{11,11}}^{5/2} G^M_{rr}
   G^M_{tt} {r}_h^2 e^{2 Z}}\nonumber\\
& & \hskip -0.7in\times  \left(3q_{4}(Z)-q_{1}(Z)-q_{2}(Z)-G^M_{11}\sqrt{G^M_{11,11}}q_5(Z)\frac{\partial_{\mu}\partial^{\mu}}{M_g^2}\right)\nonumber\\
   &&\hskip -0.7in -\frac{q_{6}(Z) \left(9 G^{M}_{\theta_{2}\theta_{2}}
   G^M_{11,11} G^M_{yy}+5 B^{IIA}_{\theta_2y}\ ^2-7 G^M_{11,11} {G^M_{\theta_{2}y}}^2\right) \sqrt{{G^M_{x^1x^1}}^3 {G^M_{11,11}}^{5/2} G^M_{rr} G^M_{tt} {r}_h^2 e^{2
   Z}}}{4 \sqrt{G^M_{11,11} \left({G^M_{\theta_{2}y}}^2-G^{M}_{\theta_{2}\theta_{2}}G^M_{11,11} G^M_{yy}\right)-B^{IIA}_{\theta_2y}\ ^2}}\Bigg)\Bigg(\frac{1}{{G^M_{11}}^2 G^M_{11,11} {G^{M}_{11,11}}^{7/4} }\Bigg)\nonumber\\
   &&\hskip -0.7in \Bigg(\psi_{1}(Z)^2\tilde{F}_{\mu\nu}\tilde{F}^{\mu\nu}\Bigg)G_E(x^1,x^2,x^3).\nonumber\\
   & &
\end{eqnarray}}
\begin{itemize}
\item
${\cal O}_d(h^0){\cal O}_\phi(h^0){\cal O}_F (h):$\\
\end{itemize}
Here:
{\small
\begin{eqnarray}
&&\hskip -0.7in{\cal O}_\phi (h^0)={G^M_{11,11}}^{-3/4}\nonumber\\
 &&\hskip -0.7in{\cal O}_{d}(h^0)=\sqrt{-{\cal A}_{\theta_2\theta_2} {G^M_{11,11}}^{2} G^M_{yy}-B^{IIA}_{\theta_2y}\ ^2+G^M_{11,11} {G^{M}_{\theta_{2} y}}^{2} }\sqrt{G^M_{11} {G^{M}_{11,11}}^{5/2} G^M_{22} G^M_{33}
   G^M_{rr} G^M_{tt} {r}_h^2 e^{2 Z}}\nonumber\\
&&\hskip -0.7in{\cal O}_{F}(h)=-\frac{2 F_{12}^2 h_{11}}{{G^M_{11}}^2 G^M_{11,11} G^M_{22}}-\frac{2 F_{12}^2 h_{11,11}}{G^M_{11} {G^M_{11,11}}^2 G^M_{22}}-\frac{2 F_{12}^2
   h_{22}}{G^M_{11} G^M_{11,11} {G^M_{22}}^2}-\frac{4 F_{12} F_{13} h_{23}}{G^M_{11} G^M_{11,11} G^M_{22} G^M_{33}}-\frac{4 F_{12}
   F_{1Z} h_{2r} e^{-Z}}{G^M_{11} G^M_{11,11} G^M_{22} G^M_{rr} {r}_h}\nonumber\\
   &&+\frac{4 F_{12} F_{23} h_{13}}{G^M_{11} G^M_{11,11} G^M_{22}
   G^M_{33}}+\frac{4 F_{12} F_{2Z} h_{1r} e^{-Z}}{G^M_{11} G^M_{11,11} G^M_{22} G^M_{rr} {r}_h}-\frac{2 F_{13}^2 h_{11}}{{G^M_{11}}^2
   G^M_{11,11} G^M_{33}}-\frac{2 F_{13}^2 h_{11,11}}{G^M_{11} {G^M_{11,11}}^2 G^M_{33}}-\frac{2 F_{13}^2 h_{33}}{G^M_{11} G^M_{11,11}
   {G^M_{33}}^2}\nonumber\\
   &&-\frac{4 F_{13} F_{1Z} h_{3r} e^{-Z}}{G^M_{11} G^M_{11,11} G^M_{33} G^M_{rr} {r}_h}-\frac{4 F_{13} F_{23}
   h_{12}}{G^M_{11} G^M_{11,11} G^M_{22} G^M_{33}}+\frac{4 F_{13} F_{3Z} h_{1r} e^{-Z}}{G^M_{11} G^M_{11,11} G^M_{33} G^M_{rr}
   {r}_h}-\frac{2 F_{1Z}^2 h_{11} e^{-2 Z}}{{G^M_{11}}^2 G^M_{11,11} G^M_{rr} {r}_h^2}-\frac{2 F_{1Z}^2 h_{11,11} e^{-2 Z}}{G^M_{11}
   {G^M_{11,11}}^2 G^M_{rr} {r}_h^2}\nonumber\\
   &&-\frac{2 F_{1Z}^2 h_{rr} e^{-2 Z}}{G^M_{11} G^M_{11,11} {G^M_{rr}}^2 {r}_h^2}-\frac{4 F_{1Z} F_{2Z}
   h_{12} e^{-2 Z}}{G^M_{11} G^M_{11,11} G^M_{22} G^M_{rr} {r}_h^2}-\frac{4 F_{1Z} F_{3Z} h_{13} e^{-2 Z}}{G^M_{11} G^M_{11,11} G^M_{33}
   G^M_{rr} {r}_h^2}-\frac{2 F_{23}^2 h_{11,11}}{{G^M_{11,11}}^2 G^M_{22} G^M_{33}}-\frac{2 F_{23}^2 h_{22}}{G^M_{11,11} {G^M_{22}}^2
   G^M_{33}}\nonumber\\
   &&-\frac{2 F_{23}^2 h_{33}}{G^M_{11,11} G^M_{22} {G^M_{33}}^2}-\frac{4 F_{23} F_{2Z} h_{3r} e^{-Z}}{G^M_{11,11} G^M_{22} G^M_{33}
   G^M_{rr} {r}_h}+\frac{4 F_{23} F_{3Z} h_{2r} e^{-Z}}{G^M_{11,11} G^M_{22} G^M_{33} G^M_{rr} {r}_h}-\frac{2 F_{2Z}^2 {h1111} e^{-2
   Z}}{{G^M_{11,11}}^2 G^M_{22} G^M_{rr} {r}_h^2}-\frac{2 F_{2Z}^2 h_{22} e^{-2 Z}}{G^M_{11,11} {G^M_{22}}^2 G^M_{rr} {r}_h^2}\nonumber\\
   &&-\frac{2 F_{2Z}^2
   h_{rr} e^{-2 Z}}{G^M_{11,11} G^M_{22} {G^M_{rr}}^2 {r}_h^2}-\frac{4 F_{2Z} F_{3Z} h_{23} e^{-2 Z}}{G^M_{11,11} G^M_{22} G^M_{33}
   G^M_{rr} {r}_h^2}-\frac{2 F_{3Z}^2 h_{11,11} e^{-2 Z}}{{G^M_{11,11}}^2 G^M_{33} G^M_{rr} {r}_h^2}-\frac{2 F_{3Z}^2 h_{33} e^{-2
   Z}}{G^M_{11,11} {G^M_{33}}^2 G^M_{rr} {r}_h^2}-\frac{2 F_{3Z}^2 h_{rr} e^{-2 Z}}{G^M_{11,11} G^M_{33} {G^M_{rr}}^2 {r}_h^2},\nonumber\\
   & &
\end{eqnarray}}
yielding:
{\small
\begin{eqnarray}
&& \hskip -0.5in{\cal L}_{{\cal O}_d(h^0){\cal O}_\phi(h^0){\cal O}_F (h)}=\sqrt{-{\cal A}_{\theta_{2}\theta_{2}} {G^{M}_{11,11}}^2 G^{M}_{yy}-B^{IIA}_{\theta_2y}\ ^2+G^{M}_{11,11} {G^{M}_{\theta_{2}y}}^2} \sqrt{G^M_{x^1x^1} {G^{M}_{11,11}}^{5/2} G^M_{x^2x^2} G^M_{x^3x^3}
   G^{M}_{rr} G^{M}_{tt} {r}_h^2 e^{2 Z}} {G^M_{11,11}}^{-3/4}\nonumber\\
   &&\hskip -0.5in\Bigg(\frac{2 e^{-2 Z} \psi_{1}'(Z)^2\rho _{\mu }^2  G_E(x^{1},x^{2},x^{3}) \left(q_{2}(Z)-q_{4}(Z)-q_{6}(Z)\right)}{G^{M}_{rr}G^M_{11} {G^{M}_{11,11}} {r}_h^2}+\frac{2 e^{-2 Z}  \psi_{1}'(Z)^2\rho _{\mu } \rho _{\nu } q_{5}(Z)
   \partial^{\mu} \partial^{\nu } G_E(x^{1},x^{2},x^{3})}{ G^{M}_{rr} G^M_{11}G^M_{11,11}M_g^2 {r}_h^2}\nonumber\\
   &&\hskip -0.5in+\frac{2 e^{-2 Z} {\partial_{\mu }\pi}^2 \phi_{0}(Z)^2 G_E(x^{1},x^{2},x^{3}) \left(q_{2}(Z)-q_{4}(Z)-q_{6}(Z)\right)}{ G^{M}_{11,11} G^{M}_{rr} G^M_{11}{r}_h^2}+\frac{2 e^{-2 Z} G^M_{11}\phi_{0}(Z)^2 q_{5}(Z) \partial_{\mu }\pi \partial_{\nu }\pi
   \partial^{\mu} \partial^{\nu } G_E(x^{1},x^{2},x^{3})}{ G^{M}_{11,11} G^{M}_{rr}G^M_{11} M_g^2 {r}_h^2}\nonumber\\
   &&\hskip -0.5in+\iota {\phi_{0}{Z}}^2 \psi_{1}(Z)\partial_{\mu}\pi\left[\pi,\rho_{\nu}\right]\frac{4e^{-2Z}}{G^M_{11}G^M_{11,11}G^M_{rr}r_{h}^2}q_5(Z)\frac{\partial^\mu\partial^\nu G_{E}(x^1,x^2,x^3)}{M_{g}^2}+\iota {\phi_{0}{Z}}^2 \psi_{1}(Z)\partial_{\mu}\pi\left[\pi,\rho^{\mu}\right]\frac{4e^{-2Z}}{G^M_{11}G^M_{11,11}G^M_{rr}r_{h}^2}(-q_4(Z)) \nonumber\\
   &&\hskip -0.5in+\tilde{F}^{\mu\nu} \tilde{F}_{\mu\nu} \psi_{1}(Z)^2 G_E(x^{1},x^{2},x^{3})
   \left(-2 q_{4}(Z)-q_{6}(Z)\right)+\frac{2  \psi_{1}(Z)^2 q_{5}(Z)}{{G^M_{11}}^2 G^M_{11,11}}
   \tilde{F}_{\mu l} {\tilde{F}_{\nu}\ ^ l} \frac{\partial^{\mu}\partial^{\nu}G_E(x^{1},x^{2},x^{3})}{M_g^2}\nonumber\\
  &&\hskip -0.5in-\frac{4 e^{-Z}\psi_{1}(Z)\psi_{1}'(Z)q_{3}(Z)}{ G^M_{11}G^M_{rr}G^{M}_{11,11}r_{h} } \rho _{\mu }  F_{\nu }\ ^{\mu }  \frac{\partial^{\nu}G_E(x^{1},x^{2},x^{3})}{M_g^2}\Bigg).
\end{eqnarray}}
Hence, one can write the following glueball-meson interaction Lagrangian up to quartic order in the meson fields:
\begin{eqnarray}
\label{interaction action full}
&&\hskip -0.5in{\rm S}_{int}= {\cal T}{Str}\int \left(\frac{1}{T_h}\right) d^{3}x\Bigg[c_{1}(\partial_\mu\pi)^{2}G_E
+c_{2}\partial_\mu\pi\partial_\nu\pi \frac{\partial^{\mu}\partial^{\nu}}{M^2}G_E\nonumber\\
&& +c_{3}\rho_{\mu}^{2}G+c_{4}\rho_\mu\rho_\nu \frac{\partial^{\mu}\partial^{\nu}}{M^2}G_E + c_{5}\tilde{F}_{\mu\nu}\tilde{F}^{\mu\nu}G_E
+c_{6}\tilde{F}_{\mu\rho}\tilde{F}_\nu^{\ \rho}\frac{\partial^{\mu}\partial^{\nu}}{M^2}G_E\nonumber\\
&&+\iota c_{7}\partial_{\mu}\pi [\pi,\rho^\mu]G_E+\iota c_{8}\partial_{\mu}\pi [\pi,\rho_\nu]\frac{\partial^{\mu}\partial^{\nu}}{M^2}G_E + c_{9}(Z)\rho_\mu \tilde{F}_{\nu}^{\ \mu}\frac{\partial^\nu G_E}{M^2}\nonumber\\
&&+c_{10}\tilde{F}_{\mu\nu}\tilde{F}^{\mu\nu}G_E +c_{11}\partial_{\mu}\pi\partial^{\mu}\pi G_E +c_{12}\rho_{\mu}\rho^{\mu} G_E +\iota c_{13}\partial_{\mu}\pi [\pi,\rho^\mu]G_E \Bigg],
\end{eqnarray}
where:
\begin{equation}
{\cal T}=\frac{-T_{D_{6}}(2\pi\alpha\prime)^2}{4}\int dyd\theta_{2}\delta\Bigg(\theta_{2}-\frac{\alpha_{\theta_{2}}}{N^{3/10}}\Bigg).
\end{equation}

At quadratic order in field strength tensor these are the only interaction terms. Terms with higher order in $\rho_{\mu}$ and $\pi$ can be obtained in the same manner by keeping higher order terms of $F$ in the DBI action. 
Assuming that in (\ref{interaction action full}), $\int_{Z=0}^\infty dZ = \int_{Z=0}^{\log \sqrt{3}b} dZ + \int_{\log \sqrt{3}b}^\infty dZ$, the coefficients $c_{i}s$ setting $q_6(r)=0$, are giver as under:
{\footnotesize
	\begin{eqnarray*}
&& \hskip -0.5in c_{1} =\int dZ\Biggl[ \frac{e^{-2 Z} \phi_0(Z)^2
			\sqrt{-{\cal A}_{\theta_2\theta_2} {G^M_{11,11}}  ^2 G^M_{yy}  -B_{\theta_2y}^2+G^M_{11,11}   {G^M_{\theta_2 y}} ^2}\sqrt{{G^M_{x^1x^1}} ^3 {G^M_{11,11}}  ^{5/2} G^M_{rr}   G^M_{tt} {r_h}^2 e^{2 Z}} }{G^M_{x^1 x^1}  {G^M_{11,11}}  ^{7/4} G^M_{rr}   {r_h}^2}\nonumber\\
		& & \hskip -0.5in \times \left(-\frac{
			q_5 (Z){m^2}}{m^2}-q_1  (Z)-q_2  (Z)+3 q_4 (Z)\right)\nonumber\\
		& & \hskip -0.5in+ \frac{2 e^{-2 Z}
			\phi_0(Z)^2 (q_2  (Z)-q_4 (Z)-q_6 (Z)) \sqrt{-{\cal A}_{\theta_2\theta_2} {G^M_{11,11}}  ^2
				G^M_{yy}  -B_{\theta_2y}^2+G^M_{11,11}   {G^M_{\theta_2 y} }^2} \sqrt{G^M_{x^1 x^1}  {G^M_{11,11}}  ^{5/2} G^M_{x^2 x^2}   G^M_{x^3 x^3}   G^M_{rr}
				G^M_{tt} {r_h}^2 e^{2 Z}}}{G^M_{x^1 x^1}  {G^M_{11,11}}  ^{7/4} G^M_{rr}   {r_h}^2}\Biggr]\nonumber\\
		& &\hskip -0.5in = -\int \frac{dZ}{216 \pi ^2   M_g^2 \alpha _{\theta _1}^3 \alpha
			_{\theta _2}^2}\Biggl\{{g_s} M \sqrt[5]{\frac{1}{N}} {N_f}^2 e^{-4 Z} (e^{4 Z}-1) \phi_0(Z)^2 \left(2 \sqrt[5]{\frac{1}{N}} \alpha _{\theta _2}^2+81 \alpha
		_{\theta _1}^2\right)\nonumber\\
		& &\hskip -0.5in \times (\log(e^Zr_h)) \left(72 a^2 {r_h} e^Z  \log(e^Zr_h)+3 a^2 +2 {r_h}^2 e^{2 Z} \right)
		M_g^2\left( {q_5}(Z) + ({q_1}(Z)-{q_2}(Z)-{q_4}(Z)+2 {q_6}(Z))\right)\Biggr\},\nonumber\\
		& &\hskip -0.5in {\rm which\ for\ }b\sim 0.6\ {\rm yields:}\nonumber\\
		& &\hskip -0.5in \Biggl[-\frac{1.16\times 10^{-7} {{\cal C}^2_{\phi_{0}}} N^{6/5} \alpha _{\theta _1} \alpha _{\theta _2}^2 c_{1_{q4}}}{M {N_f}^2 {r_h}^3}+\frac{15.9759 {{{\cal C}^{UV}_{\phi_0}}}^2 {\log (N)}^5 \sqrt[5]{N} {N_f^{UV}}^2 \log ({r_h}) ({c^{UV}_{2_{ q1}}}-3.01538 {c^{UV}_{2_{ q4}}})}{\sqrt{{g_s}}
   {M^{UV}}^2 {r_h}^6 \alpha _{\theta _1} \alpha _{\theta _2}^2}\Biggr] \nonumber
   \end{eqnarray*}}\\\\
   {\footnotesize
   \begin{eqnarray*}
			&&\hskip -0.5in c_{2} = \int dZ \Biggl[ \frac{2 e^{-2 Z} \phi _0(Z)^2 {q_5}(Z) \sqrt{-{\cal A}_{\theta_2\theta_2} {G^M_{11\ 11}}^2 G^M_{yy}\ -{B^{IIA}_{\theta_2y}}^2+G^M_{11\ 11}
				{G^M_{\theta_2 y}}^2} \sqrt{G^M_{x^1 x^1} {G^M_{11\ 11}}^{5/2} G^M_{x^2 x^2} G^M_{x^3 x^3}   G^M_{rr} G^M_{tt} {r_h}^2 e^{2 Z}}}{G^M_{x^1 x^1}
			{G^M_{11\ 11}}^{7/4} G^M_{rr} {r_h}^2} \nonumber\\
		& &\hskip -0.5in =\int  \frac{dZ}{108 \pi ^2  \alpha _{\theta _1}^3 \alpha _{\theta _2}^2}\Biggl\{{g_s} M \sqrt[5]{\frac{1}{N}} {N_f}^2 e^{-4 Z} \sqrt{e^{4 Z}-1} \phi_0(Z)^2 {q_5}(Z) \left(2 \sqrt[5]{\frac{1}{N}} \alpha _{\theta _2}^2+81
		\alpha _{\theta _1}^2\right)\nonumber\\
		& &\hskip -0.5in \times (\log(e^Zr_h)) \left(72 a^2   {r_h} e^Z (\log(e^Zr_h))+3 a^2 +2 {r_h}^2 e^{2 Z}\right)\Biggr\}\Biggr]\nonumber\\
		& &\hskip -0.5in {\rm which\ for\ }b\sim 0.6\ {\rm yields:}\nonumber\\
		& &\hskip -0.5in \Biggl[-\frac{2.32\times 10^{-7} {{\cal C}^2_{\phi_{0}}} N^{6/5} \alpha _{\theta _1} \alpha _{\theta _2}^2  c_{1_{{ q4}}}}{M {N_f}^2 {r_h}^3}-\frac{31.9518 {{\cal C}^{UV}_{\phi_{0}}}^2 {\log(N)}^5 \sqrt[5]{N} {N^{UV}_{f}}^2 \log ({r_h}) ({{c_2}^{UV}_{q1}}-3.01538 {{c_2}^{UV}_{q4}})}{\sqrt{{g^{UV}_s}}
   {M^{UV}}^2 {r_h}^6 \alpha _{\theta _1} \alpha _{\theta _2}^2}\Biggr]\nonumber
   \end{eqnarray*}}
   {\footnotesize
   \begin{eqnarray*}
		&&\hskip -0.5in c_{3} = \int dZ \Biggl[ \frac{e^{-2 Z} \psi_1'(Z)^2 \sqrt{-{\cal A}_{\theta_2\theta_2} {G^M_{11\ 11}}^2 G^M_{yy}\ -{B^{IIA}_{\theta_2y}}^2+G^M_{11\ 11}\  {G^M_{\theta_2 y}}^2}
		\sqrt{{G^M_{x^1x^1}}^3 {G^M_{11\ 11}}^{5/2} G^M_{rr} G^M_{tt} {r_h}^2 e^{2 Z}}}{G^M_{x^1x^1} {G^M_{11\ 11}}^{7/4} G^M_{rr} {r_h}^2}\nonumber\\
	& & \hskip -0.5in \times \left(-\frac{
		{q_5}(Z)m^2}{m^2}-{q_1}(Z)-{q_2}(Z)+3 {q_4}(Z)\right)\nonumber\\
	& & \hskip -0.5in +\frac{2 e^{-2 Z}
			\psi_1'(Z)^2 ({q_2}(Z)-{q_4}(Z)-{q_6}(Z)) \sqrt{-{\cal A}_{\theta_2\theta_2} {G^M_{11\ 11}}^2
				G^M_{yy}\ -{B^{IIA}_{\theta_2y}}^2+G^M_{11\ 11}\  {G^M_{\theta_2 y}}^2} \sqrt{G^M_{x^1x^1} {G^M_{11\ 11}}^{5/2} G^M_{x^2x^2} G^M_{x^3x^3}   G^M_{rr}
				G^M_{tt} {r_h}^2 e^{2 Z}}}{G^M_{x^1x^1} {G^M_{11\ 11}}^{7/4} G^M_{rr} {r_h}^2}\Biggr]\nonumber\\
		& &\hskip -0.5in = -\int  \frac{dZ}{216 \pi ^2   M_g^2 \alpha _{\theta _1}^3 \alpha
			_{\theta _2}^2}\Biggl\{{g_s} M \sqrt[5]{\frac{1}{N}} {N_f}^2 e^{-4 Z} \left(e^{4 Z}-1\right)
 \left(2 \sqrt[5]{\frac{1}{N}} \alpha _{\theta _2}^2+81 \alpha _{\theta _1}^2\right)\nonumber\\
		& &\hskip -0.5in \times
		\psi_1'(Z)^2 (\log(e^Zr_h)) \left(72 a^2 {r_h} e^Z (\log(e^Zr_h))+3 a^2 +2 {r_h}^2 e^{2 Z} \right)
	M_g^2\left( {q_5}(Z) + ({q_1}(Z)-{q_2}(Z)-{q_4}(Z)+2 {q_6}(Z))\right)\Biggr\}\nonumber\\
	& &\hskip -0.5in {\rm which\ for\ }b\sim 0.6\ {\rm yields:}\nonumber\\
	& &\hskip -0.5in \Biggl[\frac{0.68 {g_s}^2 M N^{4/5} {N_f}^2 \Biggl(\upsilon_2+\frac{\upsilon_1 g_s M^{2}(m_0^2-4)\log(r_h)}{N}\Biggr)^{3/2} {r_h} {c_{{\psi_1}}}^2 c_{1_{q4}} \log ^2({r_h})}{\alpha _{\theta _1} \alpha
   _{\theta _2}^2}+\frac{821.55 {\log(N)}^5 \sqrt[5]{N} {N^{UV}_{f}}^2 \log ({r_h}) {c^{UV}_{2_{\psi_{1}}}}^2 (1. {{c^{UV}_2}_{q1}}-3.01
   {{c^{UV}_2}_{q4}})}{\sqrt{{g^{UV}_s}} {M^{UV}}^2 {r_h}^6 \alpha _{\theta _1} \alpha _{\theta _2}^2}\Biggr]\nonumber\\
   & &
				\end{eqnarray*}
	\begin{eqnarray*}
	&&\hskip -0.5in c_{4} = \int dZ \Biggl[\frac{2 e^{-2 Z} {q_5}(Z) \psi_1'(Z)^2 \sqrt{-{\cal A}_{\theta_2\theta_2} {G^M_{11\ 11}}^2 G^M_{yy}\ -{B^{IIA}_{\theta_2y}}^2+G^M_{11\ 11}\
				{G^M_{\theta_2 y}}^2} \sqrt{G^M_{x^1x^1} {G^M_{11\ 11}}^{5/2} G^M_{x^2x^2} G^M_{x^3x^3}   G^M_{rr} G^M_{tt} {r_h}^2 e^{2
					Z}}}{{G^M_{11\ 11}}^{7/4} G^M_{x^2x^2} G^M_{rr} {r_h}^2}\Biggr]\nonumber\\
		& &\hskip -0.5in =\int \frac{dZ}{108 \pi ^2  \alpha _{\theta _1}^3 \alpha _{\theta _2}^2}\Biggl\{{g_s} M \sqrt[5]{\frac{1}{N}} {N_f}^2 e^{-4 Z} \left(e^{4 Z}-1\right) {q_5}(Z) \left(2 \sqrt[5]{\frac{1}{N}} \alpha _{\theta _2}^2+81 \alpha _{\theta
			_1}^2\right) \psi_1'(Z)^2 (\log(e^Zr_h))\nonumber\\
		& & \hskip -0.5in \times \left(3 a^2 +72 a^2 {r_h} e^Z  (\log(e^Zr_h))+2 {r_h}^2 e^{2 Z} \log
		(N)\right)\Biggr\}\nonumber\\
		& &\hskip -0.5in {\rm which\ for\ }b\sim 0.6\ {\rm yields:}\nonumber\\
		& &\hskip -0.5in \Biggl[-\frac{1.36 {g_s}^2 M N^{4/5} {N_f}^2 \Biggl(\upsilon_2+\frac{\upsilon_1 g_s M^{2}(m_0^2-4)\log(r_h)}{N}\Biggr)^{3/2} {r_h} {c_{{\psi_{1}}}}^2 c_{1_{q4}} \log ^2({r_h})}{\alpha _{\theta _1} \alpha
   _{\theta _2}^2}-\frac{1643.11 {\log(N)}^5 \sqrt[5]{N} {N^{UV}_{f}}^2 \log ({r_h}) {{c_2}^{UV}_{\psi_{1}}}^2 (1. {{c_2}^{UV}_{q1}}-3.01
   {{c_2}^{UV}_{q4}})}{\sqrt{{g^{UV}_s}} {M^{UV}}^2 {r_h}^6 \alpha _{\theta _1} \alpha _{\theta _2}^2}\Biggr]\nonumber\\
		&&\hskip -0.5in c_{5} =\int dZ\Biggl[ \frac{\psi_1(Z)^2 \sqrt{-{\cal A}_{\theta_2\theta_2} {G^M_{11\ 11}}^2 G^M_{yy}\ -{B^{IIA}_{\theta_2y}}^2+G^M_{11\ 11}\  {G^M_{\theta_2 y}}^2}
			\sqrt{{G^M_{x^1 x^1}}^3 {G^M_{11\ 11}}^{5/2} G^M_{rr} G^M_{tt} {r_h}^2 e^{2 Z}}}{2 {G^M_{x^1 x^1}}^2 {G^M_{11\ 11}}^{7/4}}\nonumber\\
& & \hskip -0.5in \times  \left(-\frac{
				{q_5}(Z)m^2}{m^2}-{q_1}(Z)-{q_2}(Z)+3 {q_4}(Z)\right)\nonumber\\
			& & \hskip -0.5in +\frac{\psi_1(Z)^2 (-2
			{q_4}(Z)-{q_6}(Z)) \sqrt{-{\cal A}_{\theta_2\theta_2} {G^M_{11\ 11}}^2 G^M_{yy}\ -{B^{IIA}_{\theta_2y}}^2+G^M_{11\ 11}\  {G^M_{\theta_2 y}}^2}
			\sqrt{G^M_{x^1x^1} {G^M_{11\ 11}}^{5/2} G^M_{x^2x^2} G^M_{x^3x^3}   G^M_{rr} G^M_{tt} {r_h}^2 e^{2 Z}}}{{G^M_{x^1x^1}}^2 {G^M_{11\ 11}}^{7/4}}\Biggr]\nonumber\\
		& &\hskip -0.5in -\int\frac{dZ}{108 \pi  M_g^2 {r_h}^2  \alpha _{\theta _1}^3 \alpha _{\theta _2}^2}\Biggl\{{g_s}^2 M N^{3/5} {N_f}^2 e^{-2 Z}  \psi_1(Z)^2 \left(81 \sqrt[5]{N} \alpha _{\theta _1}^2+2 \alpha _{\theta _2}^2\right) (\log
		({r_h})+Z)\nonumber\\
		& &\hskip -0.5in \times \left(72 a^2 {r_h} e^Z (\log(e^Zr_h))-3 a^2 +2 {r_h}^2 e^{2 Z} \right)
		 M_g^2\left( {q_5}(Z) + ({q_1}(Z)+{q_2}(Z)+{q_4}(Z)+2 {q_6}(Z))\right)\Biggr\}\nonumber\\
		& &\hskip -0.5in {\rm which\ for\ }b\sim 0.6\ {\rm yields:}\nonumber\\
		& &\hskip -0.5in  \Biggl[\frac{-55.75 {g_s}^3 M N^{9/5} {N_f}^2 \sqrt{\Biggl(\upsilon_2+\frac{\upsilon_1 g_s M^{2}(m_0^2-4)\log(r_h)}{N}\Biggr)} {c^2_{{\psi_{1}}}} c_{1_{q4}} \log ^2({r_h})}{{r_h} \alpha _{\theta _1} \alpha
   _{\theta _2}^2}\nonumber\\
   & & \hskip -0.5in +\frac{176.96 \sqrt{{g^{UV}_s}} {\log(N)}^5 N^{6/5} {N^{UV}_{f}}^2 \log ({r_h}) {c^{UV}_{2_{\psi_{1}}}}^2 (0.0196 {{c_2}^{UV}_{q4}}-0.006
   {{c_2}^{UV}_{q1}})}{{M^{UV}}^2 {r_h}^8 \alpha _{\theta _1} \alpha _{\theta _2}^2}\Biggr]\nonumber\\
   &&\hskip -0.5in c_{6} = \int dZ \Biggl[ \frac{2 \psi_1(Z)^2 {q_5}(Z) \sqrt{-{\cal A}_{\theta_2\theta_2} {G^M_{11\ 11}}^2 G^M_{yy}\ -{B^{IIA}_{\theta_2y}}^2+G^M_{11\ 11}\  {G^M_{\theta_2 y}}^2}
			\sqrt{G^M_{x^1x^1} {G^M_{11\ 11}}^{5/2} G^M_{x^2x^2} G^M_{x^3x^3}   G^M_{rr} G^M_{tt} {r_h}^2 e^{2 Z}}}{{G^M_{x^1x^1}}^2 {G^M_{11\ 11}}^{7/4}}\Biggr]\nonumber\\
		& &\hskip -0.5in =\int  \frac{dZ}{27 \pi
			{r_h}^2  \alpha _{\theta _1}^3 \alpha _{\theta _2}^2}\Biggl\{{g_s}^2 M N^{3/5} {N_f}^2 e^{-2 Z} \psi_1(Z)^2 {q_5}(Z) \left(81 \sqrt[5]{N} \alpha _{\theta _1}^2+2 \alpha _{\theta
			_2}^2\right)\nonumber\\
		& &\hskip -0.5in \times (\log(e^Zr_h)) \left(72 a^2 {r_h} e^Z  (\log(e^Zr_h))-3 a^2 +2 {r_h}^2 e^{2 Z} \right)\Biggr\}\nonumber\\
		& &\hskip -0.5in {\rm which for\ }b\sim 0.6\ {\rm yields:}\nonumber\\
		& &\hskip -0.9in \Biggl[\frac{223.007 {g_s}^3 M N^{9/5} {Nf}^2 \sqrt{\Biggl(\upsilon_2+\frac{\upsilon_1 g_s M^{2}(m_0^2-4)\log(r_h)}{N}\Biggr)} {c^2_{{\psi_{1}}}} c_{1_{q4}} \log ^2({r_h})}{{r_h} \alpha _{\theta _1} \alpha
   _{\theta _2}^2}+\frac{4.60 \sqrt{{g^{UV}_s}} {\log(N)}^5 N^{6/5} {N^{UV}_{f}}^2 \log ({r_h}) {c^{UV}_{2_{\psi_{1}}}}^2 (1. {{c_2}^{UV}_{q1}}-3.015
   {{c_2}^{UV}_{q4}})}{{M^{UV}}^2 {r_h}^8 \alpha _{\theta _1} \alpha _{\theta _2}^2}\Biggr]\nonumber\\
   & &
			\end{eqnarray*}
		\begin{eqnarray}
	\label{interaction coefficients}
	&&\hskip -0.5in   c_{7} = \int dZ \Biggl[\frac{2 e^{-2 Z} \phi_0(Z)^2 \psi_1(Z) \sqrt{-{\cal A}_{\theta_2\theta_2} {G^M_{11\ 11}}^2 G^M_{yy}\ -{B^{IIA}_{\theta_2y}}^2+G^M_{11\ 11}\
			{G^M_{\theta_2 y}}^2} \sqrt{{G^M_{x^1 x^1}}^3 {G^M_{11\ 11}}^{5/2} G^M_{rr} G^M_{tt} {r_h}^2 e^{2 Z}} }{G^M_{x^1x^1} {G^M_{11\ 11}}^{7/4} G^M_{rr} {r_h}^2}\nonumber\\
		& & \hskip -0.5in \times \left(-\frac{
			{q_5}(Z)m^2}{m^2}-{q_1}(Z)-{q_2}(Z)+3 {q_4}(Z)\right)\nonumber\\
			&&\hskip -0.5in +\frac{4 e^{-2 Z}
		\phi_0(Z)^2 \psi_1(Z) ({q_2}(Z)-{q_4}(Z)-{q_6}(Z)) \sqrt{-{\cal A}_{\theta_2\theta_2} {G^M_{11\ 11}}^2
		G^M_{yy}\ -{B^{IIA}_{\theta_2y}}^2+G^M_{11\ 11}\  {G^M_{\theta_2 y}}^2} \sqrt{G^M_{x^1 x^1} {G^M_{11\ 11}}^{5/2} G^M_{x^2 x^2} G^M_{x^3x^3}   G^M_{rr}
			G^M_{tt} {r_h}^2 e^{2 Z}}}{G^M_{x^1x^1} {G^M_{11\ 11}}^{7/4} G^M_{rr} {r_h}^2}\Biggr]\nonumber\\
	& &\hskip -0.5in = -\int \frac{dZ}{108 \pi ^2 M_g^2  \alpha _{\theta _1}^3
		\alpha _{\theta _2}^2}\Biggl\{{g_s} M \sqrt[5]{\frac{1}{N}} {N_f}^2 e^{-2 Z} \left(e^{4 Z}-1\right) \phi_0(Z)^2 \psi_1(Z) \left(2 \sqrt[5]{\frac{1}{N}} \alpha _{\theta
		_2}^2+81 \alpha _{\theta _1}^2\right)\nonumber\\
	& &\hskip -0.5in \times (\log(e^Zr_h)) \left(72 a^2 {r_h} e^Z  (\log(e^Zr_h))+3 a^2 +2 {r_h}^2 e^{2 Z}
	\log (N)\right)\nonumber\\
	& &\hskip -0.5in \times M_g^2\left( {q_5}(Z) + ({q_1}(Z)-{q_2}(Z)-{q_4}(Z)+2 {q_6}(Z))\right)\Biggr\}\nonumber\\
	& &\hskip -0.5in {\rm which\ for\ }b\sim 0.6\ {\rm yields:}\nonumber\\
	& &\hskip -0.9in \Biggl[-\frac{3.28\times 10^{-7} {{\cal C}_{\phi_{0}}}^2 N^{6/5} (\Biggl(\upsilon_2+\frac{\upsilon_1 g_s M^{2}(m_0^2-4)\log(r_h)}{N}\Biggr))^{1/4} \alpha _{\theta _1} \alpha _{\theta _2}^2
   c_{{\psi_{1}}} c_{1_{q4}}}{M {N_f}^2{r_h}^3}+\frac{2666.71 {{\cal C}^{UV}_{\phi_{0}}}^2 {\log(N)}^5 \sqrt[5]{N} {N^{UV}_{f}}^2 \log ({r_h}) {{c_2}^{UV}_{\psi_{1}}} (1. {{c_2}^{UV}_{q1}}-3.015
   {{c_2}^{UV}_{q4}})}{\sqrt{{g^{UV}_s}} {M^{UV}}^2 {r_h}^6 \alpha _{\theta _1} \alpha _{\theta _2}^2}\Biggr]\nonumber\\
	&&\hskip -0.5in c_{8} = \int dZ \Biggl[\frac{4 e^{-2 Z} \phi_0(Z)^2 \psi_1(Z) {q_5}(Z) \sqrt{-{\cal A}_{\theta_2\theta_2} {G^M_{11\ 11}}^2
			G^M_{yy}\ -{B^{IIA}_{\theta_2y}}^2+G^M_{11\ 11}\  {G^M_{\theta_2 y}}^2} \sqrt{G^M_{x^1x^1} {G^M_{11\ 11}}^{5/2} G^M_{x^2x^2} G^M_{x^3x^3}   G^M_{rr}
			G^M_{tt} {r_h}^2 e^{2 Z}}}{G^M_{x^1x^1} {G^M_{11\ 11}}^{7/4} G^M_{rr} {r_h}^2}\Biggr]\nonumber\\
	& &\hskip -0.5in = \int \frac{dZ}{54 \pi ^2 \alpha _{\theta _1}^3 \alpha _{\theta _2}^2}\Biggl\{{g_s} M \sqrt[5]{\frac{1}{N}} {N_f}^2 e^{-4 Z} \left(e^{4 Z}-1\right) \phi_0(Z)^2 \psi_1(Z) {q_5}(Z) \left(2 \sqrt[5]{\frac{1}{N}} \alpha
	_{\theta _2}^2+81 \alpha _{\theta _1}^2\right)\nonumber\\
	& &\hskip -0.5in \times (\log(e^Zr_h)) \left(72 a^2 {r_h} e^Z (\log(e^Zr_h))+3 a^2 +2 {r_h}^2
	e^{2 Z} \right)\Biggr\}\nonumber\\
	& &\hskip -0.5in {\rm which\ for\ }b\sim 0.6\ {\rm yields:}\nonumber\\
	& &\hskip -0.8in \Biggl[\frac{6.57\times 10^{-7} {{\cal C}^2_{\phi_{0}}} N^{6/5} \Biggl(\upsilon_2+\frac{\upsilon_1 g_s M^{2}(m_0^2-4)\log(r_h)}{N}\Biggr)^{1/4} \alpha _{\theta _1} \alpha _{\theta _2}^2
   {c}_{\psi_1} c_{1_{q4}}}{M {N_f}^2 {r_h}^3}-\frac{5333.42 {{\cal C}^{UV}_{\phi_{0}}}^2 {\log(N)}^5 \sqrt[5]{N} {N^{UV}_{f}}^2 \log ({r_h}) {{c_2}^{UV}_{\psi_{1}}} (1. {{c_2}^{UV}_{q1}}-3.015
   {{c_2}^{UV}_{q4}})}{\sqrt{{g^{UV}_s}} {M^{UV}}^2 {r_h}^6 \alpha _{\theta _1} \alpha _{\theta _2}^2}\Biggr]\nonumber\\
	&&\hskip -0.5in c_{9} = -\int dZ \Biggl[\frac{4 e^{-Z} \psi_1(Z) {q_3}(Z) \psi_1'(Z) \sqrt{-{\cal A}_{\theta_2\theta_2} {G^M_{11\ 11}}^2
			G^M_{yy}\ -{B^{IIA}_{\theta_2y}}^2+G^M_{11\ 11}\  {G^M_{\theta_2 y}}^2} \sqrt{G^M_{x^1x^1} {G^M_{11\ 11}}^{5/2} G^M_{x^2x^2} G^M_{x^3x^3}   G^M_{rr}
			G^M_{tt} {r_h}^2 e^{2 Z}}}{G^M_{x^1x^1} {G^M_{11\ 11}}^{7/4} G^M_{rr} {r_h}}\Biggr]\nonumber\\
	& &\hskip -0.5in = -\int \frac{dZ}{54 \pi ^2  \alpha _{\theta _1}^3 \alpha _{\theta _2}^2}\Biggl\{{g_s} M \sqrt[5]{\frac{1}{N}} {N_f}^2 {r_h} e^{-3Z} \left(e^{4 Z}-1\right) \psi_1(Z) {q_3}(Z)
	\left(2 \sqrt[5]{\frac{1}{N}} \alpha _{\theta_2}^2+81 \alpha _{\theta _1}^2\right) \psi_1'(Z) (\log(e^Zr_h))\nonumber\\
	& &\hskip -0.5in \times \left(72 a^2 {r_h} e^Z  (\log(e^Zr_h))+3 a^2+2
	{r_h}^2 e^{2 Z} \right)\Biggr\}\nonumber\\
	& &\hskip -0.5in {\rm which\ for\ }b\sim 0.6\ {\rm yields:}\nonumber\\
	& &\hskip -0.5in \Biggl[-\frac{0.000514915 {g_s}^{3/2} {\log(N)}^5 {m_0}^4 {N_f}^2 \Biggl(\upsilon_2+\frac{\upsilon_1 g_s M^{2}(m_0^2-4)\log(r_h)}{N}\Biggr) {c_{{\psi_{1}}}}^2 c_{1_{q4}} \log ({r_h})}{M^2 {\pi g_s}^2
   {r_h}^2 \alpha _{\theta _1} \alpha _{\theta _2}^2} +\frac{2709.66 {g^{UV}_s} {M^{UV}} \sqrt[5]{\frac{1}{N}} {N^{UV}_{f}}^2 {r_h}^3 \log ({r_h}) {c_1}^{UV}_{q3} {{c_2}^{UV}_{\psi_{1}}}^2}{\alpha
   _{\theta _1} \alpha _{\theta _2}^2}\Biggr],	\nonumber\\
   & &
	\end{eqnarray}}
where:
\begin{eqnarray}
\label{defs}
& & {\cal A}_{\theta_2\theta_2} \equiv \frac{9 {g_s}^{7/2} M^2 N^{11/10} {N_f}^4 e^{-2 Z} \log ^2\left({r_h} e^Z\right) \left(36 a^2 \log
	\left({r_h} e^Z\right)+{r_h} e^Z\right)^2}{2 \pi ^{5/2} {r_h}^2 \alpha _{\theta _1}^2 \alpha _{\theta _2}^4}.
\end{eqnarray}

  \section{Decay widths}

In this section, using standard techniques in scattering theory (specially in dealing with multi-particle phase-space integrals: see \cite{Savage-Phase-Space-Lecture-Notes}, \cite{Murayama-Phase-Space}\footnote{We would like to thank M.Dhuria for bringing \cite{Murayama-Phase-Space} to our attention.}), in the following sub-sections, we calculate decay widths for $G_E\rightarrow2\pi, G_E\rightarrow2\rho, \rho\rightarrow2\pi, G_E\rightarrow 4\pi^0, G_E\rightarrow\rho+2\pi$ as well as indirect four-$\pi$ decay with associated with $G_E\rightarrow\rho+2\pi\rightarrow 4\pi$ as well as $G_E\rightarrow 2\rho\rightarrow 4\pi$ assuming $M_G>2M_\rho$ for definiteness and specifically concentrating on the potential glueball candidate $f0[1710]$.

  \subsection{$G_E\rightarrow2\pi$}
  The decay width for two body decay is given as,
  \begin{eqnarray}
 &&\Gamma=\frac{S}{8 m^{2}}|{\cal M}|^{2}
 \end{eqnarray}
  where ${\cal M}$ is the amplitude for the decay, and $\bold{p}$ is the final momentum of one of the identical particles in the decay product. The relevant coupling for the $2\pi$ decay in the rest frame of the glueball is given by following terms in the interaction lagrangian
  \begin{eqnarray}
  {\cal T}\left(\frac{1}{T_h}\right){\rm Str}\left( c_{1}(\partial_\mu\pi)^{2}G_E
+ c_{2}\partial_\mu\pi\partial_\nu\pi \frac{\partial^{\mu}\partial^{\nu}}{M_g^2}G_E + c_{11}\partial_{\mu}\pi\partial^{\mu}\pi G_E\right)
\end{eqnarray}

Considering a specific adjoint index for the pion $\pi^{a}$(a=1,2,3). ${\cal M}$ for two pions $\pi^{1}$ and $\pi^{2}$ as final state particles in the rest frame of glueball is given as,
\begin{eqnarray}
\iota{\cal M}=-\iota 2\ {\cal T}\left(\frac{1}{T_h}\right)\Bigg( 2c_{1}\iota k_{1\mu}\iota k_{2}^{\mu}+2c_{2}\iota k_{1\mu}\iota k_{2\nu}\frac{\iota k_{g}^{\mu}\iota k_{g}^{\nu}}{{M_g}^{2}}\Bigg)
\end{eqnarray}
where the factor of 2 is for the symmetry of exchanging the two final state particles. Pions are massless which gives $k^{0}=|\bold{k}|=m/2$ for both the particles, so we obtain
\begin{eqnarray}
\iota{\cal M}&&=-\iota {\cal T}\left(\frac{1}{T_h}\right)\Bigg(-2c_{1}(k_{10}k_{2}^{0}+k_{1i}k_{2}^{i})+2c_{2}k_{10}k_{20}\frac{k_{g}^{0}k_{g}^{0}}{{M_g}^{2}}\Bigg)\nonumber\\
&&=-\iota {\cal T}\left(\frac{1}{T_h}\right)(k_{1}^{0}k_{2}^{0}(2\eta_{00}^{2}c_{2}-2c_{1}\eta_{00})-2c_{1}\bold{k}_{1}.\bold{k}_{2})\nonumber\\
&&=\frac{-\iota M_g^2}{4} {\cal T}\left(\frac{1}{T_h}\right)(2\eta_{00}^{2}c_{2}-2\eta_{00}c_{1}+2c_{1})\nonumber\\
&&= -\iota \frac{{\cal T}}{2}\left(\frac{1}{T_h}\right){M_g}^{2}\left(2c_{1}+c_{2}\right)
\end{eqnarray}
The decay width summed over $a=1,2,3$ is:
\begin{eqnarray}
\label{Gto2pi-decay-width}
& & \hskip -0.5in \Gamma_{G_E\rightarrow\pi\pi}=\frac{|2c_{1}+c_{2}|^{2}{M_g}^{2}}{32}{\cal T}^{2}{\left(\frac{1}{T_h}\right)}^{2}\times3\times\frac{1}{2} \approx \frac{3}{64}c_2^2m_0^2\pi^2{\cal T}^2\nonumber\\
& &  \hskip -0.5in{\rm which\ for\ } b\sim0.6:\nonumber\\
& & \hskip -0.5in = 0.003m_0\Biggl(\frac{1.834 \times 10^{-4} {{\cal C}^{2}_{\phi_0}} N^{7/5} \alpha _{\theta _1}^3 {c_{1_{q_4}}}}{M {N_f}^2 {r_h}^3}+\frac{15.379 {{\cal C}^{UV}_{\phi_0}}^2 {\log(N)}^5 {N^{UV}_f}^2
   \log ({r_h}) ({c^{UV}_{2_{q_1}}}-3.015 {c^{UV}_{2_{q_4}}})}{\sqrt{{g^{UV}_s}} {M^{UV}}^2 {r_h}^6 \alpha _{\theta _1}^3}\Biggl)\nonumber\\
   & & \hskip -0.5in \equiv 0.003m_0\times \Lambda_{G_E\rightarrow2\pi}.
\end{eqnarray}
In our paper, we have assumed $|\log r_h| = \frac{f_{r_h}}{3}\log N, 0<f_{r_h}<1$, or equivalently $r_h=N^{-\frac{f_{r_h}}{3}}$.
From \cite{PDG-2018}, the $2\pi$-decay width per unit mass associated with $f0[1710]$  is $\sim 10^{-2}$. Therefore by a convenient choise of ${\cal C}_{\phi_0}, c_{1\ q_4}, {\cal C}^{UV}_{\phi_0}, {c^{UV}_{2_{q_1}}}-3.015 {c^{UV}_{2_{q_4}}}: \Lambda_{G_E\rightarrow2\pi}\sim10$ - implying a constraint on a linear combination of ${\cal C}_{\phi_0}^2c_{1\ q_4}$ and ${{\cal C}^{UV}_{\phi_0}}^2({c^{UV}_{2_{q_1}}}-3.015 {c^{UV}_{2_{q_4}}})$ - one obtains: $\frac{\Gamma_{G_E\rightarrow2\pi}}{m_0}=10^{-2}$ - clearly an exact match with the PDG-2018 results is also similarly possible.

\subsection{$G_E\rightarrow2\rho$}

We consider the onshell decay for $G_E\rightarrow\rho\rho$.
The differential width is given by
$$d\Gamma=\frac{1}{16\pi}\Sigma_{pol}|{\cal M}|^{2}\frac{S}{m^{2}}d\Omega_{k_{1}}$$
where
$${\cal M}={\cal T}\frac{1}{T_h}\epsilon_{\alpha}(k_{1})\epsilon_{\beta}(k_{2})(A\eta^{\alpha\beta}+B^{\alpha\beta})$$
where expression for A and $B^{\alpha\beta}$ are given as
{\small
\begin{eqnarray}
\label{A-B}
& &\hskip -0.74in A =
\frac{c_6\left(k_1.k_{\rm gl}\right)\left(k_2.k_{\rm gl}\right)}{M_g^2}
 -\frac{c_9\left(k_1+k_2\right).k_{\rm gl}}{2M_g^2}
 -2 \left(c_5 + c_{10}\right) k_1.k_2+c_3+c_{12}\nonumber\\
 & & \hskip -0.74 in {\rm which\ for}\ q_6(Z)=c_{1\ q_1}=0\ {\rm yields}:\nonumber\\
 & & \hskip -0.74 in = c_3 - c_5\left(m_\rho^2 - M_g^2\right) + \frac{c_9}{2} + \frac{c_6}{4}M_g^2\nonumber\\
 & & \hskip -0.74in {\rm For}\ b\sim0.6\ {\rm dominated\ by\ }\frac{c_6}{4}M_g^2;\nonumber\\
& & \hskip -0.74in B^{\alpha\beta}= \frac{1}{2} c_6 \delta _0^{\beta } k_2^{\alpha } k_g^{\beta }+\frac{1}{2} c_6 \delta _0^{\alpha } k_1^{\beta } k_g^{\alpha }+\frac{c_9 \delta _0^{\beta } k_2^{\alpha } k_g^{\beta
   }}{2 M^2}+\frac{c_9 \delta _0^{\alpha } k_1^{\beta } k_g^{\alpha }}{2 M^2}-\frac{c_4 \delta _0^{\alpha }\delta _0^{\beta } k_g^{\alpha}k_g^{\beta }}{M^2}+\frac{c_6 k_{1}.k_{2}\delta _0^{\alpha }\delta _0^{\beta } k_g^{\alpha}k_g^{\beta }}{M^2}+2 c_5 k_2^{\alpha } k_1^{\beta }\nonumber\\
& &
\end{eqnarray}}
 Now using:
\begin{eqnarray}
   k_{1}.k_{2}&&=\frac{1}{2}(2M_{\rho}^{2}-m^{2})\nonumber\\
   &&=\frac{1}{2}\sqrt{m^{4}\lambda(M_{\rho}^{2},M_{\rho}^{2};m^{2})+4M_{\rho}^{4}}\nonumber\\
   &&|\bold{k_{1}}|=\frac{m}{2}\sqrt{\lambda(M_{\rho}^{2},M_{\rho}^{2};m^{2})}
\end{eqnarray}
we can write
\begin{eqnarray}
&&\sum_{\rm pol}={\cal T}^2\Bigg(\frac{1}{T_h}\Bigg)^2\frac{m^{4}}{4M_{\rho}^{4}}\left(A^2\lambda(M_{\rho}^{2},M_{\rho}^{2};m^{2})+8 A^2\frac{M_{\rho}^{4}}{m^{4}}+4\frac{M_{\rho}^{4}}{m^{4}}X(k_{1},k_{2},m,B)\right),
\end{eqnarray}
where:
\begin{eqnarray}
\label{X-1}
& & X=\frac{A c_4 \left(-6 M_g^2 M_{\rho }^2+M_g^4+8 M_{\rho }^4\right)}{4 M_{\rho }^4}-\frac{A c_5
	\left(M_g^2-4 M_{\rho }^2\right) \left(-6 M_g^2 M_{\rho }^2+M_g^4+8 M_{\rho }^4\right)}{2 M_{\rho }^4}\nonumber\\
& & -\frac{A
	c_6 \left(M_g^2+2 M_{\rho }^2\right) \left(-6 M_g^2 M_{\rho }^2+M_g^4+8 M_{\rho }^4\right)}{8 M_{\rho
	}^4}-\frac{A c_9 \left(-6 M_g^2 M_{\rho }^2+M_g^4+8 M_{\rho }^4\right)}{4 M_{\rho }^4}\nonumber\\
& & +\frac{c_4^2
	\left(M_g^2-4 M_{\rho }^2\right){}^2}{16 M_{\rho }^4}-\frac{c_4 c_5 \left(M_g^3-4 M_g M_{\rho
	}^2\right){}^2}{4 M_{\rho }^4}\nonumber\\
& & -\frac{c_4 c_6 \left(-6 M_g^4 M_{\rho }^2+M_g^6+32 M_{\rho }^6\right)}{16
	M_{\rho }^4}-\frac{c_4 c_9 \left(M_g^2-4 M_{\rho }^2\right){}^2}{8 M_{\rho }^4}\nonumber\\
& & +\frac{c_5^2 M_g^4
	\left(M_g^2-4 M_{\rho }^2\right){}^2}{4 M_{\rho }^4}+\frac{c_5 c_6 \left(-5 M_g^6 M_{\rho }^2-4 M_g^4
	M_{\rho }^4+36 M_g^2 M_{\rho }^6+M_g^8\right)}{8 M_{\rho }^4}\nonumber\\
& & +\frac{c_5 c_9 \left(M_g^3-4 M_g M_{\rho
	}^2\right){}^2}{4 M_{\rho }^4}+\frac{c_6^2 \left(-2 M_g^2 M_{\rho }^2+M_g^4-8 M_{\rho }^4\right){}^2}{64
	M_{\rho }^4}\nonumber\\
& & +\frac{c_6 c_9 \left(-6 M_g^4 M_{\rho }^2+M_g^6+32 M_{\rho }^6\right)}{16 M_{\rho
	}^4}+\frac{c_9^2 \left(M_g^2-4 M_{\rho }^2\right){}^2}{16 M_{\rho }^4}.
\end{eqnarray}
For $b\sim0.6$, from the expressions of the coupling constants $c_i$s, (\ref{X-1}) will be dominated by the $c_6^2, A c_6$ and $c_4c_6$ terms (if $M_g = 2 M_\rho + \epsilon, 0<\epsilon\ll M_\rho$ then the $c_4^2$ term will be further suppressed). Demanding $\Gamma_{G_E\rightarrow2\rho} = \Gamma_{G_E\rightarrow4\pi}$ for $M_g>2M_\rho$ \cite{Brunner_Hashimoto-results}, would require $c_4 = \frac{3}{4}c_6 M_g^2$; so for $M_g = m_0$ MeV $\equiv m_0 \left(\frac{r_h}{\pi\sqrt{4\pi g_s N}}\right)$, $c_4 = \frac{3 m_0^2}{4}c_6\left(\frac{r_h}{\pi\sqrt{4\pi g_s N}}\right)^2$.

\subsection{$\rho\rightarrow2\pi$}

The relevant interaction term in the action is given by:
\begin{equation}
\label{rho-pi^2-c_16}
c_{16}{\cal T}\left(\frac{1}{T_h}\right)\int d^3x\partial_\mu\pi[\pi,\rho^\mu],
\end{equation}
where:
\begin{eqnarray}
\label{c_16}
& & c_{16} = \frac{5.61\times 10^{-9} {{\cal C}_{\phi _0}}^2 \sqrt[4]{{\omega_2}} \alpha _{\theta _1} \alpha _{\theta _2}^2 c_{ \psi_1}
   N^{\frac{f_{r_h}}{3}+\frac{1}{5}}}{{g_s} M {N_f}^2}\nonumber\\
	& & -\frac{43017.7 {{{\cal C}^{UV}_{\phi _0}}}^2 {f_{r_h}} {g_s}^{UV} {M^{UV}} \sqrt[5]{\frac{1}{N}} {N_f^{UV}}^2 \log (N) {c^{UV}_{2_{\psi_1}}} N^{-\frac{2 {fr}
   h}{3}}}{\alpha _{\theta _1} \alpha _{\theta _2}^2},
\end{eqnarray}
\begin{eqnarray}
&&\omega_2\equiv \upsilon_2+\frac{\upsilon_1
   {g_s} M^2 \left({m_0}^2-4\right) \log ({r_h})}{N};\nonumber\\
\end{eqnarray}
 $M_{UV}\ll M$ and $N_{f\ UV}\ll N_f$ are the tiny values of the number of fractional $D3$-branes and flavor branes in the UV. The $\rho\rightarrow2\pi$ decay width is hence given as under:
\begin{equation}
\label{rho-to-2-rho-Gamma}
\Gamma_{\rho\rightarrow2\pi} = {\cal T}^2\left(\frac{1}{T_h}\right)^2\frac{c_{16}^2}{2}.
\end{equation}
We will demand $\Gamma_{\rho\rightarrow2\pi} = 149 MeV$ (\cite{PDG-2018}); replacing MeV by $\frac{r_h}{\pi \sqrt{4 \pi g_s N}}$, this implies a constraint on ${\cal C}_{\phi_0}^2 \left(c_{\psi_1} \right)$ and  ${\cal C}_{\phi_0}^{UV}\ ^2c_{2\ \psi_1}^{UV}$:
\begin{eqnarray}
\label{constraint-rho-2pi-decay}
& & \Biggl[\frac{5.61\times 10^{-9} {{\cal C}_{\phi _0}}^2 \sqrt[4]{{\omega_2}} \alpha _{\theta _1} \alpha _{\theta _2}^2 c_{\psi_1}
   N^{\frac{f_{r_h}}{3}+\frac{1}{5}}}{{g_s} M {N_f}^2}\nonumber\\
	& & -\frac{43017.7 {{{\cal C}^{UV}_{\phi _0}}}^2 {f_{r_h}} {g_s}^{UV} {M^{UV}} \sqrt[5]{\frac{1}{N}} {N_f^{UV}}^2 \log (N) {c^{UV}_{2_{ \psi_1}}} N^{-\frac{2 {fr}
   h}{3}}}{\alpha _{\theta _1} \alpha _{\theta _2}^2}\Biggr]^2 = \frac{298}{{\cal T}^2}\left(\frac{r_h}{\sqrt{4\pi g_s N}}\right)^3.
\end{eqnarray}

\subsection{Direct Glueball Decay to 4$\pi^{0}$s }

For coupling to four $\pi^{0}$ we need to expand the DBI action upto quartic order in $F
_{\mu \nu}$. The action restricted to quartic order, reads
\begin{eqnarray}
\label{NLO dbi action}
&&\hskip -0.5in S=-T_{D_{6}}(2\pi\alpha\prime)^{4}{\rm Str}\int d^{4}x dZ d\theta_{2} dy \delta \left(\theta_{2}-\frac{\alpha_{\theta_{2}}}{N^{3/10}}\right)\exp^{-\Phi}\sqrt{-det(\iota^{ *}(g+B))}\nonumber\\
& & \times
\Bigg\{\frac{1}{32}STr\left(g^{-1}Fg^{-1}F)Tr(g^{-1}Fg^{-1}F\right)-\frac{1}{8}STr\left(g^{-1}Fg^{-1}Fg^{-1}Fg^{-1}F\right)\Bigg\}
\end{eqnarray}
Inserting the metric fluctuations corresponding to the glueball and keeping the terms which are quartic in $\phi_{0}(Z)$ gives the interaction term
\begin{itemize}
\item
${\cal O}_d(h^0){\cal O}_F(h^0){\cal O}_\phi (h):$\\
\end{itemize}

{\small
\begin{eqnarray}
&& {\cal L}_{{\cal O}_d(h^0){\cal O}_F(h^0){\cal O}_\phi (h)} = \sqrt{-{\cal A}^M_{\theta_2\theta_2} {G^M_{11,11}}^{2} G^M_{yy}-B^{IIA}_{\theta_2y}\ ^2+G^M_{11,11} {G^M_{\theta_2 y}}^2} \sqrt{G^M_{11} {G^M_{11,11}}^{5/2} G^M_{22} G^M_{33}
   G^M_{rr} G^M_{tt} {r}_h^2 e^{2 Z}}\nonumber\\
& & \times{G^M_{11,11}}^{-3/4}\Bigg(\frac{3 e^{-4 Z} \phi_{0}(Z)^4 q_{6}(Z)\partial_\mu   \pi \partial^{\mu }\pi \partial_\nu   \pi \partial^{\nu }\pi  G_E(x^{1},x^{2},x^{3})}{16
   {G^M_{rr}}^{2} {G^M_{11,11}}^{2} {G^M_{11}}^2{r}_h^4}\Bigg)
\end{eqnarray}}

\begin{itemize}
\item
${\cal O}_\phi (h^0){\cal O}_F(h^0){\cal O}_d (h):$\\
\end{itemize}

{\small
\begin{eqnarray}
&&\hskip -0.7in {\cal L}_{{\cal O}_\phi (h^0){\cal O}_F(h^0){\cal O}_d (h)} = \Bigg(\frac{1}{2} \sqrt{-{\cal A}^{M}_{\theta_{2}\theta_{2}} {G^M_{11,11}}^{2}  G^M_{yy}-B^{IIA}_{\theta_2y}\ ^2+G^M_{11,11} {G^M_{\theta_{2}y}}^2} \sqrt{{G^M_{x^1x^1}}^3 {G^M_{11,11}}^{5/2} G^M_{rr}
   G^M_{tt} {r}_h^2 e^{2 Z}}\nonumber\\
& & \hskip -0.7in \times \left(3q_{4}(Z)-q_{1}(Z)-q_{2}(Z)-q_{5}(Z)\right)\nonumber\\
   &&\hskip -0.7in -\frac{q_{6}(Z) \left(9 {\cal A}^{M}_{\theta_{2}\theta_{2}}
   {G^M_{11,11}}^{2} G^M_{yy}+5 B^{IIA}_{\theta_2y}\ ^2-7 G^M_{11,11} {G^M_{\theta_{2}y}}^2\right) \sqrt{{G^M_{x^1x^1}}^3 {G^M_{11,11}}^{5/2} G^M_{rr} G^M_{tt} {r}_h^2 e^{2
   Z}}}{4 \sqrt{G^M_{11,11} \left({G^M_{\theta_{2}y}}^2-{\cal A}^{M}_{\theta_{2}\theta_{2}}G^M_{11,11} G^M_{yy}\right)-B^{IIA}_{\theta_2y}\ ^2}}\Bigg)\nonumber\\
& & \hskip -0.7in\times{G^M_{11,11}}^{-3/4}\Bigg(\frac{- e^{-4 Z}}{ {G^{M}_{11,11}}^{2} {G^{M}_{rr}}^{2}{G^M_{11}}^2 {r}_h^4}\Bigg)\Bigg(\phi_{0}(Z)^4 \partial_{\nu }\pi \partial^{\nu }\pi \partial_{\mu }\pi \partial^{\mu }\pi  \Bigg)G_E(x^1,x^2,x^3)
\end{eqnarray}}

\begin{itemize}
\item
${\cal O}_d(h^0){\cal O}_\phi(h^0){\cal O}_F (h):$\\
\end{itemize}

{\small
\begin{eqnarray}
   && {\cal L}_{{\cal O}_d(h^0){\cal O}_\phi(h^0){\cal O}_F (h)} = \sqrt{-{\cal A}^{M}_{\theta_{2}\theta_{2}} {G^{M}_{11,11}}^2 G^{M}_{yy}-B^{IIA}_{\theta_2y}\ ^2+G^{M}_{11,11} {G^{M}_{\theta_{2}y}}^2} \sqrt{G^M_{x^1x^1} {G^{M}_{11,11}}^{5/2} G^M_{x^2x^2} G^M_{x^3x^3}
   G^{M}_{rr} G^{M}_{tt} {r}_h^2 e^{2 Z}} \nonumber\\
   &&{G^M_{11,11}}^{-3/4}\Bigg(\frac{e^{-4 Z} {\partial_{\nu }\pi}^2{\partial_{\mu }\pi}^2 \phi_{0}(Z)^4 G_E(x^{1},x^{2},x^{3}) \left(-q_{2}(Z)+q_{4}(Z)+q_{6}(Z)\right)}{ 4{G^{M}_{11,11}}^{2} {G^{M}_{rr}}^{2} {r}_h^4{G^M_{11}}^2}\nonumber\\
   & & -\frac{ e^{-4 Z} \partial_{\sigma }\pi\partial^{\sigma }\pi \partial_{\nu }\pi\partial_{\mu }\pi
   \phi_{0}(Z)^4 q_{5}(Z) \partial^{\mu} \partial^{\nu } G_E(x^{1},x^{2},x^{3})}{ 4{G^{M}_{11,11}}^{2} {G^{M}_{rr}}^{2} M_g^2 {r}_h^4{G^M_{11}}^2}\Bigg)
\end{eqnarray}}

Putting everything together and setting $q_2(Z)=q_6(Z)=0$, one gets the following interaction Lagrangian corresponding to the direct $G_E\rightarrow4\pi$ decay:
\begin{eqnarray}
\label{interaction-G-4pi_direct}
S_{\rm int}^{G_E\rightarrow4\pi} = {\cal T}\left(\frac{1}{T_h}\right){\rm Str}\int d^3x\left(c_{14}\partial_\mu\pi\partial^\mu\pi \partial_\nu\pi\partial^\nu\pi G_E(x^{1,2,3})
+ c_{15}\partial_\sigma\pi\partial^\sigma\pi \partial_\mu\pi\partial_\nu\pi\frac{\partial^\mu\partial^\nu}{M_g^2}G_E(x^{1,2,3})\right),\nonumber\\
& &
\end{eqnarray}
where:
\begin{eqnarray}
\label{c14_c15-IR}
& & \hskip -0.3in c_{14}^{\rm IR} = \int dZ\Biggl[ -\frac{{\cal G} e^{-4 Z} \phi_0(Z)^4 \left(-\frac{{q_1}(Z)}{2}-\frac{{q_2}(Z)}{2}+\frac{3
		{q_4}(Z)}{2}-\frac{{q_5}(Z)}{2}\right)}{8 G^M_{x^1x^1}\ ^2 G^M_{11\ 11}\ ^2 G^M_{rr}\ ^2 {r_h}^4}-\frac{{\cal G} e^{-4 Z}
	\phi_0(Z)^4 ({q_2}(Z)-{q_4}(Z)-{q_6}(Z))}{4 G^M_{x^1x^1}\ ^2 G^M_{11\ 11}\ ^2 G^M_{rr}\ ^2 {r_h}^4}\Biggr]\nonumber\\
& &\hskip -0.3in {\rm which\ for}\ b\sim0.6\ {\rm yields}:\nonumber\\
& & \hskip -0.3in =\frac{6.219\times 10^{-16} {{\cal C}_{\phi _0}}^4 N^{8/5} \alpha _{\theta _1}^3 \alpha _{\theta _2}^6 c_{1_{q4}}}{{g_s}^2 M^3
   {N_f}^6 {r_h}^9 \log ^2({r_h})} \nonumber\\
& &\hskip -0.3in c_{15}^{\rm IR} = \int dZ \Biggl[-\frac{{\cal G} e^{-4 Z} \phi_0(Z)^4 {q_5}(Z)}{4 G^M_{x^1x^1}\ ^2 G^M_{11\ 11}\ ^2 G^M_{rr}\ ^2 {r_h}^4}\Biggr]\nonumber\\
& &\hskip -0.3in  = -\frac{11.224\  {{\cal C}_{\phi _0}}^4 N^{8/5}  \alpha _{\theta _1}^3 \alpha _{\theta _2}^6 c_{1_{q4}}}{{g_s}^2 M^3 {N_f}^6 {r_h}^9 \log ^2({r_h})}.
\end{eqnarray}
From (\ref{c14_c15-IR}), $c_{15}^{\rm IR}>c_{14}^{\rm IR}$. We will drop $c_{14}$ in the direct $4\pi^0$-decay of the glueball decay. One can show that the contribution from the UV: $Z\in\left[\log\left(\sqrt{3}b\right),\infty\right]$ yields:
\begin{equation}
\label{c15-UV}
c_{15}^{\rm UV} = \frac{638116. {{\cal C}_{\phi _0}^{UV}}^4 {\log (N)}^5 \sqrt[5]{N} {{N_{f}}^{UV}}^2 \log ({r_h}) (  {c_{2_{q1}}}^{UV}-3.015 {c_{2_{q4}}}^{UV})}{\sqrt{{g^{UV}_s}}
   {M^{UV}}^2 {r_h}^8 \alpha _{\theta _1} \alpha _{\theta _2}^2}.
\end{equation}
From (\ref{c14_c15-IR}) and (\ref{c15-UV}):
\begin{eqnarray}
\label{c15-IR+UV}
& & c_{15} = {1.35\times 10^{-13}} N^{21/20} \Biggl(\frac{4.72\times 10^{18} {{\cal C}_{\phi _0}^{UV}}^4 {\log (N)}^5  {{N^{UV}_{f}}}^2 \log ({r_h}) (  {c_{2_{q1}}}^{UV}-3.01538 {c_{2_{q4}}}^{UV})}{\sqrt{{g^{UV}_s}}N^{17/20}
   {M^{UV}}^2 {r_h}^8 \alpha _{\theta _1} \alpha _{\theta _2}^2}\nonumber\\
& &  -\frac{8.31\times 10^{13}\  {{\cal C}_{\phi _0}}^4 N^{11/20}  \alpha _{\theta _1}^3 \alpha _{\theta _2}^6 c_{1_{q4}}}{{g_s}^2 M^3 {N_f}^6 {r_h}^9 \log ^2({r_h})}\Biggr).
\end{eqnarray}
One can show for $f0[1710](M_g>2 M_\rho)$:
\begin{eqnarray}
\label{Gamma-direct_4pi0_i}
& & \frac{\Gamma_{G_E\rightarrow4\pi^0}}{m_0} \sim 10^{17} c_{15}^2 \sim 10^{-5}\Biggl(\frac{4.72\times 10^{18} {{\cal C}_{\phi _0}^{UV}}^4 {\log (N)}^5  {{N^{UV}_{f}}}^2 \log ({r_h}) (  {c_{2_{q1}}}^{UV}-3.015 {c_{2_{q4}}}^{UV})}{\sqrt{{g^{UV}_s}}N^{17/20}
   {M^{UV}}^2 {r_h}^8 \alpha _{\theta _1} \alpha _{\theta _2}^2}\nonumber\\
& &  -\frac{8.31\times 10^{13}\  {{\cal C}_{\phi _0}}^4 N^{11/20}  \alpha _{\theta _1}^3 \alpha _{\theta _2}^6 c_{1_{q4}}}{{g_s}^2 M^3 {N_f}^6 {r_h}^9 \log ^2({r_h})}\Biggr)^2.
\end{eqnarray}
Currently \cite{PDG-2018} does not have an entry against the experimental value of $\frac{\Gamma_{G_E\rightarrow4\pi^0}}{m_0}$. Let us say it is $\sim 10^{-5 +{\rm required}_1}, {\rm required}_1$ could be positive or negative \footnote{$\mu\times 10^n = 10^{n + \log_{10}\mu} \equiv 10^{\rm required}$.}. This implies the following constraint:
\begin{eqnarray}
\label{Constraint_direct_4pi0_decay}
& &  \Biggl. \Biggl(\frac{4.72\times 10^{18} {{\cal C}_{\phi _0}^{UV}}^4 {\log (N)}^5  {{N^{UV}_{f}}}^2 \log ({r_h}) (  {c_{2_{q1}}}^{UV}-3.015 {c_{2_{q4}}}^{UV})}{\sqrt{{g^{UV}_s}}N^{17/20}
   {M^{UV}}^2 {r_h}^8 \alpha _{\theta _1} \alpha _{\theta _2}^2}\nonumber\\
& &  -\frac{8.31\times 10^{13}\  {{\cal C}_{\phi _0}}^4 N^{11/20}  \alpha _{\theta _1}^3 \alpha _{\theta _2}^6 c_{1_{q4}}}{{g_s}^2 M^3 {N_f}^6 {r_h}^9 \log ^2({r_h})}\Biggr)^2\Biggr| _{N={10^2}}\sim 10^{{\rm required}_1}.
\end{eqnarray}
\subsection{$G_E\rightarrow\rho+2\pi$}

\begin{figure}[t]
	\begin{center}
		\begin{tabular}{c c}
			\includegraphics[width=5.5cm]{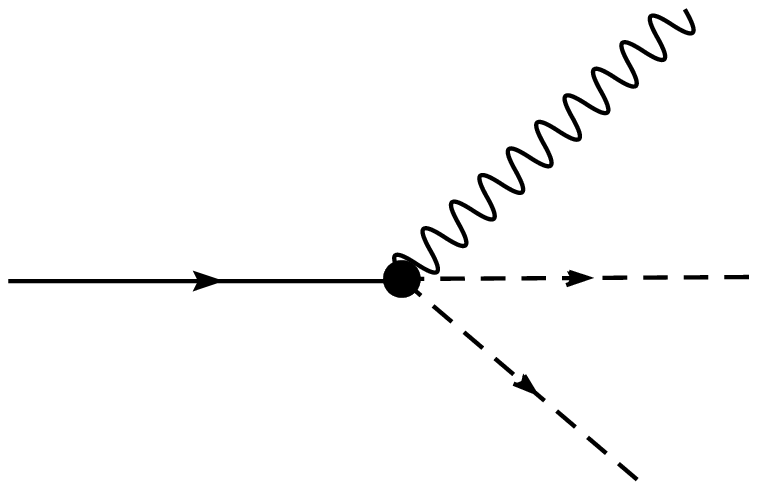}&
			\hspace{1 cm}
			\includegraphics[width=5.5cm]{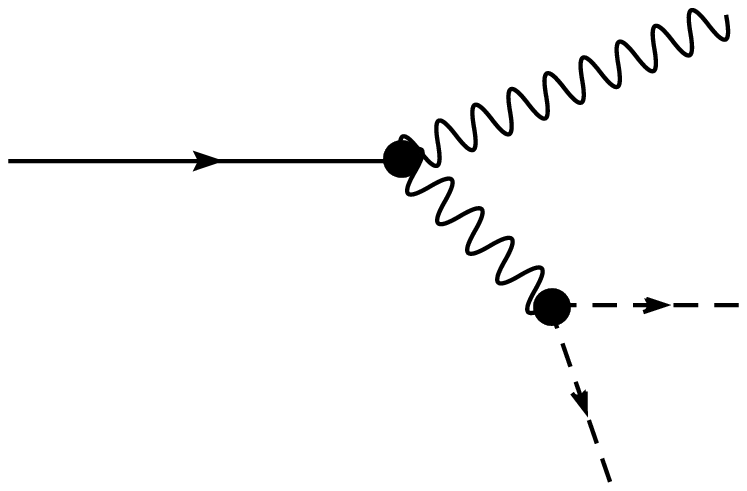}\\
			(a) & (b)
		\end{tabular}
		\caption{$G_E\rightarrow\rho+2\pi$}
	\end{center}
	
\end{figure}

One obtains:
{\footnotesize
\begin{eqnarray}
\label{Gamma-both_and_cross_terms}
& & \hskip -0.8in \Gamma_{(a)} = -3 c_8^2  L^4 {\cal T}^2\int_{k_1=0}^{\frac{\left(M_g^2 - M_\rho^2\right)}{2M_g}}\int_{\frac{\left(M_g^2 - M_\rho^2\right)}{2M_g} - k_1}^{\frac{\left(M_g^2 - M_\rho^2\right)}{2M_g}}{dk_1} {dk_2}\frac{ (k_1-k_2)^2 \left(\frac{\left(M_g-k_1-k_2\right){}^2}{M_{\rho }^2}-1\right)}{4
	k_1 k_2 {r_h}^2 M_g \sqrt{1-\frac{\left(\left(M_g-k_1-k_2\right){}^2-k_1^2-k_2^2-M_{\rho }^2\right){}^2}{4
			k_1^2 k_2^2}}}\nonumber\\
		& &\hskip -0.8in \times\theta\left(1-\frac{\left[\left(M_g-k_1-k_2\right){}^2-k_1^2-k_2^2-M_{\rho }^2\right]{}^2}{4 k_1^2 k_2^2}\right)\nonumber\\
		& &\hskip -0.8in \Gamma_{(b)} = -\frac{3 \pi ^2 c_6^2 c_{16}^2  L^8}{4M_g M_{\rho }^4} {\cal T}^4\int_{k_1=0}^{\frac{\left(M_g^2 - M_\rho^2\right)}{2M_g}}\int_{\frac{\left(M_g^2 - M_\rho^2\right)}{2M_g} - k_1}^{\frac{\left(M_g^2 - M_\rho^2\right)}{2M_g}}{dk_1} {dk_2}\frac{\theta\left(1-\frac{\left[\left(M_g-k_1-k_2\right){}^2-k_1^2-k_2^2-M_{\rho }^2\right]{}^2}{4 k_1^2 k_2^2}\right)}{\sqrt{1-\frac{\left(\left(M_g-k_1-k_2\right){}^2-k_1^2-k_2^2-M_{\rho }^2\right){}^2}{4 k_1^2 k_2^2}}}(k_1-k_2)^2\nonumber\\
		& &\hskip -0.8in \times\frac{  \left((k_1+k_2)^2-M_{\rho }^2\right){}^2
			\left(\frac{\left(M_g-k_1-k_2\right){}^2}{M_{\rho }^2}-1\right) \left(\left(M_g-k_1-k_2\right){}^2+4 k_1
			\left(M_g-k_1-k_2\right)+4 k_2 \left(M_g-k_1-k_2\right)-M_{\rho }^2\right){}^2}{ k_1 k_2
			\left(\left(-\left(M_g-k_1-k_2\right){}^2+k_1^2+2 k_1 k_2+k_2^2+2 M_{\rho }^2\right){}^2+M_{\rho }^2 \Gamma _{\rho
			}^2\right)}\nonumber\\
		& &\hskip -0.8in \Gamma_{(a)(b)^*+(a)^*(b)} = \frac{3 \pi  c_6 c_8 c_{16}  L^6}{2 M_g
			M_{\rho }^2 } {\cal T}^3\int_{k_1=0}^{\frac{\left(M_g^2 - M_\rho^2\right)}{2M_g}}\int_{\frac{\left(M_g^2 - M_\rho^2\right)}{2M_g} - k_1}^{\frac{\left(M_g^2 - M_\rho^2\right)}{2M_g}}{dk_1} {dk_2}\frac{\theta\left(1-\frac{\left[\left(M_g-k_1-k_2\right){}^2-k_1^2-k_2^2-M_{\rho }^2\right]{}^2}{4 k_1^2 k_2^2}\right)}{\sqrt{1-\frac{\left(\left(M_g-k_1-k_2\right){}^2-k_1^2-k_2^2-M_{\rho }^2\right){}^2}{4 k_1^2 k_2^2}}}\frac{(k_1-k_2)^2\left((k_1+k_2)^2-M_{\rho }^2\right)}{k_1 k_2}\nonumber\\
		& &\hskip -0.8in \times\frac{
			\left(\frac{\left(M_g-k_1-k_2\right){}^2}{M_{\rho }^2}-1\right) \left(\left(M_g-k_1-k_2\right){}^2+k_1
			\left(M_g-k_1-k_2\right)+k_2 \left(M_g-k_1-k_2\right)+3 (k_1+k_2)
			\left(M_g-k_1-k_2\right)-M_{\rho }^2\right) }{
			\left(-4 M_{\rho }^2 \left(\frac{1}{2} \left(\left(M_g-k_1-k_2\right){}^2-k_1^2-k_2^2-M_{\rho }^2\right)-k_1
			k_2\right)+4 \left(\frac{1}{2} \left(\left(M_g-k_1-k_2\right){}^2-k_1^2-k_2^2-M_{\rho }^2\right)-k_1
			k_2\right){}^2+M_{\rho }^2 \Gamma _{\rho }^2+M_{\rho }^4\right)}\nonumber\\
		& & \hskip -0.8in\times \left(M_{\rho }^2-2 \left(\frac{1}{2}
		\left(\left(M_g-k_1-k_2\right){}^2-k_1^2-k_2^2-M_{\rho }^2\right)-k_1 k_2\right)\right).
\end{eqnarray}}
Writing:
\begin{eqnarray}
\label{c_6-IR+UV}
& & \hskip -0.7in  c_6 = 10^{-5}N^{\frac{13}{10}}\Biggl(-\frac{2.23\times 10^7 {f_{r_h}}^2 {g_s}^3 M {N_f}^2 \sqrt{{\omega_2}} {c_{\psi_1}}^2 {c_{1_{\ q4}}} \log ^2\left(\sqrt[3]{N}\right)
   N^{\frac{fr_ h}{3}}}{\alpha _{\theta _1} \alpha _{\theta _2}^2}\nonumber\\
& & \hskip -0.7in -\frac{460099 {f_{r_h}} \sqrt{{g_s}} {\log (N)}^5 {N_f^{UV}}^2 \log \left(\sqrt[3]{N}\right) {{c_{2_{\ \psi_1}}}^{UV}}^2 (  {c_{2_{\ q1}}}^{UV}-3.015
   {c_{2_{\ q4}}}^{UV}) N^{\frac{8 {fr}_h}{3}-\frac{1}{10}}}{{M^{UV}}^2 \alpha _{\theta _1} \alpha _{\theta _2}^2}\Biggr),
\end{eqnarray}
working with $f0[1710]: M_g>2M_\rho$ having dropping $\upsilon_2$:
\begin{eqnarray}
\label{Gamma-3body-over-m0}
& & \hskip -0.7in \frac{\Gamma_{G_E\rightarrow\rho+2\pi}}{m_0}\sim \frac{\Gamma_{(b)}}{m_0} \sim c_6^2c_{16}^2\nonumber\\
& & \hskip -0.7in\sim 10^{-10}N^{\frac{13}{5}}\Biggl(-\frac{2.23\times 10^7 {f_{r_h}}^2 {g_s}^3 M {N_f}^2 \sqrt{{\omega_2}} {c_{\psi_1}}^2 {c_{1_{ q4}}} \log ^2\left(\sqrt[3]{N}\right)
   N^{\frac{fr_ h}{3}}}{\alpha _{\theta _1} \alpha _{\theta _2}^2}\nonumber\\
& & \hskip -0.7in -\frac{460099 {f_{r_h}} \sqrt{{g^{UV}_s}} {\log (N)}^5 {N_f^{UV}}^2 \log \left(\sqrt[3]{N}\right) {{c^{UV}_{2_{ \psi_1}}}}^2 (  {c_{2_{q1}}}^{UV}-3.0153
   {c_{2_{ q4}}}^{UV}) N^{\frac{8 {fr}_h}{3}-\frac{1}{10}}}{{M^{UV}}^2 \alpha _{\theta _1} \alpha _{\theta _2}^2}\Biggr)^2.
\end{eqnarray}
Assuming the experimental value for $\frac{\Gamma_{G_E\rightarrow\rho+2\pi}}{m_0}$ - not yet known in \cite{PDG-2018} - is $10^{-5 + {\rm required}_2}$, (\ref{Gamma-3body-over-m0}) for $N=10^2$, implies the following constaint:
\begin{eqnarray}
\label{constraint_3body_decay-glueball}
& & \Biggl.\Biggl(-\frac{2.23\times 10^7 {f_{r_h}}^2 {g_s}^3 M {N_f}^2 \sqrt{{\omega_2}} {c_{\psi_1}}^2 {c_{1_{ q4}}} \log ^2\left(\sqrt[3]{N}\right)
   N^{\frac{fr_ h}{3}}}{\alpha _{\theta _1} \alpha _{\theta _2}^2}\nonumber\\
& & \hskip -0.7in -\frac{460099 {f_{r_h}} \sqrt{{g^{UV}_s}} {\log (N)}^5 {N_f^{UV}}^2 \log \left(\sqrt[3]{N}\right) {{c^{UV}_{2_{ \psi_1}}}}^2 (  {c_{2_{q1}}}^{UV}-3.0153
   {c_{2_{ q4}}}^{UV}) N^{\frac{8 {fr}_h}{3}-\frac{1}{10}}}{{M^{UV}}^2 \alpha _{\theta _1} \alpha _{\theta _2}^2}\Biggr|_{N=10^2} = 10^{{\rm required}_2}.\nonumber\\
   &&
\end{eqnarray}

\subsection{Indirect Decay of Glueball to $4\pi$ }

The relevant interaction Lagrangian is given by:
\begin{eqnarray}
\label{interaction action}
&&\hskip -0.5in{\rm S}_{int}= {\cal T}{\rm Str}\int \left(\frac{1}{T_h}\right) d^{3}x\Bigg[c_{3}\rho_{\mu}^{2}G_E + c_{4}\rho_\mu\rho_\nu \frac{\partial^{\mu}\partial^{\nu}}{M^2}G + c_{5}\tilde{F}_{\mu\nu}\tilde{F}^{\mu\nu}G_E
+ c_{6}\tilde{F}_{\mu\rho}\tilde{F}_\nu^{\ \rho}\frac{\partial^{\mu}\partial^{\nu}}{M^2}G_E\nonumber\\
&& + \iota c_{7}\partial_{\mu}\pi [\pi,\rho^\mu]G_E + \iota c_{8}\partial_{\mu}\pi [\pi,\rho_\nu]\frac{\partial^{\mu}\partial^{\nu}}{M^2}G_E + c_{9}(Z)\rho_\mu \tilde{F}_{\nu}^{\ \mu}\frac{\partial^\nu G_E}{M^2}\nonumber\\
&&+c_{10}\tilde{F}_{\mu\nu}\tilde{F}^{\mu\nu}G_E +c_{11}\partial_{\mu}\pi\partial^{\mu}\pi G_E +c_{12}\rho_{\mu}\rho^{\mu} G_E +\iota c_{13}\partial_{\mu}\pi [\pi,\rho^\mu]G_E \Bigg].
\end{eqnarray}
At LO order glueball decay into four pions is a successive decay process which involve two process. First process is  $G_E\rightarrow \rho \rho$ in which each
$\rho$ meson further decays into two $\pi$ each and, the second process is $G_E\rightarrow \pi\pi\rho$ in which $\rho$ meson further decays into two $\pi$. The LO order decay amplitude of a glueball into four pions involves two pairs of pions with different isospin index.If  {\cal M} is the amplitude for $G_E\rightarrow 2\pi^{a}2\pi^{b}$ where $a\neq b$ then without any loss of generality we can set $a=1$ and $b=2$. The total decay rate is given by:
\begin{eqnarray}
&&\Gamma =\frac{3}{4}\frac{1}{2M}\int d\Phi_{4}|M|^{2}
\end{eqnarray}
where the factor of 3/4 is due to a factor of 3 for the three different pairs of isospin and 4 is due to the symmetry factor of two pairs of identical particles.
The full four body phase space in 2+1 dimension is given by
\begin{eqnarray}
\int d\Phi_{4}=\prod_{i=1}^{4} \frac{d^{2}\vec{k_i}}{(2\pi)^2 2E_i}(2\pi)^3\delta(k^\mu -k_1^\mu -k_2^\mu -k_3^\mu -k_4^\mu)
\end{eqnarray}

The amplitude corresponding to process $G_E\rightarrow \rho\pi\pi\rightarrow\pi\pi\pi\pi$ (Fig. 2(a)) is given as under:
\begin{eqnarray}
&& {\cal M}_{(a)}={\cal T}^{2}\Bigg(\frac{\pi L^{2}}{r_h} \Bigg)^{2}8 c_{8}c_{16} \Bigg( \Delta_{\rho}^{\mu\mu\prime}(k_{2}+k_{4})(-k_{(1)\mu}k_{(2)\mu\prime}+k_{(1)\mu}k_{(4)\mu\prime}+k_{(3)\mu}k_{(2)\mu\prime}-k_{(3)\mu}k_{(4)\mu\prime})\nonumber\\
&&+\Delta_{\rho}^{\mu\mu\prime}(k_{3}+k_{4})(-k_{(1)\mu}k_{(3)\mu\prime}+k_{(1)\mu}k_{(4)\mu\prime}+k_{(2)\mu}k_{(3)\mu\prime}-k_{(2)\mu}k_{(4)\mu\prime})\nonumber\\
&&+\Delta_{\rho}^{\mu\mu\prime}(k_{1}+k_{2})(-k_{(1)\mu}k_{(3)\mu\prime}+k_{(4)\mu}k_{(1)\mu\prime}+k_{(2)\mu}k_{(3)\mu\prime}-k_{(2)\mu}k_{(4)\mu\prime})\nonumber\\
&&+\Delta_{\rho}^{\mu\mu\prime}(k_{1}+k_{3})(-k_{(1)\mu}k_{(2)\mu\prime}+k_{(1)\mu}k_{(4)\mu\prime}+k_{(3)\mu}k_{(2)\mu\prime}-k_{(3)\mu}k_{(4)\mu\prime})\Bigg),
\end{eqnarray}
where $\Delta_{\rho}^{\mu\mu\prime}$ corresponds to the vector meson propagator  given as under:
\begin{eqnarray}
& & \Delta_{\rho}^{\mu\mu\prime}(k_{i}+k_{j})=\frac{\delta _0^{\mu  }\delta _0^{\nu } \left(-\delta _{\nu }^{\mu '}-\frac{\left(k_i+k_j\right){}^{\mu '} \left(k_{{i}}+k_{{j}}\right)_{\nu}}{M_{\rho
   }^2}\right)}{\left(k_i+k_j\right){}^2+i \Gamma _{\rho } M_{\rho }^2-M_{\rho }^2}.
\end{eqnarray}

 The amplitude corresponding to the second process $G_E\rightarrow \rho\rho \rightarrow \pi\pi\pi\pi$ (Fig. 2(b)) is given as under:
 \begin{eqnarray}
 &&{\cal M}_{(b)}=-\frac{16 \pi ^3 {c_{16}}^2 L^6 {\cal T}^3}{r_h ^3}
 \Biggl(\left({A\eta }^{\sigma \gamma}+B^{\sigma \gamma }\right) \Delta_{\rho}\left(k_3+k_4\right)_{\sigma }^{\mu } \Delta_{\rho }\nonumber\\
 & & \times\left(k_1+k_2\right)_{\gamma}^{\mu '} \left(-k_{{\mu 1}} k_{{\mu 3}'}+k_{{\mu 1}}
   k_{{\mu 4}'}+k_{{\mu 2}} k_{{\mu 3}'}-k_{{\mu 2}} k_{{\mu 4}'}\right)+k_2\longleftrightarrow k_3\Biggr),
\end{eqnarray}
where the expression for $A$ and $B^{\mu\nu}$ are same as given in the section for $G_E\rightarrow \rho\rho$ decay with appropriate momentum substitution. For $f0[1710]:M_g>2M_\rho$,  (b) dominates.

So the total decay width can be approximated by
\begin{eqnarray}
\label{Gamma-indirect-4pi-decay-glueball-i}
&& \hskip -0.7in \frac{\Gamma_{G_E\rightarrow2\rho\rightarrow4\pi}}{m_0} \sim c_6^2c_{16}^4\nonumber\\
& & \hskip -0.7in \sim 10^{-10}N^{\frac{13}{5}}\Biggl(-\frac{2.23007\times 10^7 {f_{r_h}}^2 {g_s}^3 M {N_f}^2 \sqrt{{\omega_2}} {c_{\psi_1}}^2 {c_{1_{ q4}}} \log ^2\left(\sqrt[3]{N}\right)
   N^{\frac{fr_ h}{3}}}{\alpha _{\theta _1} \alpha _{\theta _2}^2}\nonumber\\
& & \hskip -0.7in -\frac{460099. {fr_h} \sqrt{{g^{UV}_s}} {\log (N)}^5 {N_f^{UV}}^2 \log \left(\sqrt[3]{N}\right) {{c^{UV}_{2_{ \psi_1}}}}^2 (  {c_{2_{q1}}}^{UV}-3.015
   {c_{2_{ q4}}}^{UV}) N^{\frac{8 {fr}_h}{3}-\frac{1}{10}}}{{M^{UV}}^2 \alpha _{\theta _1} \alpha _{\theta _2}^2}\Biggr)^2.
\end{eqnarray}
Assuming the experimental value of $\frac{\Gamma_{G_E\rightarrow2\rho\rightarrow4\pi}}{m_0}$ - not currently known \cite{PDG-2018} - is $10^{-5 + {\rm required}_3}$, one obtains, for $N=10^2$ the same constraint as (\ref{Constraint_direct_4pi0_decay}) with ${\rm required}_2$ replaced by ${\rm required}_3$; we expect ${\rm required}_2\sim{\rm required}_3$.


\begin{figure}[t]
	\begin{center}
		\begin{tabular}{c c}
			\includegraphics[width=5.5cm]{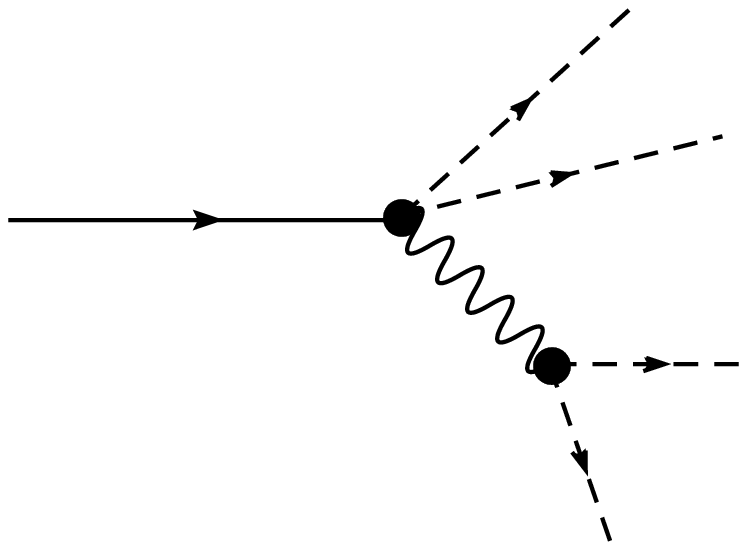}&
			\hspace{1 cm}
			\includegraphics[width=5.5cm]{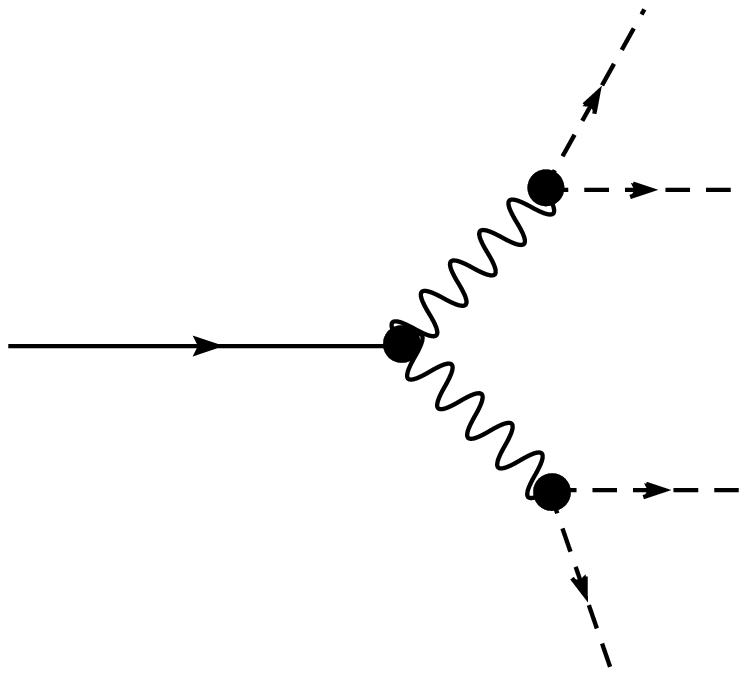}\\
			(a) & (b)
		\end{tabular}
		\caption{$G_E\rightarrow4\pi$}
	\end{center}
	
\end{figure}

\section{Summary and Discussion}

We studied (exotic) scalar glueball $0_E^{++}$-meson interaction and (exotic) scalar glueball decays at tree level wherein the glueballs corresponded to metric fluctuations of the M theory uplift of \cite{metrics}'s UV-complete type IIB holographic dual of large-N thermal QCD at finite coupling - MQGP limit of \cite{MQGP} - and the mesons corresponded to gauge fluctuations on the world-volume of type IIA flavor $D6$-branes (involving pull-back apart from that of the type IIA B, the perturbed type IIA metric corresponding to a circle reduction of the aforementioned perturbed M theory metric). The following is a summary of the main results of this paper, all of which correspond to finite gauge/string coupling on the gauge/gravity side.
\begin{itemize}
	\item
	We obtained $0_E^{++}-\rho,\pi$ interaction Lagrangian linear in the exotic scalar glueball and up to quartic in $\pi$s wherein the coefficients are given by radial integrals of components of the M theory metric that corresponds to the uplift of \cite{MQGP}'s SYZ type IIA mirror of \cite{metrics}, and perturbations thereof. This is rather gratifying as one is able to get the coupling constants from the underlying fundamental M theory.
	
	\item
	Assuming $M_G > 2M_\rho$, the following is a summary of how our calculations with appropriate choice of constants of integration appearing in the solutions to the metric fluctuations and meson radial profile functions, can be made to match PDG results on scalar glueball decay widths, exactly.
	\begin{enumerate}

\item
From (\ref{norm_psi}) using (\ref{V_V1_V2-defs}) the normalization condition for $\psi_1(Z)$ implies the following quadratic constaint on $c_{\psi_1}$ and $c_{2\ \psi_1}^{\rm UV}$:
\begin{eqnarray}
\label{psi_1_normalization}
& & V\Biggl(\frac{5\times10^{-5}}{\alpha_{\theta_1}\alpha_{\theta_2}^2}g_s^2MN^{\frac{4}{5}}N_f^2\sqrt{|\omega_2|}\left(1433.4 + b^2(-2067.37 + \omega_2)\right)(c_{\psi_1})^2\log r_h\nonumber\\
 & & + \frac{244.91 \log r_hg_s^{\rm UV 2}M^{\rm UV}N^{\frac{4}{5}}N_f^{\rm UV}\ ^2c_{2\ \psi_1}^{\rm UV}\ ^2}{\alpha_{\theta_1}\alpha_{\theta_2}^2}\Biggr)=1.
\end{eqnarray}

\item
From (\ref{norm_scalar}) using (\ref{V_V1_V2-defs}) the normalization condition for $\phi_0(Z)$ implies the following quadratic constraint on ${\cal C}_{\phi_0}$ and ${\cal C}_{\phi_0}^{\rm UV}$:
\begin{eqnarray}
\label{phi_0_normalization}
& & \frac{V}{2}\Biggl(\frac{5.51\times10^{-9}{\cal C}_{\phi_0}^2N^{\frac{1}{5}}(0.03 + 0.042 b^2)\alpha_{\theta_1}\alpha_{\theta_2}^2}{g_s r_h^2\log r_h M N_f^2}\nonumber\\
& & + \frac{793.58 {\cal C}_{\phi_0}^{\rm UV\ 2}g_s^{\rm UV}M^{\rm UV}N_f^{\rm UV}r_h^2\log r_h}{N^{\frac{1}{5}}\alpha_{\theta_1}\alpha_{\theta_2}^2}\Biggr)=1.
\end{eqnarray}

		\item
	From (\ref{Gto2pi-decay-width}),	${\cal C}_{\phi_0}^2 c_{1\ q_4}$ and $\left({\cal C}_{\phi_0}\ ^{\rm UV}({\cal C}_{\phi_0})\right)^2 \left({c_{2_{q1}}}^{\rm UV}-3.015{c_{2_{q4}}}^{\rm UV}\right)$  can be adjusted to reproduce the PDG value of $\Gamma_{G_E\rightarrow2\pi}$ {\it exactly}.
		
		\item
		Requiring $\Gamma_{G_E\rightarrow2\rho} = \Gamma_{G_E\rightarrow4\pi}$ yields: $c_4 \approx \frac{3}{4}c_6 m_0^2\left(\frac{r_h}{\pi\sqrt{4\pi g_s N}}\right)^2$, the glueball mass written as
		$M_g = m_0 \frac{r_h}{\pi\sqrt{4 \pi g_s N}}$ \cite{Sil_Yadav_Misra-EPJC-17}. Assuming $\omega_2\equiv {\cal O}(1)$, writing $c_{1\ q_4} = N^{-\alpha_4} (\alpha_4\geq1), M_g = m_\rho \frac{r_h}{\pi\sqrt{4 \pi g_s N}}$ and setting $m_0=1710, m_\rho=775$ one sees theat the aforementioned relation between $c_4$ and $c_6$ implies: $g_s = \frac{N^{\frac{2(-9+35f_{r_h}+15\alpha_4)}{75}}\left(\frac{N_f^{\rm UV}}{M^{\rm UV}}\right)^{\frac{4}{5}}\left(\frac{\left({c_{2_{\ \psi_1}}}^{UV}\ \right)^4}{\left({\cal O}(1){c_{2_{\ q1}}}^{UV} - {\cal O}(1){c_{2_{\ q4}}}^{UV}\right)^2}\right)^{\frac{1}{5}}}{{\cal O}(1)\left(f_{r_h}^2M^2N_f^4\omega_2c_{\psi_1}^4\right)^{\frac{1}{5}}}$, which can be made to be finite as part of the MQGP limit.

		\item
		The combination of constants of integration appearing in the solutions to the EOMS of $\phi_0(Z), \psi_1(Z)$ in the IR and UV, using (\ref{psi_1_normalization}): ${\cal C}_{\phi_0}^2c_{\ \psi_1}$ and $\left({\cal C}_{\phi_0}\ ^{\rm UV}({\cal C}_{\phi_0})\right)^2 c_{2_{\ \psi_1}}^{\rm UV}(c_{\ \psi_1})$, can be adjusted to reproduce the PDG value of $\Gamma_{\rho\rightarrow2\pi}$ {\it exactly}.

		\item
	From (\ref{constraint-rho-2pi-decay}), (\ref{Constraint_direct_4pi0_decay}) and (\ref{Gamma-indirect-4pi-decay-glueball-i}), and also using (\ref{psi_1_normalization}) as well as (\ref{phi_0_normalization}), we note that	the combination of constants of integration appearing in the solutions to the EOMS of $\phi_0(Z), \psi_1(Z)$ and $q_{1,2,3,4,5,6}(Z)$ in the IR and UV:\\
		-- involving  ${\cal C}_{\phi_0}^4 c_{1\ q_4}$ and $\left({\cal C}_{\phi_0}\ ^{\rm UV}({\cal C}_{\phi_0})\right)^4 \left({c_{2_{q1}}}^{\rm UV}-3.015{c_{2_{q4}}}^{\rm UV}\right)$ 	appearing in $\Gamma_{G_E\rightarrow4\pi^0}$ \\
		-- involving $c_{1\ q_4}\left(c_{\psi_1}\right)^2$ and $c_{2\ \psi_1}\ ^{\rm UV}(c_{\psi_1})\left({c_{2_{q1}}}^{\rm UV}-3.015{c_{2_{q4}}}^{\rm UV}\right)$ appearing in
		$\Gamma_{G_E\rightarrow\rho+2\pi} \approx \Gamma_{G_E\rightarrow2\rho\rightarrow4\pi}$\\
can be tuned and equality of these two combinations can be effected such that one can reproduce the PDG value of $\Gamma_{G_E\rightarrow4\pi^0} = \Gamma_{G_E\rightarrow\rho+2\pi} \approx \Gamma_{G_E\rightarrow2\rho\rightarrow4\pi}$ {\it exactly}.
		
	\end{enumerate}	
\end{itemize}

\section*{Acknowledgment}

VY is supported by a Junior Research Fellowship (JRF) from the University Grants Commission,
Govt. of India. AM was partly supported by IIT Roorkee. He
would also like to thank the theory group at McGill University, and Keshav Dasgupta in particular, for the wonderful hospitality and support during the final stages of the work and for discussions related to Appendix A. We would like to thank Karunava Sil for valuable participation in the earlier stages of this work, and Sharad Mishra in verifying some of the results in {\bf 6.1} as part of his Masters project. We would like to thank M.Dhuria for bringing \cite{Murayama-Phase-Space} to our attention. AM would like to dedicate this paper to the memory of his Ph.D. advisor, the late, Professor D.S.Koltun, a nuclear theorist, who initiated him into Hadronic Physics via Chiral Perturbation Theory.

\appendix
\section{M theory Metric Components}
\setcounter{equation}{0} \seceqaa

Near  $\theta_{1}=\alpha_{\theta_{1}}N^{\frac{-1}{5}}$, $\theta_{2}=\alpha_{\theta_{2}}N^{\frac{-3}{10}}$, $\phi_{1,2}=0/2\pi$ and $\psi=0/4\pi$, defining the local $T^3(x,y,z)$ coordinates as:
{\footnotesize
	\begin{eqnarray}
	\label{}
	&& x=\sqrt{h_2}4^{1/4}\pi^{1/4}g_{s}^{1/4}N^{1/20}\alpha_{\theta_{1}}\phi_{1},\ \ y=\sqrt{h_4}4^{1/4}\pi^{1/4}g_{s}^{1/4}N^{1/20}\alpha_{\theta_{2}}\phi_{2},\ \ z=\sqrt{h_1}4^{1/4}\pi^{1/4}g_{s}^{1/4}N^{1/20}\psi,\nonumber\\
\end{eqnarray}}
$h_{1,2,4}$ are defined in \cite{metrics}, and defining:
{
\begin{eqnarray}
\label{def-A}
& & f \equiv 1 - \frac{r_h^4}{r^4}\nonumber\\
&&A(r)\equiv\frac{1}{4} \left(\frac{3}{\pi }\right)^{2/3} \left(-{N_f} \log \left(9 b^2 r^4 {r_h}^2+r^6\right)+\frac{8 \pi }{{g_s}}-4 N_f \log\left(\frac{\alpha_{\theta_1}\alpha_{\theta_2}}{4\sqrt{\log N}}\right)\right){}^{2/3}
	\nonumber\\
	& &  \times  \left(1-\frac{32 \pi  b {g_s} M^2 {N_f} {r_h}^2 \gamma(1 + \log r_h)}{N \left(9 b^2 {r_h}^2+r^2\right)
		\left(-{N_f} \log \left(9 b^2 r^4 {r_h}^2+r^6\right)+\frac{8 \pi }{{g_s}}-4 N_f \log\left(\frac{\alpha_{\theta_1}\alpha_{\theta_2}}{4\sqrt{\log N}}\right)\right)}\right),\nonumber\\
\end{eqnarray}}
 the M-theory metric components used in Sections {\bf 3} - {\bf 6}, are given by:
 {
	\begin{eqnarray*}
	&&G^{M}_{00}=-f(r)\frac{A(r)}{\sqrt{h}}\nonumber\\
	&&G^{M}_{ii}=\frac{A(r)}{\sqrt{h}}\ \  i=1,2,3\nonumber\\
	&& G^{M}_{rr}=\sqrt{h}\frac{A(r)}{f(r)}\nonumber\\
	&&G^{M}_{\theta_{1,2}\theta_{1,2}}=0,\ \ G^{M}_{\theta_{1}\theta_{2}}=0,\ \ G^{M}_{xr}=0,\ \ G^{M}_{r\theta_{1,2}}=0\nonumber\\
	&&G^{M}_{11x}=0,\ \ G^{M}_{11r}=0,\ \ G^{M}_{11\theta_{1,2}}=0\nonumber\\
	&&G^{M}_{11,11}=\frac{16 \pi ^{4/3} \left(\frac{64 \pi  b {g_s} M^2 {N_f} {r_h}^2 \gamma(1 + \log r_h)}{N \left(9 b^2 {r_h}^2+r^2\right)
			\left(-{N_f} \log \left(9 b^2 r^4 {r_h}^2+r^6\right)+\frac{8 \pi }{{g_s}}-4 N_f \log\left(\frac{\alpha_{\theta_1}\alpha_{\theta_2}}{4\sqrt{\log N}}\right)\right)}+1\right)}{3 \sqrt[3]{3} \left(-{N_f} \log \left(9 b^2 r^4
		{r_h}^2+r^6\right)+\frac{8 \pi }{{g_s}}-4 N_f \log\left(\frac{\alpha_{\theta_1}\alpha_{\theta_2}}{4\sqrt{\log N}}\right)\right){}^{4/3}}\nonumber\\
		&&G^{M}_{x\theta_{1}}=A(r)\frac{1}{972 \pi ^{5/4} r^2 \alpha _{\theta _1}^3 \alpha _{\theta _2}^2}\Biggl\{{g_s}^{3/4} M N^{7/20} \left(-243 \sqrt{3} \alpha _{\theta _1}^3+4 \sqrt{2} \alpha _{\theta _2}^2+81 \sqrt{2} \sqrt[5]{N} \alpha _{\theta
		_1}^2\right)\nonumber\\
	& & \times \Biggl({g_s} {N_f} \left(3 a^2-r^2\right) \log (N) (2 \log (r)+1)+\log (r) \Biggl(4 {g_s} {N_f} \left(r^2-3 a^2\right) \log
	\left(\frac{1}{4} \alpha _{\theta _1} \alpha _{\theta _2}\right)\nonumber\\
& & -24 \pi  a^2+r^2 (8 \pi -3 {g_s} {N_f})\Biggr)
	\nonumber\\
	& & +2 {g_s} {N_f}
	\left(r^2-3 a^2\right) \log \left(\frac{1}{4} \alpha _{\theta _1} \alpha _{\theta _2}\right)+18 {g_s} {N_f} \left(r^2-3 a^2 (6 r+1)\right) \log
	^2(r)\Biggr)\Biggr\}\nonumber\\
&&G^{M}_{x\theta_{2}}=-A(r)\frac{{g_s}^{7/4} M N^{9/20} {N_f} \log (r) \left(36 a^2 \log (r)+r\right) \left(486 \sqrt{6} \alpha _{\theta _1}^3+11 \alpha _{\theta _2}^2-324
		\sqrt[5]{N} \alpha _{\theta _1}^2\right)}{972 \sqrt{2} \pi ^{5/4} r \alpha _{\theta _1}^2 \alpha _{\theta _2}^3}\nonumber\\
	&&G^{M}_{y\theta_{1}}=A(r)\frac{1}{72 \sqrt{3} \pi ^{5/4} N^{7/20} r^2 \alpha _{\theta _1}
		\alpha _{\theta _2}}\Biggl\{{g_s}^{3/4} M \left(67 \alpha _{\theta _2}^2+81 \sqrt[5]{N} \alpha _{\theta _1}^2\right)\nonumber\\
& & \times \Biggl[{g_s} {N_f} \left(3 a^2-r^2\right) \log
	(N) (2 \log (r)+1)+\nonumber\\
	& & \log (r) \left(4 {g_s} {N_f} \left(r^2-3 a^2\right) \log \left(\frac{1}{4} \alpha _{\theta _1} \alpha _{\theta _2}\right)-24
	\pi  a^2+r^2 (8 \pi -3 {g_s} {N_f})\right)\nonumber\\
	& & +2 {g_s} {N_f} \left(r^2-3 a^2\right) \log \left(\frac{1}{4} \alpha _{\theta _1} \alpha
	_{\theta _2}\right)+18 {g_s} {N_f} \left(r^2-3 a^2 (6 r+1)\right) \log ^2(r)\Biggr]\Biggr\}\nonumber\\
&&G^{M}_{y\theta_{2}}=A(r)\frac{\sqrt{2} \sqrt[4]{\pi } \sqrt[4]{{g_s}} \sqrt[4]{N} \alpha _{\theta _2} \left(3 \sqrt[10]{N} \alpha _{\theta _1}-7 {h5} \alpha _{\theta
			_2}\right)}{27 \alpha _{\theta _1}^3}\nonumber\\
\end{eqnarray*}
\begin{eqnarray}
\label{M-th components}
			&&G^{M}_{z\theta_{1}}=-A(r)\frac{1}{324 \sqrt{2} \pi ^{5/4} \sqrt[20]{N} r^2 \alpha _{\theta
			_1} \alpha _{\theta _2}^2}\Biggl\{{g_s}^{3/4} M \left(49 \alpha _{\theta _2}^2+81 \sqrt[5]{N} \alpha _{\theta _1}^2\right)\nonumber\\
& & \times \Biggl[{g_s} {N_f} \left(3 a^2-r^2\right) \log
	(N) (2 \log (r)+1)\nonumber\\
	& & +\log (r) \left(4 {g_s} {N_f} \left(r^2-3 a^2\right) \log \left(\frac{1}{4} \alpha _{\theta _1} \alpha _{\theta _2}\right)-24
	\pi  a^2+r^2 (8 \pi -3 {g_s} {N_f})\right)\nonumber\\
	& & +2 {g_s} {N_f} \left(r^2-3 a^2\right) \log \left(\frac{1}{4} \alpha _{\theta _1} \alpha
	_{\theta _2}\right)+18 {g_s} {N_f} \left(r^2-3 a^2 (6 r+1)\right) \log ^2(r)\Biggr]\Biggr\}\nonumber\\
	&&G^{M}_{z\theta_{2}}=-A(r)\frac{{g_s}^{7/4} M {N_f} \log (r) \left(36 a^2 \log (r)+r\right) \left(324 \sqrt[4]{N} \alpha _{\theta _1}^2+169 \sqrt[20]{N} \alpha _{\theta
			_2}^2\right)}{648 \sqrt{2} \pi ^{5/4} r \alpha _{\theta _2}^3}\nonumber\\
	&&G^{M}_{xy}=A(r)\Bigg\{\frac{2 \sqrt{\frac{2}{3}} N^{7/10}}{9 \alpha _{\theta _1}^2 \alpha _{\theta _2}}-\frac{\sqrt{\frac{2}{3}} \sqrt{N} \left(243 \sqrt{6} \alpha _{\theta
			_1}^3+118 \alpha _{\theta _2}^2\right)}{729 \alpha _{\theta _1}^4 \alpha _{\theta _2}}\Bigg\}\nonumber\\
	&&G^{M}_{xz}=-A(r)\frac{2 N^{4/5} \left(-243 \sqrt{6} \alpha _{\theta _1}^3+8 \alpha _{\theta _2}^2+162 \sqrt[5]{N} \alpha _{\theta _1}^2\right)}{6561 \alpha _{\theta
			_1}^4 \alpha _{\theta _2}^2}\nonumber\\
	&&G^{M}_{yz}=A(r)\Bigg\{\frac{14 \sqrt{\frac{2}{3}} \sqrt[10]{N} \alpha _{\theta _2}}{243 \alpha _{\theta _1}^2}-\frac{\sqrt{\frac{2}{3}} N^{3/10}}{3 \alpha _{\theta _2}}\Bigg\}
	\end{eqnarray}}

\section{Schr\"{o}dinger-Like Potential for the Radial Profile Function for $\rho$ Mesons}
\setcounter{equation}{0} \seceqbb

The Schr\"{o}dinger-like equation (\ref{near_Z=0-EOM-redefined_psi1}) satisfied by $g(Z)\equiv\sqrt{{\cal V}_1(Z)}\psi_1(Z)$ will have a potential given by:
{\scriptsize
	\begin{eqnarray}
\label{V-psi1}
	& & \hskip -0.8in V(Z) = \frac{1}{4 \left(-1+e^{4 Z}\right)^2}\Biggl\{\frac{1}{{r_h}^2}\Biggl\{3 \left(-1+e^{4 Z}\right) \Biggl(-4 m_0^2 + \frac{1}{\Delta_3}\nonumber\\
& &\hskip -0.8in \times\Biggl\{e^{-2 Z} \left(6 \left(1+e^{4 Z}\right) {g_s} {N_f} (Z+\log
   ({r_h}))^2-2 \left(4 \pi  \left(1+e^{4 Z}\right)+{g_s} {N_f} \left(e^{4 Z} (\log N -3)+\log N +3\right)\right) (Z+\log
   ({r_h}))-\left(-1+e^{4 Z}\right) ({g_s} {N_f} \log N +4 \pi )\right)\nonumber\\
   & &\hskip -0.8in \times \Biggl[216 e^Z {g_s}^2 {N_f}^2 {r_h}
   (Z+\log ({r_h}))^3-18 {g_s} {N_f} \left(32 e^Z \pi  {r_h}+{g_s} {N_f} \left(4 e^Z {r_h} (2 \log
   (N)+3)-1\right)\right) (Z+\log ({r_h}))^2\nonumber\\
   & &\hskip -0.8in +6 \left({g_s}^2 \left(4 e^Z {r_h} \log ^2 N+\left(24 e^Z {r_h}-2\right) \log
   (N)+3\right) {N_f}^2+8 {g_s} \pi  \left(4 e^Z {r_h} (\log N +3)-1\right) {N_f}+64 e^Z \pi ^2 {r_h}\right) (Z+\log
   ({r_h}))-32 \pi ^2 \left(12 e^Z {r_h}-1\right)\nonumber\\
   & &\hskip -0.8in -8 {g_s} {N_f} \pi  \left(\left(24 e^Z {r_h}-2\right) \log
   (N)+3\right)-{g_s}^2 {N_f}^2 \left(\left(24 e^Z {r_h}-2\right) \log ^2(N)+6 \log N -9\right)\Biggr]\Biggr\}\nonumber\\
   & &\hskip -0.8in +\frac{1}{\Delta_2}\Biggl\{2 e^{-2 Z} \Biggl[\Biggl(-6 \left(-1+e^{4
   Z}\right) {g_s} {N_f} (Z+\log ({r_h}))^2+2 \left(4 \pi  \left(-1+e^{4 Z}\right)+{g_s} {N_f} \left(e^{4 Z} (\log
   (N)-6)-\log N -6\right)\right)\nonumber\\
    & & \hskip -0.8in\times(Z+\log ({r_h}))+{g_s} {N_f} \left(2 \log N +e^{4 Z} (2 \log N -3)+3\right)+8 \left(1+e^{4
   Z}\right) \pi \Biggr)\nonumber\\
   & &\hskip -0.8in \times \left(-72 e^Z {g_s} {N_f} {r_h} (Z+\log ({r_h}))^2+3 \left(32 e^Z \pi  {r_h}+{g_s}
   {N_f} \left(8 e^Z {r_h} \log N -1\right)\right) (Z+\log ({r_h}))+{g_s} {N_f} (\log N -3)+4 \pi \right)\nonumber\\
   & & \hskip -0.8in-({g_s}
   {N_f} \log N -3 {g_s} {N_f} (Z+\log ({r_h}))+4 \pi )\nonumber\\
   & & \hskip -0.8in\times \Biggl[-36 e^Z \left(-9+e^{4 Z}\right) {g_s} {N_f}
   {r_h} (Z+\log ({r_h}))^3+12 \left(4 e^Z \left(-9+e^{4 Z}\right) \pi  {r_h}+{g_s} {N_f} \left(e^{5 Z} {r_h} (\log
   (N)-18)-9 e^Z {r_h} (\log N +6)+2\right)\right) (Z+\log ({r_h}))^2\nonumber\\
   & &\hskip -0.8in +8 \left(-27 e^Z \left(-1+e^{4 Z}\right) {g_s} {N_f}
   {r_h}+4 \pi  \left(18 e^Z {r_h}+6 e^{5 Z} {r_h}-1\right)+{g_s} {N_f} \left(18 e^Z {r_h}+6 e^{5 Z}
   {r_h}-1\right) \log N \right)\nonumber\\
    & & \hskip -0.8in\times(Z+\log ({r_h}))+16 \pi  \left(-6 e^Z {r_h}+6 e^{5 Z} {r_h}+1\right)+{g_s} {N_f}
   \left(4 \left(-6 e^Z {r_h}+6 e^{5 Z} {r_h}+1\right) \log N -3 \left(3+e^{4 Z}\right)\right)\Biggr]\Biggr]\Biggr\}\Biggr) a^2\Biggr\}\nonumber\\
   & & \hskip -0.8in+4 e^{2 Z} \left(-1+e^{4
   Z}\right) m_0^2+\nonumber\\
   & & \hskip -0.8in\frac{\left(-6 \left(1+e^{4 Z}\right) {g_s} {N_f} (Z+\log ({r_h}))^2+2 \left(4 \pi  \left(1+e^{4
   Z}\right)+{g_s} {N_f} \left(e^{4 Z} (\log N -3)+\log N +3\right)\right) (Z+\log ({r_h}))+\left(-1+e^{4 Z}\right)
   ({g_s} {N_f} \log N +4 \pi )\right)^2}{\Delta_1^2}\nonumber\\
   & & \hskip -0.8in-\frac{1}{\Delta_1}\Biggl\{4 \left(-1+e^{4 Z}\right) \Biggl[-6 \left(-1+e^{4 Z}\right) {g_s} {N_f} (Z+\log ({r_h}))^2\nonumber\\
   & & \hskip -0.8in+2
   \left(4 \pi  \left(-1+e^{4 Z}\right)+{g_s} {N_f} \left(e^{4 Z} (\log N -6)-\log N -6\right)\right) (Z+\log
   ({r_h}))+{g_s} {N_f} \left(2 \log N +e^{4 Z} (2 \log N -3)+3\right)+8 \left(1+e^{4 Z}\right) \pi \Biggr]\Biggr\}\Biggr\},\nonumber\\
	& & \hskip -0.8in
	\end{eqnarray}}
where:
\begin{equation*}
\Delta_n\equiv (Z+\log ({r_h}))
   ({g_s} {N_f} \log N -3 {g_s} {N_f} (Z+\log ({r_h}))+4 \pi )^n.
\end{equation*}

\end{document}